\documentclass[a4paper,twocolumn,11pt,accepted=2024-03-20]{quantumarticle}
\pdfoutput=1
\usepackage[utf8]{inputenc}
\usepackage[english]{babel}
\usepackage[T1]{fontenc}
\usepackage{amsmath,amssymb}
\usepackage[breaklinks]{hyperref}
\usepackage{tikz}
\usepackage{lipsum}
\usepackage[numbers,sort&compress]{natbib}
\usepackage[normalem]{ulem}
\usetikzlibrary{automata}
\usetikzlibrary{quantikz}
\usetikzlibrary{trees}

\begin{document}

\title{Quantum Monte Carlo simulations for financial risk analytics:
  scenario generation for equity, rate, and credit risk factors}

\author{Titos Matsakos}
\affiliation{
  \!\!Financial Risk Analytics, Credit \& Risk Solutions, Market Intelligence,
  S\&P Global, 25 Ropemaker St, London, EC2Y~9LY, UK}
\orcid{0000-0003-0447-2147}
\email{titos.matsakos@spglobal.com}
\author{Stuart Nield}
\affiliation{
  \!\!Financial Risk Analytics, Credit \& Risk Solutions, Market Intelligence,
  S\&P Global, 25 Ropemaker St, London, EC2Y~9LY, UK}

\maketitle

\begin{abstract}
Monte Carlo (MC) simulations are widely used in financial risk management,
from estimating value-at-risk (VaR) to pricing over-the-counter derivatives.
However, they come at a significant computational cost due to the number of
scenarios required for convergence.
If a probability distribution is available, Quantum Amplitude Estimation (QAE)
algorithms can provide a quadratic speed-up in measuring its properties as
compared to their classical counterparts.
Recent studies have explored the calculation of common risk measures and the
optimisation of QAE algorithms by initialising the input quantum states with
pre-computed probability distributions.
If such distributions are not available in closed form, however, they need
to be generated numerically, and the associated computational cost may limit the
quantum advantage.
In this paper, we bypass this challenge by incorporating scenario generation ---
i.e. simulation of the risk factor evolution over time to generate probability
distributions --- into the quantum computation; we refer to this process as Quantum MC (QMC) simulations.
Specifically, we assemble quantum circuits that implement stochastic models for
equity (geometric Brownian motion), interest rate (mean-reversion models), and
credit (structural, reduced-form, and rating migration credit models) risk
factors.
We then integrate these models with QAE to provide end-to-end examples for both
market and credit risk use cases.
\end{abstract}

\section{Introduction}

Several papers have been published recently that demonstrate how quantum
algorithms can significantly speed up the solution of typical numerical problems
in quantitative finance (for reviews, see \cite{Orus+2019, Egger+2020b,
Gomez+2022, Herman+2022, WilkensMoorhouse2023, Intallura+2023, Dalzell+2023}).
In the area of financial risk management, applications include calculating
value-at-risk and expected shortfall \cite{WoernerEgger2019, Egger+2020a,
Kaneko+2021}, pricing of options \cite{Rebentrost+2018, Stamatopoulos+2020,
CarreraVazquezWoerner2021, Chakrabarti+2021, Doriguello+2021} and collateralised
debt obligations \cite{Tang+2020}, credit valuation adjustments
\cite{Alcazar+2022, HanRebentrost2022}, and calculating Greeks
\cite{Stamatopoulos+2022}.
While the current state of quantum computing hardware is years away from being
used in production systems, theoretical advances on the development and
implementation of quantum algorithms are necessary to prepare financial
institutions for the transition.
The so-called Noisy Intermediate-Scale Quantum (NISQ) technology is considered
to outperform current classical computers in the limit of $50$-$100$ qubits
(``Intermediate-Scale'') and shallow circuit depths (``Noisy'', i.e. due to gate
errors stacking up) \cite{Preskill2018}.
In this paper we consider fault-tolerant quantum devices; namely, we focus on
the representation of stochastic models by quantum gates ignoring the effect of
quantum errors.

\subsection{Monte Carlo simulations}
  \label{sec:cmc}

Financial risk management applications estimate the likelihood and size of
potential losses due to hypothetical future events, such as the adverse move of
equity prices, interest rates, credit spreads, or foreign exchange rates (market
risk), the default of a debt instrument (credit risk), and the default of a
counterparty (counterparty credit risk).
The risk is generally represented by a statistical measure that depends on the
properties of the underlying risk factors.
A typical calculation consists of the following three stages: i) generating a
probability distribution $P(\mathbf{X}_t)$ of the risk factor vector
$\mathbf{X}_t$ at a future time $t$, ii) defining a risk measure
$F(\mathbf{X}_t)$ as a function of $\mathbf{X}_t$, and iii) employing a
statistical method to estimate its value.\footnote{
  The same steps apply for the valuation of many types of exchange-traded or
  over-the-counter derivatives, in which case $F(\mathbf{X}_t)$ is the pricing
  function.}
For example, consider the value $V_t$ of a portfolio consisting of $q$ shares of
a given stock, with $S_0$ its price at $t = 0$.
Here, $\mathbf{X}_t$ is one-dimensional with its only element being $S_t$.
The price and volatility of the stock can be used as inputs to model the
evolution of $S_t$ and generate a distribution $P(S_t)$.
Suppose we define our risk measure as the expected value of the portfolio at
time $t$, $E(V_t) = qE(S_t)$; then, we can estimate $E(V_t)$ with either
analytical or numerical methods.
Apart from special cases where closed-form expressions are available, the
probabilistic nature of the risk factor evolution often requires repeated random
sampling followed by statistical estimation --- this type of numerical approach
is called Monte Carlo (MC) simulation.

Financial risk management calculations that rely on MC simulations often require
10,000 to 1,000,000 experiments to achieve the desired precision.
This relatively-slow convergence can be understood by considering the following
example: suppose we want to estimate the probability $p$ of a random variable
$X$ that takes the value $X = 1$ with $p$ and the value $X = 0$ with $1 - p$.
In an MC simulation the experiment is repeated $N$ times, obtaining $N_1$
observations of the outcome $X = 1$ and $N_0$ of the outcome $X = 0$, with
$N_1 + N_0 = N$.
This is a binomial distribution with the expected value of the number of
occurrences of $X = 1$ being $\bar{N}_1 = pN$ and the standard deviation being
$\delta N_1 = [p(1-p)N]^{1/2}$.
Since $N_1$ is known but not $p$, we can invert the expressions to approximate
$p$ based on the observed outcomes:
\begin{align}
p &\simeq \bar{p} = \frac{N_1}{N}\,, \\
\delta p &\simeq \frac{\delta N_1}{N}
  = \left[\frac{\bar{p}(1 - \bar{p})}{N}\right]^{1/2}
  \propto \left(\frac{1}{N}\right)^{1/2}\,.
\label{eq:dp}
\end{align}
From the second equation we can infer that the precision of the estimate is
inversely proportional to the square root of $N$.

\subsection{Quantum Amplitude Estimation algorithms}

Equation~(\ref{eq:dp}) implies that, in order to improve the precision by a
decimal digit, classical algorithms require 100 times more experiments;
therefore, typical values of $N$ are between $10,000$--$1,000,000$.
In quantum computing (see App.~\ref{app:qc} for an introduction and
definitions), it has been shown that Quantum Amplitude Estimation (QAE)
algorithms can achieve a quadratic speed-up as compared to classical algorithms
\cite{Brassard+2002}.
For the example of Sect.~\ref{sec:cmc}, we can estimate the probability $p$ with
a quantum computer by encoding the probability distribution of the random
variable in the quantum state of a qubit:
\begin{align}
\ket{\psi}
&= \sqrt{1-p}\ket{0} + \sqrt{p}\ket{1} \nonumber\\
&= \cos(\theta/2)\ket{0} + \sin(\theta/2)\ket{1}\,,
\label{eq:p_superposition}
\end{align}
such that the state $\ket{1}$ is measured with probability $p$.
However, repeating this quantum experiment multiple times (each one called a
shot) and measuring the outcomes does not bypass the classical constraint: the
precision would again be proportional to the square root of the number of shots
as in Eq.~\eqref{eq:dp}.
Instead, the quadratic gain of QAE algorithms is achieved by leveraging quantum
interference.
Specifically, consider an input qubit initialised in the superposition of
Eq.~\eqref{eq:p_superposition} and a quantum register of $n$ output qubits.
The QAE algorithm is based on phase kickback and quantum amplitude estimation
\cite{Brassard+2002}.
Essentially, the probability $p$ --- i.e. the angle $\theta$
(Eq.~\ref{eq:p_superposition}) --- is imprinted as a phase $\phi = \pm k\theta$
onto the output qubits, where $k$ takes the values $2^0,2^2,...,2^{n-1}$ for the
output qubits labelled $0,1,...,n-1$, respectively.
This operation is then followed by an inverse quantum Fourier transform which,
through interference, converts these phases to a binary number; this enables to
directly read off the angle $\theta$ and hence the value of $p$.
With this approach, the estimated value of $p$ and its error are (see
Sect.~\ref{sec:p_estimation}):
\begin{align}
p &\simeq \bar{p} = \sin^2\left(\frac{\theta}{2}\right)\,, \label{eq:psin2}\\
\delta p &\simeq \sin\theta\frac{\pi}{2N}
  \propto \frac{1}{N}\,,
\end{align}
where $N = 2^n$ is the number of possible outcomes, and, notably, $\delta p$
scales as the inverse of $N$.

\subsection{Paper motivation and structure}

The majority of QAE papers on financial risk have studied quantum circuits with
pre-computed probability distributions of the random variables,
$P(\mathbf{X}_t)$.
However, unless we can exactly encode them in a quantum state, algorithms that
involve Monte Carlo integration for state preparation \cite{GroverRudolph2002}
may not be sufficiently efficient to preserve the quadratic speed-up
\cite{Herbert2021}.
When such distributions are readily available, there are approaches to
efficiently load them on a quantum register \cite[e.g.][]{Zoufal+2019,
LiKais2021, StamatopoulosZeng2023}, as well as to simplify state preparation and
optimise circuits which can help reduce circuit complexity
\cite{CarreraVazquezWoerner2021, McArdle2022}.
However, if $P(\mathbf{X}_t)$ is not given in closed form, an MC simulation is
needed to generate the distribution numerically with a classical computer.
In these cases, even though QAE algorithms can measure its properties with a
quadratic speed-up, the computational cost to generate $P(\mathbf{X}_t)$ ---
associated with the slow convergence of classical algorithms --- is paid
upfront.

This challenge can be bypassed by incorporating scenario generation into the
quantum circuit.
In a single shot, the exact distribution of a random variable can be represented
as a superposition of quantum states.
By avoiding loading pre-computed distributions, this approach also reduces the
dependence on classical computers, including the data transfer between classical
and quantum systems.
In this context, this paper focuses on the quantum implementation of stochastic
risk models to generate risk factor scenarios --- we refer to this process as
QMC simulations.\footnote{
  The term ``QMC simulations'' is not related to ``Quantum Monte Carlo methods''
  used to study complex quantum systems in Physics.}
The resulting distributions are then combined with QAE gates to provide
end-to-end example circuits.
We study the main risk factor groups --- equities, rates, and credit --- but not
exchange rates as they rely on similar models and/or can be derived from
interest rates.
Depending on the complexity of the stochastic differential equations that
describe the evolution of risk factors, theoretical studies have shown that MC
simulations can be further optimised by leveraging quantum-accelerated
multilevel MC methods \cite{Montanaro2015, Giles2015, An+2021}.
Here, we limit our scope to assembling QMC circuits for the foundational
stochastic models --- i.e. geometric Brownian motion, mean-reversion models,
Poisson processes, and multinomial trees --- as most risk models are based on
these or a variation thereof.\footnote{
  For an overview of risk models, as well as their limitations and extensions,
  see \cite{Hull2021}.}

The paper is structured as follows.
Section~\ref{sec:qmc} presents the main steps of QAE algorithms and their
implementation in quantum circuits.
In Sect.~\ref{sec:scenario_generation}, we assemble quantum circuits to perform
QMC simulations, namely to model the evolution of equity
(Sect.~\ref{sec:equity}), interest rate (Sect.~\ref{sec:ir}), and credit
(Sect.~\ref{sec:credit}) risk factors.
Section~\ref{sec:discussion} estimates the number of qubits and the resulting
circuit depth needed in typical financial risk use cases.
We conclude in Sect.~\ref{sec:conclusions}.

\subsection{Overview of QAE circuits}
  \label{sec:qmc}

\subsubsection{The circuit}

To estimate a statistical measure $F(X_t)\in[F_\mathrm{min},F_\mathrm{max}]$ of
a risk factor $X_t$, a QAE circuit would typically consist of:
\begin{enumerate}
\item $m$ ``risk factor'' (rf) qubits to model the distribution of $X_t$ (the
initial state of which is denoted with $\ket{0}_\mathrm{rf}^{\otimes m}$),
\item one ``risk measure'' (rm) qubit to encode the normalised value of the risk
measure\footnote{
  E.g. $f(X_t) = (F(X_t)-F_\mathrm{min})/(F_\mathrm{max}-F_\mathrm{min})$.}
$f(X_t)\in[0,1]$ in the angle $\theta\in[0,\pi]$ (the initial state of which is
denoted with $\ket{0}_\mathrm{rm}$),
\item $n$ output (out) qubits to imprint multiples of $\theta$ onto their phases
(the initial state of which is denoted with
$\ket{0}_\mathrm{out}^{\otimes n}$).\footnote{
  We use the notation
  $\ket{0}\otimes\ket{0} = \ket{0}^{\otimes2} = \ket{0}\ket{0} = \ket{00}$.}
\end{enumerate}
Without loss of generality, in the following example we choose $f(X_t) = p$,
i.e. the risk measure to be a probability of an outcome that depends on $X_t$,
i.e. $f(X_t) = p$.

The general structure of a QMC/QAE quantum circuit is:
\begin{center}
\begin{quantikz}[column sep=0.23cm]
\lstick{$\ket{0}_\mathrm{rf}^{\otimes m}$}
  & \gate{\mathcal{D}}\qwbundle[alternate]{}\gategroup[2, steps=2,
    style={dashed, rounded corners, inner xsep=0}]{QMC}
  & \ctrlbundle{1}
  & \qwbundle[alternate]{}\gategroup[3, steps=3,
    style={dashed, rounded corners, inner xsep=0}]{QAE}
  & \gate[2]{\prod\mathcal{Q}}\qwbundle[alternate]{}
  & \qwbundle[alternate]{}
  & \qwbundle[alternate]{} \\
\lstick{$\ket{0}_\mathrm{rm}\,\,$}
  & \qw
  & \gate{\mathcal{M}}\qw
  & \qw
  & \qw
  & \qw
  & \qw \\
\lstick{$\ket{0}_\mathrm{out}^{\otimes n}$}
  & \qwbundle[alternate]{}
  & \qwbundle[alternate]{}
  & \gate{\mathrm{QFT}}\qwbundle[alternate]{}
  & \ctrlbundle{-1}
  & \gate{\mathrm{QFT}^\dagger}\qwbundle[alternate]{}
  & \meter[]{}\qwbundle[alternate]{}
\end{quantikz}
\end{center}
where $\mathcal{D}$ is the gate that generates the input distribution using $m$
``risk factor'' qubits (the focus of this paper), $\mathcal{M}$ the controlled
gate that encodes the risk measure into the angle $\theta$ of the risk measure
qubit, $\prod\mathcal{Q}$ the repeated application of the controlled gate
$\mathcal{Q}$ to imprint $\theta$ onto the phases of the $n$ output qubits, and
$\mathrm{QFT}$/$\mathrm{QFT}^\dagger$ the quantum Fourier transformation and its
inverse to measure the phase of the output qubits with interference.
These circuit components are described below in more detail, see
Table~\ref{tab:notation} for state notation.

\begin{table}
\begin{center}
\begin{tabular}{l|l}
State                           & Description                              \\
\hline
$\ket{b_{m-1}...b_1b_0}$        & $m$ qubits as a binary, e.g. \ket{101}   \\
$\ket{j}$, $\ket{x}$, $\ket{z}$ & qubits as an integer, e.g. $\ket{5}$ \\
$\ket{\psi}$                    & superposition of states                  \\
$\ket{\psi}_\mathrm{in}$        & input qubit(s)                           \\
$\ket{\psi}_\mathrm{out}$       & output qubit(s)                          \\
$\ket{\psi}_\mathrm{rf}$        & ``risk factor'' qubit(s) (input)         \\
$\ket{\psi}_\mathrm{rm}$        & ``risk measure'' qubit (input)           \\
$\ket{\psi}_\mathrm{st}$        & ``state'' qubit(s)                       \\
$\ket{\psi}_\mathrm{c}$         & ``count'' qubit(s)                       \\
$\ket{\psi}_\mathrm{anc}$       & ``ancilla'' qubit(s)
\end{tabular}
\end{center}
\caption{
  Notation of states.
  \label{tab:notation}}
\end{table}

\subsubsection{$\mathcal{D}$: preparing the distribution $P(X_t)$}

The $m$ qubits, $\ket{0}_\mathrm{rf}^{\otimes m}$, can model a discrete
probability distribution of $2^m$ possible outcomes, each one with a probability
$|a_j|^2$, with $j \in \{0,1,...,2^m-1\}$.
We use the gate $\mathcal{D}$ to load the distribution of $X_t$ onto the ``risk
factor'' qubits, which can be depicted as:
\begin{center}
\begin{quantikz}
\lstick{$\ket{0}_\mathrm{rf}^{\otimes m}\,$}
  & \gate{\mathcal{D}}\qwbundle[alternate]{}
  & \qwbundle[alternate]{}\rstick{$\ket{\psi}_\mathrm{rf}$}
\end{quantikz}
\end{center}
and expressed mathematically as:
\begin{align}
&\ket{\psi}_\mathrm{rf} = \mathcal{D}\ket{0}_\mathrm{rf}^{\otimes m} \nonumber\\
&= \sum_{j=0}^{2^m-1}a_j\ket{b_{jm-1}...b_{j1}b_{j0}}_\mathrm{rf}
= \sum_{j=0}^{2^m-1}a_j\ket{j}_\mathrm{rf}\,,
\label{eq:ket_j}
\end{align}
where $j$ is an integer in the decimal number system representing the binary
number $b_{jm-1}...b_{j1}b_{j0}$, and $b_{jl} \in \{0,\,1\}$ is the $l$-th digit
of the $j$-th state.
If the distribution is pre-computed, it can be loaded by leveraging Quantum Generative Adversarial Networks \cite{Zoufal+2019} or Fourier expansion
\cite{LiKais2021}, among other approaches \cite{McArdle2022,
StamatopoulosZeng2023}.
Here, the distributions will be generated with quantum gates that implement
stochastic models for the risk factor evolution.

\subsubsection{$\mathcal{M}$: calculating the risk measure $f(X_t) = p$}

With the distribution $P(X_t)$ encoded in the state of the ``risk factor'' 
qubits, $\ket{\psi}_\mathrm{rf}$, the next step is to encode a risk measure
$f(X_t)$ --- for example a probability such as $p = P(X_t < K)$ where
$K \in [X_\mathrm{min},\,X_\mathrm{max}]$ --- in the state of the ``risk
measure'' qubit, $\ket{0}_\mathrm{rm}$.
The value of $p$ can be captured in the angle $\theta$ of a qubit with the help
of a controlled gate $\mathcal{M}$ that reads the distribution
$\ket{\psi}_\mathrm{rf}$ and encodes $p \in [0,1] \to \theta \in [0,\pi]$ into
$\ket{0}_\mathrm{rm}$.
This can be represented as:
\begin{center}
\begin{quantikz}
\lstick{$\ket{\psi}_\mathrm{rf}\,$}
  & \ctrlbundle{1}
  & \qwbundle[alternate]{}\rstick{$\ket{\psi}_\mathrm{rf}$} \\
\lstick{$\ket{0}_\mathrm{rm}$}
  & \gate{\mathcal{M}}\qw
  & \qw\rstick{$\ket{\psi}_\mathrm{rm}$}
\end{quantikz}
\end{center}
and described by the expression:
\begin{align}
  \ket{\psi}_\mathrm{in}
  &= \mathcal{M}\ket{\psi}_\mathrm{rf}\ket{0}_\mathrm{rm} \nonumber\\
  &= \sqrt{1-p}\ket{\psi_0}_\mathrm{rf}\ket{0}_\mathrm{rm}
    + \sqrt{p}\ket{\psi_1}_\mathrm{rf}\ket{1}_\mathrm{rm} \nonumber\\
  &= \sqrt{1-p}\ket{\psi_0}_\mathrm{in} + \sqrt{p}\ket{\psi_1}_\mathrm{in}\,,
\label{eq:psi_in}
\end{align}
where $\sqrt{1-p} = \cos(\theta/2)$, $\sqrt{p} = \sin(\theta/2)$, and we have
simplified notation by writing
$\ket{\psi_0}_\mathrm{in} = \ket{\psi_0}_\mathrm{rf}\ket{0}_\mathrm{rm}$ and
$\ket{\psi_1}_\mathrm{in} = \ket{\psi_1}_\mathrm{rf}\ket{1}_\mathrm{rm}$.

\subsubsection{$\prod\mathcal{Q}$ and QFT: estimation of $p$}
  \label{sec:p_estimation}

\paragraph{Amplitude amplification}

\begin{figure}[t]
  \centering
  \includegraphics{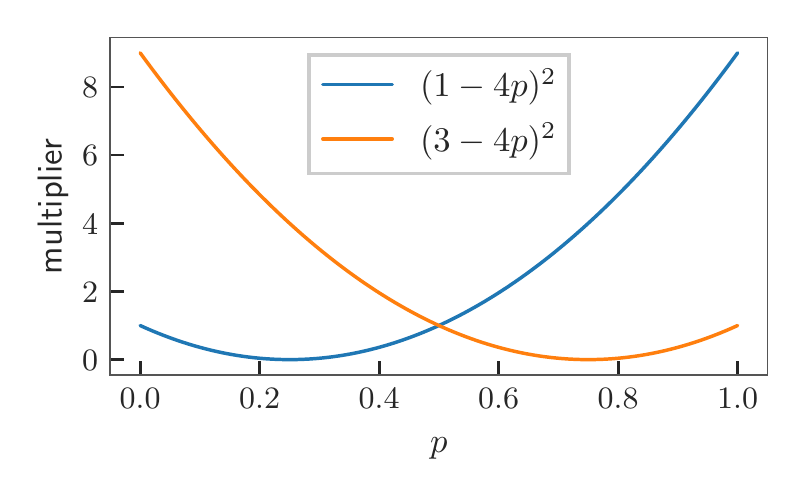}
  \caption{
    The amplification of the probabilities of the states
    $\ket{\psi_0}_\mathrm{in}$ (blue) and $\ket{\psi_1}_\mathrm{in}$ (orange)
    when $\mathcal{Q}$ acts on $\psi$ (Eq.~\ref{eq:Qpsi}).}
  \label{fig:multipliers}
\end{figure}
The next part of the circuit is the controlled $\mathcal{Q}$ gate, an operator
which is based on the Grover search algorithm \cite{Grover1996}.
$\mathcal{Q}$ consists of two reflections,
$\ket{\psi_0}_\mathrm{in} \to -\ket{\psi_0}_\mathrm{in}$ and
$\ket{\psi}_\mathrm{in} \to -\ket{\psi}_\mathrm{in}$:
\begin{align}
\mathcal{Q} &= \mathcal{Q}_\psi\mathcal{Q}_{\psi_0} \label{eq:Q}\\
  &= \big(\openone - 2\ket{\psi}_\mathrm{in}\bra{\psi}_\mathrm{in}\big)
    \big(\openone - 2\ket{\psi_0}_\mathrm{in}\bra{\psi_0}_\mathrm{in}\big)\,,
    \nonumber
\end{align}
which amplify/deamplify the amplitudes of $\ket{\psi_0}_\mathrm{in}$ and
$\ket{\psi_1}_\mathrm{in}$ depending on the value of $p$ (see
App.~\ref{app:Qpsi}):
\begin{align}
&\mathcal{Q}\ket{\psi}_\mathrm{in} = \label{eq:Qpsi}\\
&=(1-4p)\sqrt{1-p}\ket{\psi_0}_\mathrm{in}
  + (3-4p)\sqrt{p}\ket{\psi_1}_\mathrm{in} \nonumber\,.
\end{align}
Fig.~\ref{fig:multipliers} shows the multipliers $(1-4p)^2$ and $(3-4p)^2$ which
increase/decrease the probabilities of measuring $\ket{0}_\mathrm{rm}$ and
$\ket{1}_\mathrm{rm}$, respectively.\footnote{
  For $p < 1/2$, the probability of measuring the state $\ket{1}_\mathrm{rm}$ is
  amplified and that of the state $\ket{0}_\mathrm{rm}$ deamplified, and vice
  versa for $p > 1/2$.}
A key property of $\mathcal{Q}$ is that it leaves the states:
\begin{equation}
\ket{\psi_\pm}_\mathrm{in}
  = \frac{1}{\sqrt{2}}\big(\ket{\psi_1}_\mathrm{in}
    \pm i \ket{\psi_0}_\mathrm{in}\big)\,,
\end{equation}
unchanged when applied $k$ times, but introduces a global phase (see
App.~\ref{app:Qpsi_pm}):
\begin{align}
\mathcal{Q}^k\ket{\psi_\pm}_\mathrm{in}
  &= e^{\pm ik\theta}\ket{\psi_\pm}_\mathrm{in}\,.
\label{eq:Qk}
\end{align}

\paragraph{Phase kickback}

We can exploit this property by implementing a controlled gate $\mathcal{Q}$
acting on $\ket{\psi}_\mathrm{in}$, which we express as (see
App.~\ref{app:psi_psi_pm}):
\begin{align}
&\ket{\psi}_\mathrm{in} \nonumber\\
&= -i\frac{1}{\sqrt{2}}
  \left(e^{i\theta/2}\ket{\psi_+}_\mathrm{in}
  - e^{-i\theta/2}\ket{\psi_-}_\mathrm{in}\right)\,.
\end{align}
The target of this control gate is the output qubits, after first rotating their
initial state, $\ket{0...00}_\mathrm{out}$, with the $H$ operator such that it
becomes $\ket{+...++}_\mathrm{out}$.
This is equivalent to a quantum Fourier transformation, which we can represent
as:
\begin{center}
\begin{quantikz}
\lstick{$\ket{0}_\mathrm{out}^{\otimes n}$}
  & \gate{\mathrm{QFT}}\qwbundle[alternate]{}
  & \rstick{$\ket{+}_\mathrm{out}^{\otimes n}$}\qwbundle[alternate]{}
\end{quantikz}
\end{center}
and mathematically express as:
\begin{align}
\mathrm{QFT}\ket{0}^{\otimes n}_\mathrm{out}
&= \bigotimes_{l=0}^{n-1} H\ket{0}_\mathrm{out}^l
= \ket{+}^{\otimes n}_\mathrm{out}\,.
\end{align}
Then, we can leverage phase kickback to imprint the multiples of the angle
$\theta$ onto their phases.
Based on Eq.~\eqref{eq:Qk}, the phase of an output qubit
$\ket{+}_\mathrm{out}^l = \frac{1}{\sqrt{2}}(\ket{0} + \ket{1})$ will become
$\frac{1}{\sqrt{2}}(\ket{0} + e^{ik\theta}\ket{1})$ for
$\ket{\psi_+}_\mathrm{in}$, and
$\frac{1}{\sqrt{2}}(\ket{0} + e^{-ik\theta}\ket{1})$ for
$\ket{\psi_-}_\mathrm{in}$, where we set $k = 2^l$.
Namely, $k = 2^0$ for the first output qubit, $k = 2^1$ for the second, ..., and
$k = 2^{n-1}$ for the last ($n$-th).
The repeated application of the controlled gate $\mathcal{Q}$ can be represented
as:
\begin{center}
\begin{quantikz}[column sep=0.3cm]
\lstick{$\ket{\psi}_\mathrm{in}\,\,\,\,\,\,$}
  & \gate{\mathcal{Q}}\qwbundle[alternate]{}
  & \gate{\mathcal{Q}^2}\qwbundle[alternate]{}
  & \qwbundle[alternate]{}\cdots
  & \gate{\mathcal{Q}^{2^{n-1}}}\qwbundle[alternate]{}
  & \qwbundle[alternate]{}\rstick{$\ket{\psi}_\mathrm{in}$} \\
\lstick{$\ket{+}_\mathrm{out}^0\,\,$}
  & \ctrl{-1}
  & \qw
  & \qw\cdots
  & \qw
  & \qw\rstick{$\ket{\psi}_\mathrm{out}^0$}\\
\lstick{$\ket{+}_\mathrm{out}^1\,\,$}
  & \qw
  & \ctrl{-2}
  & \qw\cdots
  & \qw
  & \qw\rstick{$\ket{\psi}_\mathrm{out}^1$}\\
\wave&&&&& \\
\lstick{$\ket{+}_\mathrm{out}^{n-1}$}
  & \qw
  & \qw
  & \qw\cdots
  & \ctrl{-4}
  & \qw\rstick{$\ket{\psi}_\mathrm{out}^{n-1}$}
\end{quantikz}
\end{center}
and the state of the output qubits can be written as (see
App.~\ref{app:phase_kickback}):
\begin{align}
&\prod_{l=0}^{n-1}\mathcal{Q}^{2^l}
  \ket{\psi}_\mathrm{in}\ket{+}_\mathrm{out}^{\otimes n} = \nonumber\\
&\qquad = \ket{\psi}_\mathrm{in}\left\{\bigotimes_{l=0}^{n-1}
  \left[\frac{1}{\sqrt{2}}
    \left(\ket{0} + e^{\pm i2^l\theta}\ket{1}\right)\right]\right\}
  \nonumber\\
&\qquad = \ket{\psi}_\mathrm{in}
  \left(\frac{1}{2^{n/2}}\sum_{x=0}^{2^n-1}e^{\pm ix\theta}\ket{x}\right)\,,
\label{eq:piq}
\end{align}
where $x = \sum_{l=0}^{n-1}2^lb_l$ is the binary number $b_{n-1}...b_1b_0$
expressed in the decimal number system.

\paragraph{Interference}

The state $\sum_{x=0}^{2^n-1}a_x\ket{x}$, where
$a_x = (1/2^{n/2})e^{\pm ix\theta}$, can be transformed to the standard basis
$\sum_{z=0}^{2^n-1}a_z\ket{z}$ with an inverse quantum Fourier transform, which
depicted by the gate QFT$^\dagger$:
\begin{center}
\begin{quantikz}
\lstick{$\sum_xa_x\ket{x}$}
  & \gate{\mathrm{QFT}^\dagger}\qwbundle[alternate]{}
  & \rstick{$\sum_za_z\ket{z}$}\qwbundle[alternate]{}
\end{quantikz}
\end{center}
and mathematically expressed by (see App.~\ref{app:QFT}):
\begin{align}
\ket{\psi}_\mathrm{out}
&=\mathrm{QFT}^\dagger\frac{1}{2^{n/2}}\sum_{x=0}^{2^n-1}e^{\pm ix\theta}\ket{x}
  \nonumber\\
&=\frac{1}{2^n}
  \sum_{z=0}^{2^n-1}\sum_{x=0}^{2^n-1}e^{ix(\pm\theta-2\pi z/2^n)}\ket{z}\,,
\label{eq:psi_out}
\end{align}
which consists of a superposition of the states $\ket{z}$, where $z$ is an
integer in the decimal number system.

\paragraph{Measurement}

The measurement of the state $\sum_za_z\ket{z}$ will result in the collapse
of the superposition and will give one of the possible $z$ values, $z_0$.
This can be represented by:
\begin{center}
\begin{quantikz}
\lstick{$\sum_za_z\ket{z}$}\qwbundle[alternate]{}
  & \meter[]{}\qwbundle[alternate]{}
  & \rstick{$\ket{z_0}$}\qwbundle[alternate]{}
\end{quantikz}
\end{center}
If $\theta$ has a value such that an integer $z_0\in[0, 2^{n-1}]$ exists that
makes $\pm\theta - 2\pi z/2^n$ a multiple of $2\pi$, namely,
$z_0 = 2^n\theta/2\pi$ or $z_0 = 2^n(2\pi-\theta)/2\pi$, then
Eq.~\eqref{eq:psi_out} implies that the measured state is one of (see
App.~\ref{app:az}):
\begin{align}
\ket{z_0} &= \ket{2^n\theta/2\pi}\,, \label{eq:z0_1}\\
\ket{z_0} &= \ket{2^n(2\pi-\theta)/2\pi}\,.
\label{eq:z0_2}
\end{align}
Then, it is straightforward to calculate the probability $p$ by substituting
$\theta$ in Eq.~(\ref{eq:p_superposition}) with the measured value of $z_0$:
\begin{align}
p &= \sin^2\left(\frac{\theta}{2}\right)
= \sin^2\left(\frac{z_0}{2^n}\pi\right)\,.
\label{eq:pqmc}
\end{align}
If there is no $z_0$ such that $\pm\theta - 2\pi z_0/2^n$ is a multiple of
$2\pi$, the closest integer $z_0$ is measured with a probability
$|a_{z_0}|^2 \sim 20\%$ for either $\theta$ or $2\pi - \theta$, which gives a
total probability of approximately $40\%$ to get the closest value of $p$ (see
App.~\ref{app:az2}).

The precision with which $\theta$ is estimated is
$\delta\theta \simeq 2\pi/2^n$; therefore, the precision of $p$ is (see
App.~\ref{app:dp}):
\begin{align}
\delta p
&\simeq \sin\theta\frac{\pi}{N} \propto\frac{1}{N}\,,
\label{eq:dpqmc}
\end{align}
where $N = 2^n$ is the total number of possible outcomes.
Notably, the error in QAE decreases proportionally to $1/N$
(Eq.~\ref{eq:dpqmc}), much faster than the $1/N^{1/2}$ scaling of classical
algorithms (Eq.~\ref{eq:dp}).

Apart from the basic QAE algorithm described above which is based on quantum
phase estimation (QPE), recent studies have explored other variants of the QAE
family that do not require QPE \cite{Suzuki+2020, Grinko+2021, Plekhanov+2022}.

\section{Quantum gates and circuits for scenario generation}
  \label{sec:scenario_generation}

\subsection{Equity risk factors}
  \label{sec:equity}

Consider an equity price, $S_t$, that follows the stochastic differential
equation:
\begin{align}
dS_t = \mu S_t dt + \sigma S_t dW_t\,,
\end{align}
where $\mu$ is the drift, $\sigma$ the volatility, $t$ the time, and $W_t$ a
Wiener process.
It\^o's lemma implies that
\begin{align}
d\ln S_t = \left(\mu - \frac{\sigma^2}{2}\right)dt + \sigma dW_t\,,
\end{align}
and thus $S_t$ is log-normally distributed,
\begin{align}
S_{t+dt} = S_t e^{\left(\mu - \sigma^2/2\right)dt + \sigma dW_t}\,,
\end{align}
with expected value and variance:
\begin{align}
E(S_{t+dt}) &= S_te^{\mu dt} \,, \\
V\!ar(S_{t+dt}) &= E(S_{t+dt}^2) - E(S_{t+dt})^2 \nonumber\\
  &= S_t^2e^{2\mu dt + \sigma^2 dt} - S_t^2e^{2\mu dt}\,.
\end{align}
The path of the price can be modelled with a binomial tree with $m+1$ nodes at
times $t = \{0,\,\delta t,\,2\delta t,\,...,\,T\}$, respectively, where
$\delta t = T/m$ is the time interval.
At each node, the price $S_t$ can either go up by a factor $u$,
$S_{t+\delta t}^\mathrm{u} = S_tu$, or down by a factor $d$,
$S_{t+\delta t}^\mathrm{d} = S_td$, with probabilities $q$ and $1-q$,
respectively.
\begin{center}
\begin{tikzpicture}[->]
\node[state] at (0, 0)  (S0)  {$S_0$};
\node[state] at (3, 1)  (Su)  {$S_{\delta t}^\mathrm{u}$};
\node[state] at (3, -1) (Sd)  {$S_{\delta t}^\mathrm{d}$};
\node[state] at (6, 2)  (Suu) {$S_{2\delta t}^\mathrm{uu}$};
\node[state] at (6, 0)  (S)   {$S_{2\delta t}$};
\node[state] at (6, -2) (Sdd) {$S_{2\delta t}^\mathrm{dd}$};
\path (S0) edge node[pos=0.5, above left] {$q$}   (Su);
\path (S0) edge node[pos=0.7, below left] {$1-q$} (Sd);
\path (Su) edge node[pos=0.5, above left] {$q$}   (Suu);
\path (Su) edge node[pos=0.4, above right] {$1-q$} (S);
\path (Sd) edge node[pos=0.5, below right] {$q$}   (S);
\path (Sd) edge node[pos=0.7, below left] {$1-q$} (Sdd);
\end{tikzpicture}
\end{center}
By requiring the discrete model to have the same mean and variance as the
continuous model, we obtain the following expressions for $u$ and $q$,
respectively (see App.~\ref{app:binomial_tree}):
\begin{align}
q &= \frac{ue^{\mu \delta t} - 1}{u^2 - 1}\,, \\
u &= \frac{1}{d} = e^{\sigma\sqrt{\delta t}}\,.
\end{align}
Binomial trees are common in pricing equity derivatives \cite{Cox+1979}.

\subsubsection{$\mathcal{D}_\mathrm{eq}$: the distribution $P(S_t)$}

Since at each timestep the stock price has two possible outcomes, we can model a
transition from time period $t$ to $t+\delta t$ with a qubit in
superposition, $\ket{\psi}_\mathrm{rf}^{t + \delta t}$, such that the states
$\ket{0}$ and $\ket{1}$ represent downwards and upwards moves, respectively.
Therefore, a scenario consisting of $m$ timesteps can be modelled with $m$
qubits:
\begin{align}
\ket{\psi}_\mathrm{rf}
  &= \ket{\psi}_\mathrm{rf}^{\delta t}\ket{\psi}_\mathrm{rf}^{2\delta t}...
  \ket{\psi}_\mathrm{rf}^{m\delta t}\,,
\end{align}
where the superscripts $\delta t$, $2\delta t$, ..., $m\delta t$ are labels of
individual qubits.
For $m = 2$, the binomial tree is:
\begin{center}
\begin{tikzpicture}
\node[state] at (0, 0)  (S0)  {$\ket{\emptyset}_\mathrm{rf}^{2\delta t}\!
  \ket{\emptyset}_\mathrm{rf}^{\delta t}$};
\node[state] at (2, 1.2)  (Su)  {$\ket{\emptyset}_\mathrm{rf}^{2\delta t}\!
  \ket{1}_\mathrm{rf}^{\delta t}$};
\node[state] at (2, -1.2) (Sd)  {$\ket{\emptyset}_\mathrm{rf}^{2\delta t}\!
  \ket{0}_\mathrm{rf}^{\delta t}$};
\node[state] at (4, 2.4)  (Suu) {$\ket{1}_\mathrm{rf}^{2\delta t}\!
  \ket{1}_\mathrm{rf}^{\delta t}$};
\node[state] at (4, 0)  (S)   {$\ket{0}_\mathrm{rf}^{2\delta t}\!
  \ket{1}_\mathrm{rf}^{\delta t}
  \atop \ket{1}_\mathrm{rf}^{2\delta t}\ket{0}_\mathrm{rf}^{\delta t}$};
\node[state] at (4, -2.4) (Sdd) {$\ket{0}_\mathrm{rf}^{2\delta t}\!
  \ket{0}_\mathrm{rf}^{\delta t}$};
\node[state] at (6, 2.4)  (2)   {$\ket{j=2}$};
\node[state] at (6, 0)  (1)   {$\ket{j=1}$};
\node[state] at (6, -2.4) (0)   {$\ket{j=0}$};
\path[->] (S0)  edge node[pos=0.5, above left]  {} (Su);
\path[->] (S0)  edge node[pos=0.7, below left]  {} (Sd);
\path[->] (Su)  edge node[pos=0.5, above left]  {} (Suu);
\path[->] (Su)  edge node[pos=0.3, above right] {} (S);
\path[->] (Sd)  edge node[pos=0.5, below right] {} (S);
\path[->] (Sd)  edge node[pos=0.7, below left]  {} (Sdd);
\path[-]  (Suu) edge                               (2);
\path[-]  (S)   edge                               (1);
\path[-]  (Sdd) edge                               (0);
\end{tikzpicture}
\end{center}
where we use the notation $\ket{\emptyset} = \ket{0}$ to denote a qubit that has
not yet been through a gate.

At each time $t$, the probability $q$ can be encoded into the angle
$\theta_\mathrm{u}$ of a ``risk factor'' qubit with a $y$-rotation gate,
$R_y(\theta_\mathrm{u})$, i.e. $\ket{\psi}_\mathrm{rf}^t
= \cos(\theta_\mathrm{u}/2)\ket{0} + \sin(\theta_\mathrm{u}/2)\ket{1}$ with
$q = \sin^2(\theta_\mathrm{u}/2)$:
\begin{center}
\begin{quantikz}
\lstick{$\ket{0}_\mathrm{rf}^{\delta t}\,\,\,\,$}
  & \gate{R_y(\theta_\mathrm{u})}
    \gategroup[4, steps=1, style={dashed, rounded corners}]
    {$\mathcal{D}_\mathrm{eq}$}
  & \qw\rstick{$\ket{\psi}_\mathrm{rf}^{\delta t}$} \\
\lstick{$\ket{0}_\mathrm{rf}^{2\delta t}\,\,$}
  & \gate{R_y(\theta_\mathrm{u})}
  & \qw\rstick{$\ket{\psi}_\mathrm{rf}^{2\delta t}$} \\
\wave&& \\
\lstick{$\ket{0}_\mathrm{rf}^{m\delta t}$}
  & \gate{R_y(\theta_\mathrm{u})}
  & \qw\rstick{$\ket{\psi}_\mathrm{rf}^{m\delta t}$}
\end{quantikz}
\end{center}
The inverse of this gate, $\mathcal{D}_\mathrm{eq}^\dagger$, consists of the
same rotation gates but with a negative angle, $-\theta_\mathrm{u}$:
\begin{center}
\begin{quantikz}
\lstick{$\ket{\psi}_\mathrm{rf}^{\delta t}\,\,\,\,$}
  & \gate{R_y(-\theta_\mathrm{u})}
    \gategroup[4, steps=1, style={dashed, rounded corners}]
    {$\mathcal{D}_\mathrm{eq}^\dagger$}
  & \qw\rstick{$\ket{0}_\mathrm{rf}^{\delta t}$} \\
\lstick{$\ket{\psi}_\mathrm{rf}^{2\delta t}\,\,$}
  & \gate{R_y(-\theta_\mathrm{u})}
  & \qw\rstick{$\ket{0}_\mathrm{rf}^{2\delta t}$} \\
\wave&& \\
\lstick{$\ket{\psi}_\mathrm{rf}^{m\delta t}$}
  & \gate{R_y(-\theta_\mathrm{u})}
  & \qw\rstick{$\ket{0}_\mathrm{rf}^{m\delta t}$}
\end{quantikz}
\end{center}

\subsubsection{$\mathcal{M}_\mathrm{eq}$: risk measures $F(S_t)$}

The choice of the risk measure depends on the use case; here, we consider as
examples the probabilities of observing the maximum and minimum
values.\footnote{
  While min and max are not typical risk measures of equity price distributions,
  such functions appear in the payoff of equity options \cite{Hull2021} and thus
  their quantum implementation is important for pricing derivatives.}
For some risk measures, $m_\mathrm{anc}$ additional (``ancilla'') qubits are
required to store the intermediate values of the calculation.
We ensure that all assembled gates leave the state of these qubits to their
initial state $\ket{0}_\mathrm{anc}^{\otimes m_\mathrm{anc}}$;\footnote{
  Reverting the ancilla qubits back to their original states with a quantum gate
  rather than a ``reset'' instruction ensures the gate is reversible.
  This is because the QAE algorithm relies on the application of
  $\prod\mathcal{Q}$, which is based on the operator of Eq.~(\ref{eq:qpsi})
  (discussed later on) that includes the inverse of both the $\mathcal{D}$ and
  $\mathcal{M}$ gates.}
therefore, we do not explicitly show them when not needed.

\paragraph{Maximum}

The gate that calculates the probability of $S_T$ taking its maximum value,
$F(S_T) = P(S_\mathrm{max}) = P(\ket{1}_\mathrm{rf}^{\otimes m}) = q^m$, can be
assembled with a sequence of \textsc{and} (Toffoli) gates, such that the ``risk
measure'' qubit flips to the state $\ket{1}_\mathrm{rm}$ if all ``risk factor''
qubits are in the state $\ket{1}_\mathrm{rf}^{\otimes m}$; namely,
$\ket{\psi}_\mathrm{rm} = (1-q^m)\ket{0} + q^m\ket{1}$.
In its simplest form, an additional $m-2$ ``ancilla'' qubits are required to
store the result of the \textsc{and} operators, which are then applied a second
time to revert the ``ancilla'' qubits back to their original
$\ket{0}_\mathrm{anc}^{\otimes m_\mathrm{anc}}$ state.
\begin{center}
\begin{quantikz}[column sep=0.1cm]
\lstick{$\ket{\psi}_\mathrm{rf}^{\delta t}\quad\quad\,\,$}
  & \ctrl{1}\gategroup[12, steps=9,
    style={dashed, rounded corners, inner xsep=0}]{$\mathcal{M}_\mathrm{max}$}
  & \qw
  & \cdots\qw
  & \qw
  & \qw
  & \qw
  & \cdots\qw
  & \qw
  & \ctrl{1}
  & \qw\rstick{$\ket{\psi}_\mathrm{rf}^{\delta t}$} \\
\lstick{$\ket{\psi}_\mathrm{rf}^{2\delta t}\quad\,\,\,\,\,$}
  & \ctrl{6}
  & \qw
  & \cdots\qw
  & \qw
  & \qw
  & \qw
  & \cdots\qw
  & \qw
  & \ctrl{6}
  & \qw\rstick{$\ket{\psi}_\mathrm{rf}^{2\delta t}$} \\
\lstick{$\ket{\psi}_\mathrm{rf}^{3\delta t}\quad\,\,\,\,\,$}
  & \qw
  & \ctrl{5}
  & \cdots\qw
  & \qw
  & \qw
  & \qw
  & \cdots\qw
  & \ctrl{5}
  & \qw
  & \qw\rstick{$\ket{\psi}_\mathrm{rf}^{3\delta t}$} \\
\wave&&&&&&&&&& \\
\lstick{$\ket{\psi}_\mathrm{rf}^{(m-1)\delta t}$}
  & \qw
  & \qw
  & \cdots\qw
  & \ctrl{6}
  & \qw
  & \ctrl{6}
  & \cdots\qw
  & \qw
  & \qw
  & \qw\rstick{$\ket{\psi}_\mathrm{rf}^{(m-1)\delta t}$} \\
\lstick{$\ket{\psi}_\mathrm{rf}^{m\delta t}\,\,\,\,\,\,\,\,\,$}
  & \qw
  & \qw
  & \cdots\qw
  & \qw
  & \ctrl{5}
  & \qw
  & \cdots\qw
  & \qw
  & \qw
  & \qw\rstick{$\ket{\psi}_\mathrm{rf}^{m\delta t}$} \\
\lstick{$\ket{0}_\mathrm{rm}\quad\,\,\,\,\,\,\,$}
  & \qw
  & \qw
  & \cdots\qw
  & \qw
  & \targ{}
  & \qw
  & \cdots\qw
  & \qw
  & \qw
  & \qw\rstick{$\ket{\psi}_\mathrm{rm}$} \\
\lstick{$\ket{0}_\mathrm{anc}^1\quad\,\,\,\,\,$}
  & \targ{}
  & \ctrl{1}
  & \cdots\qw
  & \qw
  & \qw
  & \qw
  & \cdots\qw
  & \ctrl{1}
  & \targ{1}
  & \qw\rstick{$\ket{0}_\mathrm{anc}^1$} \\
\lstick{$\ket{0}_\mathrm{anc}^2\quad\,\,\,\,\,$}
  & \qw
  & \targ{}
  & \cdots\qw
  & \qw
  & \qw
  & \qw
  & \cdots\qw
  & \targ{}
  & \qw
  & \qw\rstick{$\ket{0}_\mathrm{anc}^2$} \\
\wave&&&&&&&&&& \\
\lstick{$\ket{0}_\mathrm{anc}^{m-3}\quad\,\,$}
  & \qw
  & \qw
  & \cdots\qw
  & \ctrl{1}
  & \qw
  & \ctrl{1}
  & \cdots\qw
  & \qw
  & \qw
  & \qw\rstick{$\ket{0}_\mathrm{anc}^{m-3}$} \\
\lstick{$\ket{0}_\mathrm{anc}^{m-2}\quad\,\,$}
  & \qw
  & \qw
  & \cdots\qw
  & \targ{}
  & \ctrl{-4}
  & \targ{}
  & \cdots\qw
  & \qw
  & \qw
  & \qw\rstick{$\ket{0}_\mathrm{anc}^{m-2}$}
\end{quantikz}
\end{center}
Note that if we apply $\mathcal{M}_\mathrm{max}$ twice, the ``risk measure''
qubit returns to its initial state $\ket{0}_\mathrm{rm}$, while the states of all
other qubits remain unchanged:
$\mathcal{M}_\mathrm{max}^\dagger\mathcal{M}_\mathrm{max} = \openone$; thus,
$\mathcal{M}_\mathrm{max}^\dagger = \mathcal{M}_\mathrm{max}$.

\paragraph{Minimum}

The probability of measuring the minimum value,
$F(S_T) = P(S_\mathrm{min}) = P(\ket{0}_\mathrm{rf}^{\otimes m}) = (1-q)^m$, can
be constructed following the same logic.
We first flip all ``risk factor'' qubits and then apply \textsc{and} operators
such that only the state $\ket{0}_\mathrm{rf}^{\otimes m}$ will lead to the
state $\ket{1}_\mathrm{rm}$ of the ``risk measure'' qubit:
$\ket{\psi}_\mathrm{rm} = [1-(1-q)^m]\ket{0} + (1-q)^m\ket{1}$.
The corresponding quantum gate can be assembled by leveraging
$\mathcal{M}_\mathrm{max}$:
\begin{center}
\begin{quantikz}
\lstick{$\ket{\psi}_\mathrm{rf}^{\delta t}\,\,\,\,\,$}
  & \gate{X}\gategroup[7, steps=3, style={dashed, rounded corners}]
    {$\mathcal{M}_\mathrm{min}$}
  & \gate[7]{\mathcal{M}_\mathrm{max}}
  & \gate{X}
  & \qw\rstick{$\ket{\psi}_\mathrm{rf}^{\delta t}$} \\
\wave&&&&& \\
\lstick{$\ket{\psi}_\mathrm{rf}^{m\delta t}\,$}
  & \gate{X}
  & \qw
  & \gate{X}
  & \qw\rstick{$\ket{\psi}_\mathrm{rf}^{m\delta t}$} \\
\lstick{$\ket{0}_\mathrm{rm}\,\,\,\,$}
  & \qw
  & \qw
  & \qw
  & \qw\rstick{$\ket{\psi}_\mathrm{rm}$} \\
\lstick{$\ket{0}_\mathrm{anc}^1\,\,\,$}
  & \qw
  & \qw
  & \qw
  & \qw\rstick{$\ket{0}_\mathrm{anc}^1$} \\
\wave&&&&& \\
\lstick{$\ket{0}_\mathrm{anc}^{m-2}$}
  & \qw
  & \qw
  & \qw
  & \qw\rstick{$\ket{0}_\mathrm{anc}^{m-2}$}
\end{quantikz}
\end{center}
Here too, we notice that the inverse gate is
$\mathcal{M}_\mathrm{min}^\dagger = \mathcal{M}_\mathrm{min}$.

\paragraph{Other risk measures}

Gates for other risk measures can be assembled following similar logic, e.g. for
value-at-risk and expected shortfall see Refs.~\cite{WoernerEgger2019,
Egger+2020a}.
In Sect.~\ref{sec:credit} we show an example of probability distributions that
involve inequalities, e.g. $P(S_T \geq u^{j_T}S_0)$, where $j_T$ is a specified
number of upwards moves.
Generally, any risk measure can be calculated by using quantum gates for 
arithmetic operations and comparisons \cite{Vedral+1996, OliveiraRamos2007}.

\subsubsection{Phase estimation}

To assemble the gate $\mathcal{Q} = \mathcal{Q_\psi}\mathcal{Q}_{\psi0}$, we
represent the operator $\mathcal{Q}_{\psi0}
= \openone - 2\ket{\psi_0}_\mathrm{in}\bra{\psi_0}_\mathrm{in}$ with a gate that
flips the sign of the state
$\ket{\psi_0}_\mathrm{in} = \ket{\psi_0}_\mathrm{rf}\ket{0}_\mathrm{rm}$:
\begin{center}
\begin{quantikz}
\lstick{$\ket{\psi_0}_\mathrm{rf}\,\,$}
  & \qwbundle[alternate]{}
    \gategroup[2, steps=3, style={dashed, rounded corners}]{$Q_{\psi0}$}
  & \qwbundle[alternate]{}
  & \qwbundle[alternate]{}
  & \qwbundle[alternate]{}\rstick{\quad$\ket{\psi_0}_\mathrm{rf}$} \\
\lstick{$\ket{0}_\mathrm{rm}$\,\,\,\,}
  & \gate{X}\qw
  & \gate{Z}\qw
  & \gate{X}\qw
  & \qw\rstick{$-\ket{0}_\mathrm{rm}$}
\end{quantikz}
\end{center}
\begin{align}
\mathcal{Q}_{\psi0}\ket{\psi_0}_\mathrm{in}
  &= -\ket{\psi_0}_\mathrm{in} \,,\\
\mathcal{Q}_{\psi0}\ket{\psi_1}_\mathrm{in}
  &= \ket{\psi_1}_\mathrm{in}\,.
\end{align}
For $\mathcal{Q}_\psi$, we decompose the operator as (see
App.~\ref{app:q_decomp}):
\begin{align}
\mathcal{Q}_\psi
&= \mathcal{M}\mathcal{D}Q_{00}\mathcal{D}^\dagger\mathcal{M}^\dagger\,,
\label{eq:qpsi}
\end{align}
where $\mathcal{Q}_{00} = \openone - 2\ket{0}_\mathrm{in}\bra{0}_\mathrm{in}$ is
a reflection of the initial state
$\ket{0}_\mathrm{in} = \ket{0}_\mathrm{rf}^{\otimes m}\ket{0}_\mathrm{rm}$:
\begin{align}
\mathcal{Q}_{00}\ket{0}_\mathrm{in}
  &= -\ket{0}_\mathrm{in} \,,\\
\mathcal{Q}_{00}\ket{\psi \neq 0}_\mathrm{in}
  &= \ket{\psi}_\mathrm{in}\,.
\end{align}
To apply a reflection only if all input qubits are in the state $\ket{0}$, we
first flip them with \textsc{not} gates, we then operate with a sequence of
\textsc{and} gates storing the result in the last ``ancilla'' qubit, and finally
we apply a controlled $Z$ gate to flip the sign of the ``risk measure'' qubit if
the last ``ancilla'' qubit was in the state $\ket{1}$.
Operating with the \textsc{and} and \textsc{not} gates again brings all qubits
to their initial state (apart from the sign of the ``risk measure'' qubit
depending on the initial state).
\begin{center}
\begin{quantikz}[column sep=0.2cm]
\lstick{$\ket{0}_\mathrm{rf}^{\delta t}\quad\,\,$}
  & \gate{X}\gategroup[8, steps=5,
    style={dashed, rounded corners, inner xsep=0}]{$\mathcal{Q}_{00}$}
  & \gate[8]{\mathcal{\,A\,}}
  & \qw
  & \gate[8]{\mathcal{A}^\dagger}
  & \gate{X}
  & \qw\rstick{\quad$\ket{0}_\mathrm{rf}^{\delta t}$} \\
\lstick{$\ket{0}_\mathrm{rf}^{2\delta t}\,\,\,\,\,$}
  & \gate{X}
  & \qw
  & \qw
  & \qw
  & \gate{X}
  & \qw\rstick{\quad$\ket{0}_\mathrm{rf}^{2\delta t}$} \\
\wave&&&&&&&&&& \\
\lstick{$\ket{0}_\mathrm{rf}^{m\delta t}\,\,\,$}
  & \gate{X}
  & \qw
  & \qw
  & \qw
  & \gate{X}
  & \qw\rstick{\quad$\ket{0}_\mathrm{rf}^{m\delta t}$} \\
\lstick{$\ket{0}_\mathrm{rm}\,\,\,\,\,$}
  & \gate{X}
  & \qw
  & \gate{Z}
  & \qw
  & \gate{X}
  & \qw\rstick{$-\ket{0}_\mathrm{rm}$} \\
\lstick{$\ket{0}_\mathrm{anc}^1\,\,\,\,$}
  & \qw
  & \qw
  & \qw
  & \qw
  & \qw
  & \qw\rstick{\quad$\ket{0}_\mathrm{anc}^1$} \\
\wave&&&&&&&&&& \\
\lstick{$\ket{0}_\mathrm{anc}^{m-1}$}
  & \qw
  & \qw
  & \ctrl{-3}
  & \qw
  & \qw
  & \qw\rstick{\quad$\ket{0}_\mathrm{anc}^{m-1}$}
\end{quantikz}
\end{center}
where the $\mathcal{A}$ and $\mathcal{A}^\dagger$ gates consist of the
\textsc{and} gates:
\begin{center}
\begin{quantikz}[column sep=0.18cm]
\lstick{$\ket{\psi}_\mathrm{rf}^{\delta t}\,\,\,\,$}
  & \ctrl{1}\gategroup[12, steps=5,
    style={dashed, rounded corners, inner xsep=0}]{$\mathcal{A}$}
  & \qw
  & \cdots\qw
  & \qw
  & \qw
  & \qw\gategroup[12, steps=5,
    style={dashed, rounded corners, inner xsep=0}]{$\mathcal{A}^\dagger$}
  & \qw
  & \cdots\qw
  & \qw
  & \ctrl{1}
  & \qw\rstick{$\ket{\psi}_\mathrm{rf}^{\delta t}$} \\
\lstick{$\ket{\psi}_\mathrm{rf}^{2\delta t}\,\,$}
  & \ctrl{5}
  & \qw
  & \cdots\qw
  & \qw
  & \qw
  & \qw
  & \qw
  & \cdots\qw
  & \qw
  & \ctrl{5}
  & \qw\rstick{$\ket{\psi}_\mathrm{rf}^{2\delta t}$} \\
\lstick{$\ket{\psi}_\mathrm{rf}^{3\delta t}\,\,$}
  & \qw
  & \ctrl{5}
  & \cdots\qw
  & \qw
  & \qw
  & \qw
  & \qw
  & \cdots\qw
  & \ctrl{5}
  & \qw
  & \qw\rstick{$\ket{\psi}_\mathrm{rf}^{3\delta t}$} \\
\wave&&&&&&&&&&& \\
\lstick{$\ket{\psi}_\mathrm{rf}^{m\delta t}$}
  & \qw
  & \qw
  & \cdots\qw
  & \ctrl{5}
  & \qw
  & \qw
  & \ctrl{5}
  & \cdots\qw
  & \qw
  & \qw
  & \qw\rstick{$\ket{\psi}_\mathrm{rf}^{m\delta t}$} \\
\lstick{$\ket{0}_\mathrm{rm}\,\,\,\,$}
  & \qw
  & \qw
  & \cdots\qw
  & \qw
  & \ctrl{5}
  & \ctrl{5}
  & \qw
  & \cdots\qw
  & \qw
  & \qw
  & \qw\rstick{$\ket{\psi}_\mathrm{rm}$} \\
\lstick{$\ket{0}_\mathrm{anc}^1\,\,\,$}
  & \targ{}
  & \ctrl{1}
  & \cdots\qw
  & \qw
  & \qw
  & \qw
  & \qw
  & \cdots\qw
  & \ctrl{1}
  & \targ{1}
  & \qw\rstick{$\ket{0}_\mathrm{anc}^1$} \\
\lstick{$\ket{0}_\mathrm{anc}^2\,\,\,$}
  & \qw
  & \targ{}
  & \cdots\qw
  & \qw
  & \qw
  & \qw
  & \qw
  & \cdots\qw
  & \targ{}
  & \qw
  & \qw\rstick{$\ket{0}_\mathrm{anc}^2$} \\
\wave&&&&&&&&&&& \\
\lstick{$\ket{0}_\mathrm{anc}^{m-2}$}
  & \qw
  & \qw
  & \cdots\qw
  & \ctrl{1}
  & \qw
  & \qw
  & \ctrl{1}
  & \cdots\qw
  & \qw
  & \qw
  & \qw\rstick{$\ket{0}_\mathrm{anc}^{m-2}$} \\
\lstick{$\ket{0}_\mathrm{anc}^{m-1}$}
  & \qw
  & \qw
  & \cdots\qw
  & \targ{}
  & \ctrl{1}
  & \ctrl{1}
  & \targ{}
  & \cdots\qw
  & \qw
  & \qw
  & \qw\rstick{$\ket{0}_\mathrm{anc}^{m-1}$} \\
\lstick{$\ket{0}_\mathrm{anc}^m\,\,\,$}
  & \qw
  & \qw
  & \cdots\qw
  & \qw
  & \targ{}
  & \targ{}
  & \qw
  & \cdots\qw
  & \qw
  & \qw
  & \qw\rstick{$\ket{0}_\mathrm{anc}^m$}
\end{quantikz}
\end{center}

\subsubsection{Measurement}

The QFT gate prepares the output qubits in the state $\ket{+}$:
\begin{center}
\begin{quantikz}
\lstick{$\ket{0}_\mathrm{out}^0\,\,$}
  & \gate{H}\gategroup[3, steps=1,
    style={dashed, rounded corners, inner xsep=0}]{QFT}
  & \qw\rstick{$\ket{+}_\mathrm{out}^0\,$} \\
\wave&& \\
\lstick{$\ket{0}_\mathrm{out}^{n-1}$}
  & \gate{H}
  & \qw\rstick{$\ket{+}_\mathrm{out}^{n-1}$}
\end{quantikz}
\end{center}
\begin{figure*}
\begin{center}
\begin{quantikz}[column sep=0.1cm]
\lstick{$\ket{\psi}_\mathrm{out}^0\,\,\,$}
  & \swap{7}\gategroup[8,steps=31,
      style={dashed, rounded corners, inner xsep=0,inner ysep=0.2pt}]
      {QFT$^\dagger$}
  & \qw
  & \cdots\qw
  & \qw
  & \gate{H}
  & \ctrl{1}
  & \qw
  & \cdots\qw
  & \ctrl{3}
  & \qw
  & \cdots\qw
  & \qw
  & \ctrl{4}
  & \qw
  & \cdots\qw
  & \qw
  & \qw
  & \ctrl{6}
  & \qw
  & \cdots\qw
  & \qw
  & \qw
  & \qw
  & \ctrl{7}
  & \qw
  & \cdots\qw
  & \qw
  & \qw
  & \cdots\qw
  & \qw
  & \qw
  & \qw \\
\lstick{$\ket{\psi}_\mathrm{out}^1\,\,\,$}
  & \qw
  & \swap{5}
  & \cdots\qw
  & \qw
  & \qw
  & \phase{}
  & \gate{H}
  & \cdots\qw
  & \qw
  & \ctrl{2}
  & \cdots\qw
  & \qw
  & \qw
  & \ctrl{3}
  & \cdots\qw
  & \qw
  & \qw
  & \qw
  & \ctrl{5}
  & \cdots\qw
  & \qw
  & \qw
  & \qw
  & \qw
  & \ctrl{6}
  & \cdots\qw
  & \qw
  & \qw
  & \cdots\qw
  & \qw
  & \qw
  & \qw \\
\wave&&&&&&&&&&&&&&&&&&&&&&&&&&&&&&&&&& \\
\lstick{$\ket{\psi}_\mathrm{out}^{\frac{n}{2}-1}$}
  & \qw
  & \qw
  & \cdots\qw
  & \swap{1}
  & \qw
  & \qw
  & \qw
  & \cdots\qw
  & \phase{}
  & \phase{}
  & \cdots\qw
  & \gate{H}
  & \qw
  & \qw
  & \cdots\qw
  & \ctrl{1}
  & \qw
  & \qw
  & \qw
  & \cdots\qw
  & \ctrl{3}
  & \qw
  & \qw
  & \qw
  & \qw
  & \cdots\qw
  & \ctrl{4}
  & \qw
  & \cdots\qw
  & \qw
  & \qw
  & \qw \\
\lstick{$\ket{\psi}_\mathrm{out}^{n/2}\,\,$}
  & \qw
  & \qw
  & \cdots\qw
  & \swap{}
  & \qw
  & \qw
  & \qw
  & \cdots\qw
  & \qw
  & \qw
  & \cdots\qw
  & \qw
  & \phase{}
  & \phase{}
  & \cdots\qw
  & \phase{}
  & \gate{H}
  & \qw
  & \qw
  & \cdots\qw
  & \qw
  & \ctrl{2}
  & \qw
  & \qw
  & \qw
  & \cdots\qw
  & \qw
  & \ctrl{3}
  & \cdots\qw
  & \qw
  & \qw
  & \qw \\
\wave&&&&&&&&&&&&&&&&&&&&&&&&&&&&&&&& \\
\lstick{$\ket{\psi}_\mathrm{out}^{n-2}$}
  & \qw
  & \swap{}
  & \cdots\qw
  & \qw
  & \qw
  & \qw
  & \qw
  & \cdots\qw
  & \qw
  & \qw
  & \cdots\qw
  & \qw
  & \qw
  & \qw
  & \qw
  & \qw
  & \qw
  & \phase{}
  & \phase{}
  & \cdots\qw
  & \phase{}
  & \phase{}
  & \gate{H}
  & \qw
  & \qw
  & \cdots\qw
  & \qw
  & \qw
  & \cdots\qw
  & \ctrl{1}
  & \qw
  & \qw \\
\lstick{$\ket{\psi}_\mathrm{out}^{n-1}$}
  & \swap{}
  & \qw
  & \cdots\qw
  & \qw
  & \qw
  & \qw
  & \qw
  & \cdots\qw
  & \qw
  & \qw
  & \cdots\qw
  & \qw
  & \qw
  & \qw
  & \qw
  & \qw
  & \qw
  & \qw
  & \qw
  & \cdots\qw
  & \qw
  & \qw
  & \qw
  & \phase{}
  & \phase{}
  & \cdots\qw
  & \phase{}
  & \phase{}
  & \cdots\qw
  & \phase{}
  & \gate{H}
  & \qw
\end{quantikz}
\end{center}
\caption{The inverse quantum Fourier transform gate, QFT$^{\dagger}$.
  \label{fig:invqft}}
\end{figure*}
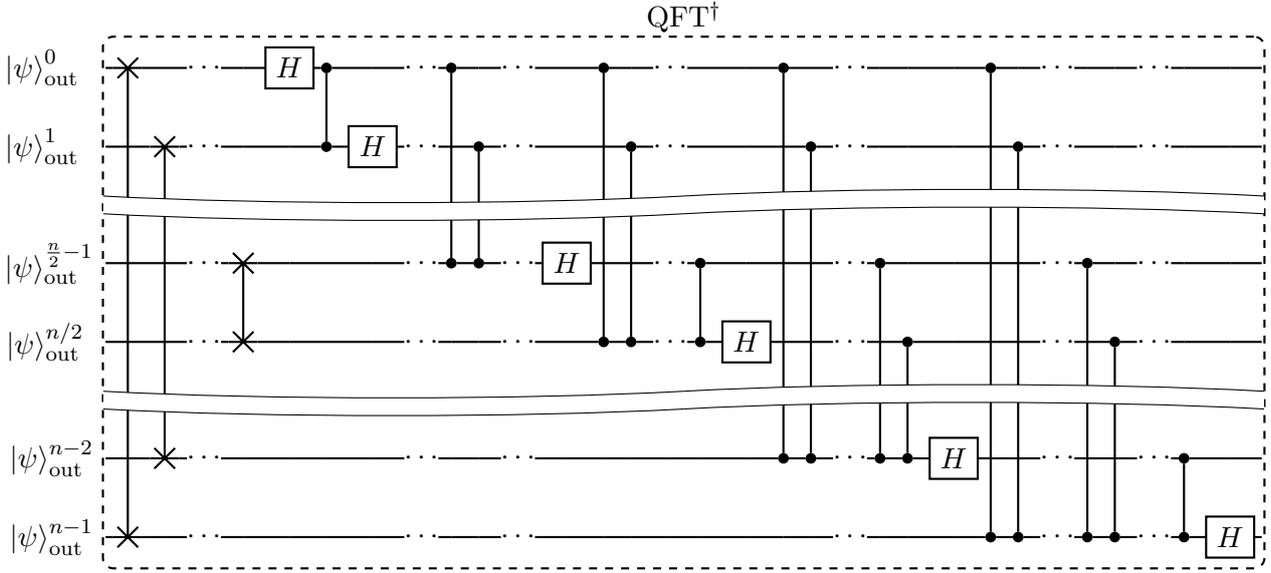
The QFT$^\dagger$ gate converts the phases $\pm k\phi$ to an integer
$z\in[0,1,...,2^n-1]$; Fig.~\ref{fig:invqft} shows the implementation based on
\cite{QiskitBook2023}, where the controlled rotations between qubits $l_1$ and
$l_2$ are:
\begin{center}
\begin{quantikz}
\lstick{$\ket{\psi}_\mathrm{out}^{l_1}$}
  & \ctrl{2}
  & \qw\rstick{} \\
\wave&& \\
\lstick{$\ket{\psi}_\mathrm{out}^{l_2}$}
  & \phase{}
  & \qw\rstick{}
\end{quantikz}
=
\begin{quantikz}
\lstick{$\ket{\psi}_\mathrm{out}^{l_1}$}
  & \ctrl{2}
  & \qw\rstick{} \\
\wave&& \\
\lstick{$\ket{\psi}_\mathrm{out}^{l_2}$}
  & \gate{R_z\left(\frac{-\pi}{2^{l_2-l_1}}\right)}
  & \qw\rstick{}
\end{quantikz}
\end{center}

See App.~\ref{app:ex_equity} for an example of a complete quantum circuit that
estimates $P(S_\mathrm{max})$ for $m = 2$ and $n = 3$.
The accompanying figures show a visualisation of the qubit states and their
transformations on the Bloch sphere, as well as the corresponding probability
distributions when measured.

\subsubsection{Results}

\begin{table}
\begin{center}
\begin{tabular}{c|c}
Parameter                                & Value        \\
\hline
$m$                                      & $6$          \\
$n$                                      & $1$-$9$      \\
$T$                                      & $1$          \\
$\mu$                                    & $8\%$        \\
$\sigma$                                 & $20\%$       \\
$u$                                      & $\sim$$1.09$ \\
$q$                                      & $\sim$$0.56$ \\
$\theta_\mathrm{u}\frac{180^\circ}{\pi}$ & $\sim$$97.1^\circ$
\end{tabular}
\end{center}
\caption{
  List of the binomial tree parameters for the equity risk factor evolution.
  \label{tab:eq_params}}
\end{table}
To analyse the convergence of the estimate of the risk measure we assemble 9
quantum circuits, each one with a different number of output qubits,
$n\in[1, 2, ..., 9]$, and adopt the parameters listed in
Table~\ref{tab:eq_params} (the value of $S_0$ is not needed).
\begin{figure}
  \centering
  \includegraphics{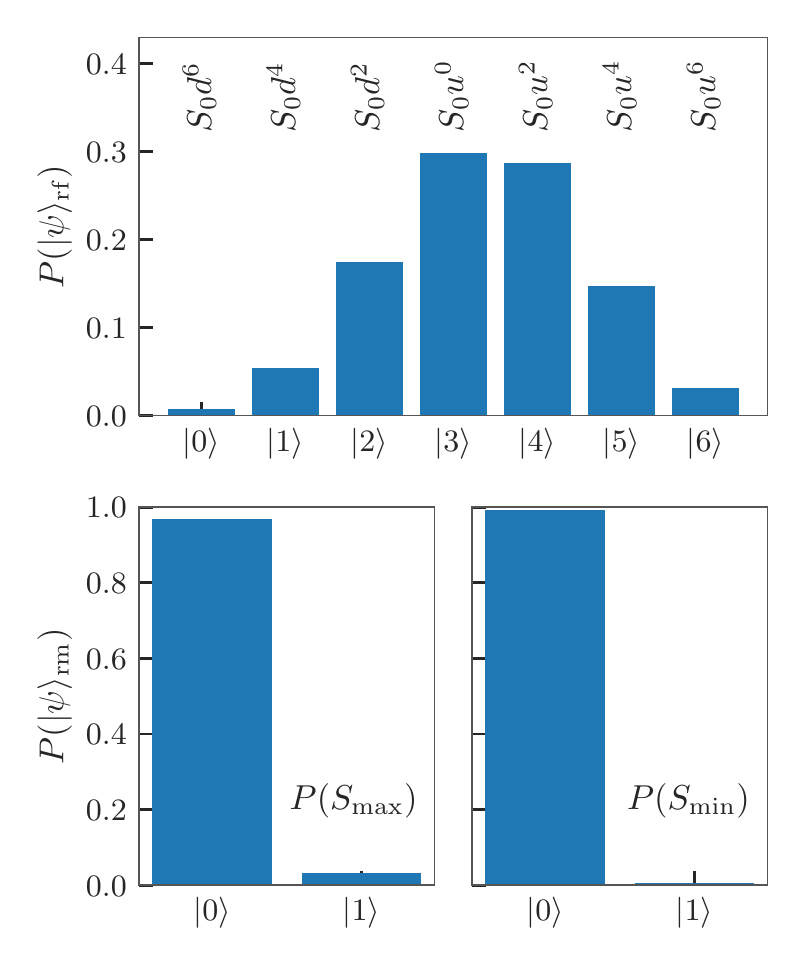}
  \caption{
    Top: the distribution of $S_T$ for $m = 6$ qubits as generated with the
    operator $\mathcal{D}_\mathrm{eq}$.
    Bottom: the probabilities $P(S_\mathrm{max})$ (left) and $P(S_\mathrm{min})$
    (right) as encoded with the operators $\mathcal{M}_\mathrm{max}$ and
    $\mathcal{M}_\mathrm{min}$, respectively, into the ``risk measure'' qubit.}
  \label{fig:eq_rf}
\end{figure}
The top panel of Fig.~\ref{fig:eq_rf} shows the probability distribution
$P(S_T)$ obtained after applying the gate $\mathcal{D}_\mathrm{eq}$ on the
``risk factor'' qubits.
The distribution is represented by the states $\ket{j}$ (see
Eq.~\ref{eq:ket_j}), which consist of all possible $2^m$ combinations of the
``risk factor'' qubits when $j$ of them are in the state $\ket{1}$ (i.e. $j$
upwards price moves).
Since $q > 1/2$, the distribution is positively skewed:
$P(S_T \geq u^2S_0) > P(S_T \leq d^2S_0)$.
The gates $\mathcal{M}_\mathrm{max}$ and $\mathcal{M}_\mathrm{min}$ essentially
measure the probabilities of the states $\ket{6}$ and $\ket{0}$, respectively,
which are encoded in the ``risk measure'' qubit; see bottom panel of
Fig.~\ref{fig:eq_rf}.

\begin{figure}
  \centering
  \includegraphics{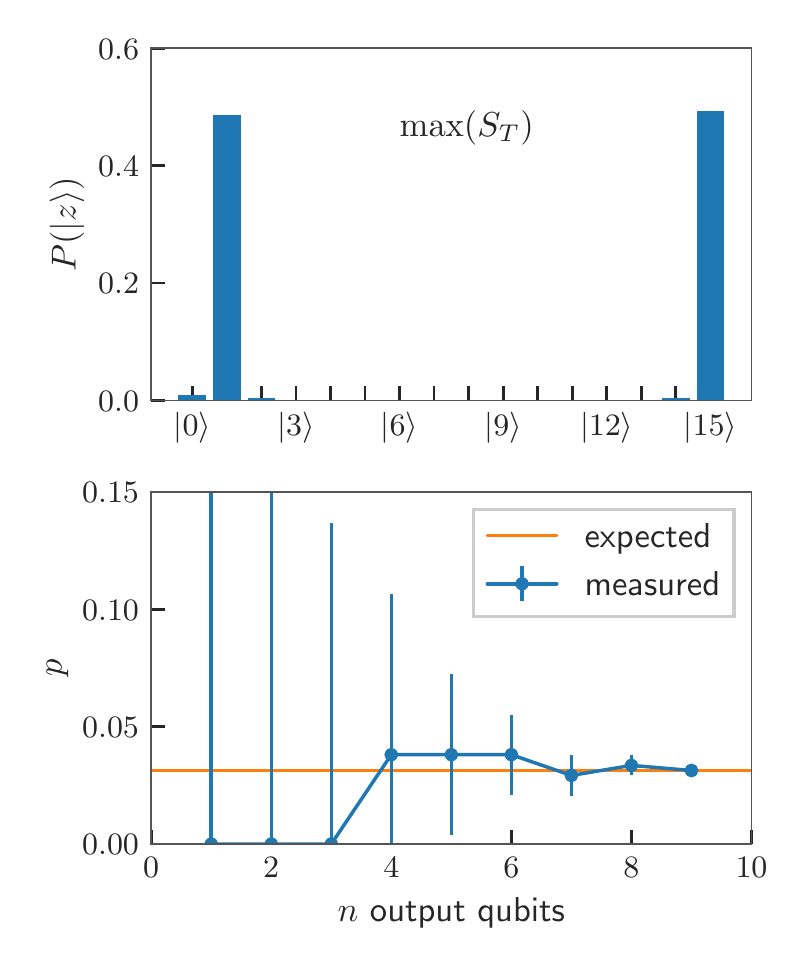}
  \includegraphics{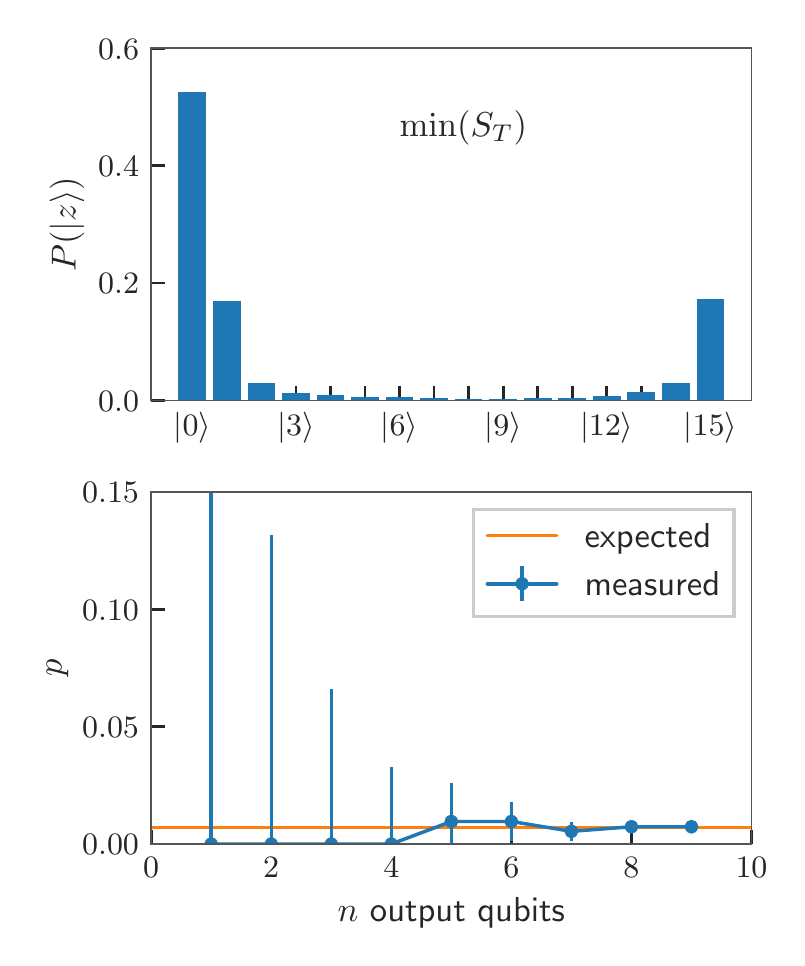}
  \caption{
    Top two panels: the probability distribution $P(S_\mathrm{max})$ (top) when
    measuring $z_0$ with 10,000 shots using $n = 4$ qubits, and the convergence
    of $P(S_\mathrm{max}) \to p = q^m$ (with $\delta p \to 0$, bottom), as a
    function of the number of output qubits.
    The measured value of $p$ (blue dots) is calculated using
    Eq.~\eqref{eq:pqmc} after measuring the result of one shot, the error
    $\delta p$ (blue lines) is calculated from Eq.~\eqref{eq:dpqmc}, and the
    expected value is shown with an orange line.
    Bottom two panels: the probability distribution $P(S_\mathrm{min})$ (top)
    and $P(S_\mathrm{min}) \to p = (1-q)^m$ (bottom).}
  \label{fig:eq_results}
\end{figure}
The bar charts of Fig.~\ref{fig:eq_results} show the distribution $P(\ket{z})$
obtained by measuring the output qubits (here $n = 4$ and 10,000 shots) for the
two risk measures, $P(S_\mathrm{max})$ (top) and $P(S_\mathrm{min})$ (bottom).
The left and right peaks correspond to $z = 2^n\theta/2\pi$ and
$z = 2^n(2\pi-\theta)/2\pi$, respectively, originating from the two signs of the
phase kickback, $\pm k\theta$.
The line charts of Fig.~\ref{fig:eq_results} show the convergence of the 
estimated probability $p$ (Eq.~\ref{eq:pqmc}), $P(S_\mathrm{max})$ (top) and
$P(S_\mathrm{min})$ (bottom), and error $\delta p$ (Eq.~\ref{eq:dpqmc}), as a
function of the number of output qubits, $n$.
Here, the measured state of the output qubits is obtained from one shot.
The expected values of the probabilities are $P(S_\mathrm{max}) = q^m$ and
$P(S_\mathrm{min}) = (1-q)^m$, respectively.

Equity risk factors can also be based on trinomial trees, the implementation of
which is presented in Sect.~\ref{sec:ir}.

\subsection{Interest rate risk factors}
  \label{sec:ir}

The evolution of interest rates can be simulated with short-rate mean reversion
models, the most basic of which is the Vasicek model \cite{Vasicek1977}:
\begin{align}
dr_t &= a(b - r_t)dt + \sigma dW_t\,,
\end{align}
where $r_t$ is the instantaneous interest rate at time $t$, $b$ the long-term
mean, $a$ the speed of reversion, $\sigma$ the volatility,  and $W_t$ a Wiener
process.
The expected value and variance are asymptotically constant for
$\delta t\to \infty$, whereas for finite $\delta t$ the expected value depends
on $r_t$:
\begin{align}
E(r_{t + \delta t}) &= r_te^{-a\delta t} + b\left(1 - e^{-a\delta t}\right)\,,\\
V\!ar(r_{t + \delta t})
  &= \frac{\sigma^2}{2a}\left(1 - e^{-2a\delta t}\right)\,.
\end{align}

Such models are often discretised with trinomial trees.
For example, consider a simple tree that is bounded at both low and high
interest rates, and, at each timestep, $r_t$ can take one of three possible
values: $r_t^\mathrm{h} = b + \delta r$ (high), $r_t^\mathrm{m} = b$ (mid), and
$r_t^\mathrm{l} = b - \delta r$ (low), with $\delta r$ a constant.
\begin{center}
\begin{tikzpicture}[->, node distance = 2cm]
\node[state](1){$r_0^\mathrm{m}$};
\node[state](2)[right of=1]{$r_{\delta t}^\mathrm{m}$};
\node[state](3)[above of=2]{$r_{\delta t}^\mathrm{h}$};
\node[state](4)[below of=2]{$r_{\delta t}^\mathrm{l}$};
\node[state](5)[right of=2]{$r_{2\delta t}^\mathrm{m}$};
\node[state](6)[above of=5]{$r_{2\delta t}^\mathrm{h}$};
\node[state](7)[below of=5]{$r_{2\delta t}^\mathrm{l}$};
\node[state](8)[right of=5]{$r_{3\delta t}^\mathrm{m}$};
\node[state](9)[above of=8]{$r_{3\delta t}^\mathrm{h}$};
\node[state](10)[below of=8]{$r_{3\delta t}^\mathrm{l}$};
\path (1) edge [swap] node {}(2);
\path (1) edge [swap] node {}(3);
\path (1) edge [swap] node {}(4);
\path (2) edge [swap] node {}(5);
\path (2) edge [swap] node {}(6);
\path (2) edge [swap] node {}(7);
\path (3) edge [swap] node {}(5);
\path (3) edge [swap] node {}(6);
\path (3) edge [swap] node {}(7);
\path (4) edge [swap] node {}(5);
\path (4) edge [swap] node {}(6);
\path (4) edge [swap] node {}(7);
\path (5) edge [swap] node {}(8);
\path (5) edge [swap] node {}(9);
\path (5) edge [swap] node {}(10);
\path (6) edge [swap] node {}(8);
\path (6) edge [swap] node {}(9);
\path (6) edge [swap] node {}(10);
\path (7) edge [swap] node {}(8);
\path (7) edge [swap] node {}(9);
\path (7) edge [swap] node {}(10);
\end{tikzpicture}
\end{center}
The transition probabilities $q_{t \to t + \delta t}$ from a node $r_t$ to a
node $r_{t+\delta t}$ depend both on the $t$ and $t + \delta t$ nodes, and thus
$q_{t \to t + \delta t}$ is an array of $3\times3 = 9$ values.
These probabilities can be derived by equating the expected value and variance
of the continuous and discrete models, respectively
(see App.~\ref{app:ir_calcs}).

\subsubsection{$\mathcal{D}_\mathrm{ir}$: the distribution $P(r_t)$}

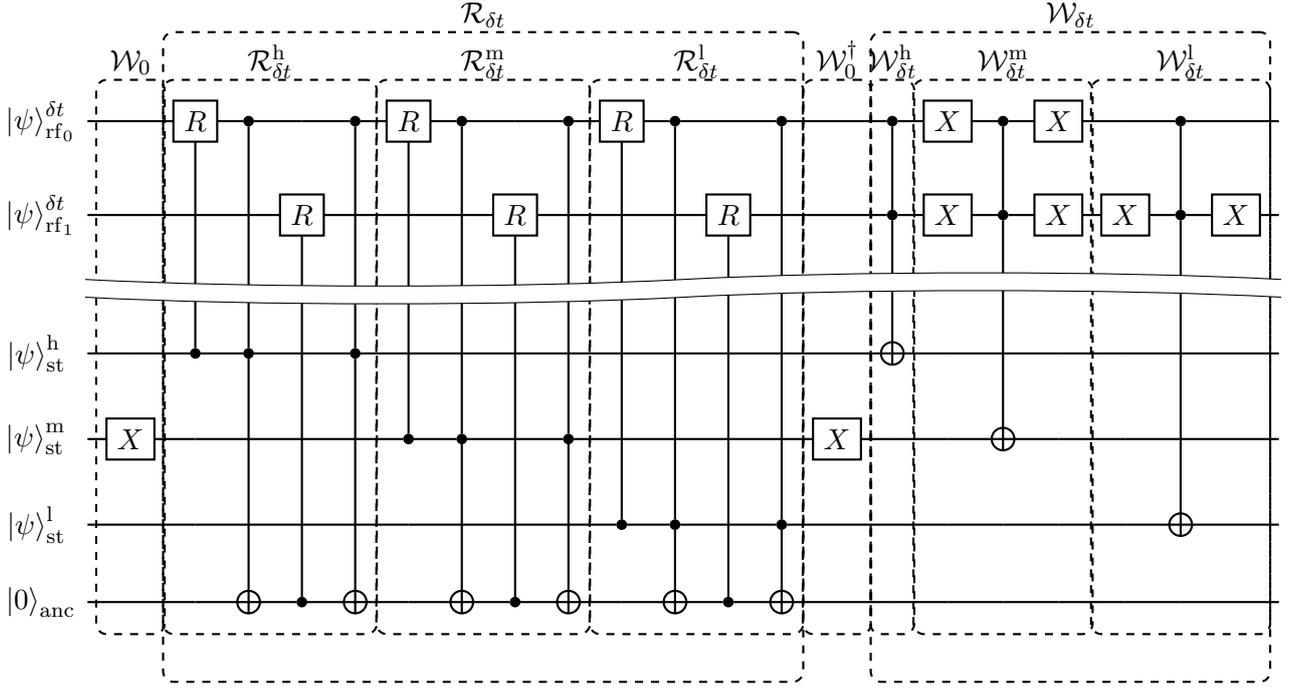
\begin{figure*}
\begin{center}
\begin{quantikz}[column sep=0.25cm, row sep=0.7cm]
\lstick{$\ket{\psi}_{\mathrm{rf}_0}^{\delta t}$}
  & \qw\gategroup[7, steps=1,
    style={dashed, rounded corners, inner xsep=0}]{$\mathcal{W}_0$}
  & \gate{R}\gategroup[7, steps=4,
    style={dashed, rounded corners, inner xsep=0}]
      {$\mathcal{R}_{\delta t}^\mathrm{h}$}\gategroup[7, steps=12,
      style={dashed, rounded corners, inner xsep=0, inner ysep=22pt}]
      {$\mathcal{R}_{\delta t}$}
  & \ctrl{3}
  & \qw
  & \ctrl{3}
  & \gate{R}\gategroup[7, steps=4,
    style={dashed, rounded corners, inner xsep=0}]
    {$\mathcal{R}_{\delta t}^\mathrm{m}$}
  & \ctrl{4}
  & \qw
  & \ctrl{4}
  & \gate{R}\gategroup[7, steps=4,
    style={dashed, rounded corners, inner xsep=0}]
    {$\mathcal{R}_{\delta t}^\mathrm{l}$}
  & \ctrl{5}
  & \qw
  & \ctrl{5}
  & \qw\gategroup[7, steps=1,
    style={dashed, rounded corners, inner xsep=0}]{$\mathcal{W}_0^\dagger$}
  & \ctrl{1}\gategroup[7, steps=1,
    style={dashed, rounded corners, inner xsep=0}]{
      $\mathcal{W}_{\delta t}^\mathrm{h}$}\gategroup[7, steps=7,
      style={dashed, rounded corners, inner xsep=0, inner ysep=22pt}]
      {$\mathcal{W}_{\delta t}$}
  & \gate{X}\gategroup[7, steps=3,
    style={dashed, rounded corners, inner xsep=0}]
      {$\mathcal{W}_{\delta t}^\mathrm{m}$}
  & \ctrl{1}
  & \gate{X}
  & \qw\gategroup[7, steps=3,
    style={dashed, rounded corners, inner xsep=0}]
      {$\mathcal{W}_{\delta t}^\mathrm{l}$}
  & \ctrl{1}
  & \qw
  & \qw \\
\lstick{$\ket{\psi}_{\mathrm{rf}_1}^{\delta t}$}
  & \qw
  & \qw
  & \qw
  & \gate{R}
  & \qw
  & \qw
  & \qw
  & \gate{R}
  & \qw
  & \qw
  & \qw
  & \gate{R}
  & \qw
  & \qw
  & \ctrl{2}
  & \gate{X}
  & \ctrl{3}
  & \gate{X}
  & \gate{X}
  & \ctrl{4}
  & \gate{X}
  & \qw \\
\wave&&&&&&&&&&&&&&&&&&&&&& \\
\lstick{$\ket{\psi}_\mathrm{st}^\mathrm{h}\,\,$}
  & \qw
  & \ctrl{-3}
  & \ctrl{3}
  & \qw
  & \ctrl{3}
  & \qw
  & \qw
  & \qw
  & \qw
  & \qw
  & \qw
  & \qw
  & \qw
  & \qw
  & \targ{}
  & \qw
  & \qw
  & \qw
  & \qw
  & \qw
  & \qw
  & \qw \\
\lstick{$\ket{\psi}_\mathrm{st}^\mathrm{m}\,\,$}
  & \gate{X}
  & \qw
  & \qw
  & \qw
  & \qw
  & \ctrl{-4}
  & \ctrl{2}
  & \qw
  & \ctrl{2}
  & \qw
  & \qw
  & \qw
  & \qw
  & \gate{X}
  & \qw
  & \qw
  & \targ{}
  & \qw
  & \qw
  & \qw
  & \qw
  & \qw \\
\lstick{$\ket{\psi}_\mathrm{st}^\mathrm{l}\,\,$}
  & \qw
  & \qw
  & \qw
  & \qw
  & \qw
  & \qw
  & \qw
  & \qw
  & \qw
  & \ctrl{-5}
  & \ctrl{1}
  & \qw
  & \ctrl{1}
  & \qw
  & \qw
  & \qw
  & \qw
  & \qw
  & \qw
  & \targ{}
  & \qw
  & \qw \\
\lstick{$\ket{0}_\mathrm{anc}$}
  & \qw
  & \qw
  & \targ{}
  & \ctrl{-5}
  & \targ{}
  & \qw
  & \targ{}
  & \ctrl{-5}
  & \targ{}
  & \qw
  & \targ{}
  & \ctrl{-5}
  & \targ{}
  & \qw
  & \qw
  & \qw
  & \qw
  & \qw
  & \qw
  & \qw
  & \qw
  & \qw
\end{quantikz}
\end{center}
\caption{
  The decomposition of the ``read'' $\mathcal{R}_t$ and ``write''
  $\mathcal{W}_t$ operators for the first timestep, assuming that $r_0 = b$
  (i.e. $\mathcal{W}_0\ket{000}_\mathrm{st} = \ket{010}_\mathrm{st}^0$).
  From left to right the $R$ gates are   $R_y(\theta_0^\mathrm{hm})$,
  $R_y(\theta_1^\mathrm{hh})$, $R_y(\theta_0^\mathrm{mm})$,
  $R_y(\theta_1^\mathrm{mh})$, $R_y(\theta_0^\mathrm{lm})$, and
  $R_y(\theta_1^\mathrm{lh})$.
  \label{fig:ir_D1}}
\end{figure*}

Given that at each timestep $t$ there are 3 possible outcomes we need two
qubits, $\ket{0}_{\mathrm{rf}_1}^t\ket{0}_{\mathrm{rf}_0}^t$, to encode the
probabilities $q$; therefore, $m$ timesteps require $2m$ ``risk factor'' qubits.
At each timestep, we use the first qubit (rf$_0$) from the pair to represent
whether the interest rate transitions to the mid level:
$r_t \to r^m_{t + \delta t}$.
Specifically, the state $\ket{0}_{\mathrm{rf}_0}^t$ describes a transition to
the mid node and the state $\ket{1}_{\mathrm{rf}_0}^t$ a transition to a
different node.
In the latter case, we use the second qubit (rf$_1$) to model whether $r_t$
transitions to the high ($r_t \to r^h_{t + \delta t}$, described by
$\ket{1}_{\mathrm{rf}_1}^t$) or low ($r_t \to r^l_{t + \delta t}$, described by
$\ket{0}_{\mathrm{rf}_1}^t$) levels, respectively.
Because the transition probabilities depend on the level of the interest rate,
the ``risk factor'' state needs to be read at each timestep before encoding the
probabilities for the next transition.
This can be facilitated by including $3$ additional ``state'' qubits ---
alongside the $2m$ ``risk factor'' qubits --- to store the value of $r_t$ at
each timestep; i.e.
$\ket{001}_\mathrm{st}^t$, $\ket{010}_\mathrm{st}^t$, and
$\ket{100}_\mathrm{st}^t$, which represent the high, mid, and low interest rate
levels, respectively.
Assuming $r_0$ is at the mid level, the tree is:
\begin{center}
\begin{tikzpicture}[->, node distance = 3cm]
\node[state] at (0, 0) (1)
  {$\ket{\emptyset\emptyset}_\mathrm{rf}^{2\delta t}
    \ket{\emptyset\emptyset}_\mathrm{rf}^{\delta t}
    \atop \ket{010}_\mathrm{st}^0$};
\node[state] at (3, 0) (2)
  {$\ket{\emptyset\emptyset}_\mathrm{rf}^{2\delta t}
    \ket{00}_\mathrm{rf}^{\delta t}
    \atop \ket{010}_\mathrm{st}^{\delta t}$};
\node[state] at (3, 2.5) (3)
  {$\ket{\emptyset\emptyset}_\mathrm{rf}^{2\delta t}
    \ket{11}_\mathrm{rf}^{\delta t}
    \atop \ket{001}_\mathrm{st}^{\delta t}$};
\node[state] at (3, -2.5) (4)
  {$\ket{\emptyset\emptyset}_\mathrm{rf}^{2\delta t}
    \ket{01}_\mathrm{rf}^{\delta t}
    \atop \ket{100}_\mathrm{st}^{\delta t}$};
\node[state] at (6, 0) (5)
  {$\ket{00}_\mathrm{rf}^{2\delta t}\ket{\psi}_\mathrm{rf}^{\delta t}
    \atop \ket{010}_\mathrm{st}^{2\delta t}$};
\node[state] at (6, 2.5) (6)
  {$\ket{11}_\mathrm{rf}^{2\delta t}\ket{\psi}_\mathrm{rf}^{\delta t}
    \atop \ket{001}_\mathrm{st}^{2\delta t}$};
\node[state] at (6, -2.5) (7)
  {$\ket{01}_\mathrm{rf}^{2\delta t}\ket{\psi}_\mathrm{rf}^{\delta t}
    \atop \ket{100}_\mathrm{st}^{2\delta t}$};
\path (1) edge [swap] node {}(2);
\path (1) edge [swap] node {}(3);
\path (1) edge [swap] node {}(4);
\path (2) edge [swap] node {}(5);
\path (2) edge [swap] node {}(6);
\path (2) edge [swap] node {}(7);
\path (3) edge [swap] node {}(5);
\path (3) edge [swap] node {}(6);
\path (3) edge [swap] node {}(7);
\path (4) edge [swap] node {}(5);
\path (4) edge [swap] node {}(6);
\path (4) edge [swap] node {}(7);
\end{tikzpicture}
\end{center}
The quantum state at the first timestep is:
\begin{align}
&\ket{\psi}_\mathrm{rf}^{\delta t}\ket{\psi}_\mathrm{st}^{\delta t}
= \sqrt{q_\mathrm{mh}}\ket{11}_\mathrm{rf}^{\delta t}
    \ket{001}_\mathrm{st}^{\delta t} \\
&\quad\quad + \sqrt{q_\mathrm{mm}}\ket{00}_\mathrm{rf}^{\delta t}
    \ket{010}_\mathrm{st}^{\delta t}
  + \sqrt{q_\mathrm{ml}}\ket{01}_\mathrm{rf}^{\delta t}
    \ket{100}_\mathrm{st}^{\delta t}\,, \nonumber
\end{align}
where the first and second subscripts of the transition probabilities denote the
start and end nodes.
To assemble the gate $\mathcal{D}_\mathrm{ir}$ we put together a ``read''
operator $\mathcal{R}_t
= \mathcal{R}_t^\mathrm{l}\mathcal{R}_t^\mathrm{m}\mathcal{R}_t^\mathrm{h}$,
which reads the state $\ket{\psi}_\mathrm{st}^t$ and encodes the transition
probabilities to $\ket{\psi}_\mathrm{rf}^t$.
As an example, the operator $\mathcal{R}_{\delta t}^\mathrm{m}$ is:
\begin{center}
\begin{quantikz}
\lstick{$\ket{\psi}_{\mathrm{rf}_0}^{\delta t}$}
  & \gate{R_y(\theta_0^\mathrm{mm})}\gategroup[7, steps=4,
    style={dashed, rounded corners, inner xsep=0}]
      {$\mathcal{R}_{\delta t}^\mathrm{m}$}
  & \ctrl{4}
  & \qw
  & \ctrl{4}
  & \qw \\
\lstick{$\ket{\psi}_{\mathrm{rf}_1}^{\delta t}$}
  & \qw
  & \qw
  & \gate{R_y(\theta_1^\mathrm{mh})}
  & \qw
  & \qw \\
\wave&&&&& \\
\lstick{$\ket{\psi}_\mathrm{st}^\mathrm{h}\,\,$}
  & \qw
  & \qw
  & \qw
  & \qw
  & \qw \\
\lstick{$\ket{\psi}_\mathrm{st}^\mathrm{m}\,\,$}
  & \ctrl{-4}
  & \ctrl{2}
  & \qw
  & \ctrl{2}
  & \qw \\
\lstick{$\ket{\psi}_\mathrm{st}^\mathrm{l}\,\,$}
  & \qw
  & \qw
  & \qw
  & \qw
  & \qw \\
\lstick{$\ket{0}_\mathrm{anc}$}
  & \qw
  & \targ{}
  & \ctrl{-5}
  & \targ{}
  & \qw
\end{quantikz}
\end{center}
where
\begin{align}
\theta_0^\mathrm{mm} &= 2\arcsin\sqrt{q_\mathrm{mm}} \\
\theta_1^\mathrm{mh} &= 2\arcsin\sqrt{q_\mathrm{mh}/(1-q_\mathrm{mm})}\,.
\end{align}
Here, the controlled $R_y(\theta_0^\mathrm{hm})$ gate encodes the probability
$q_\mathrm{mm}$ --- namely the likelihood that the rate will remain at the mid
level $r_{\delta t} = r_0 = b$ --- into $\ket{\psi}_{\mathrm{rf}_0}^{\delta t}$.
The second gate, \textsc{and}, checks whether both $r_0 = b$ and
$r_{\delta t} \neq b$ are true --- namely whether the interest rate changed ---
and writes the result to an ``ancilla'' qubit.
The third gate, controlled $R_y(\theta_1^\mathrm{mh})$, checks the result stored
in the ``ancilla'' qubit, and if it is $\ket{1}$ it encodes the conditional
probability of $r_{\delta t}$ transitioning to the high interest rate value
given that it did not remain at the mid level:
$P(r_{\delta t} = b + \delta r | r_{\delta t} \neq b)
= P(r_{\delta t} = b + \delta r)/P(r_{\delta t} \neq b)
= q_\mathrm{mh}/(1-q_\mathrm{mm})$.
The fourth gate, \textsc{and}, ensures the ``ancilla'' qubit is in its original
state.
The logic for the $\mathcal{R}_{\delta t}^\mathrm{h}$ and
$\mathcal{R}_{\delta t}^\mathrm{l}$ gates is similar, see their decomposition in
Fig.~\ref{fig:ir_D1} which shows an example of the entire ``read'' gate at the
first timestep, $\mathcal{R}_{\delta t}$.

The ``write'' gates $\mathcal{W}_t$ read the ``risk factor'' qubit pair
$\ket{\psi}_\mathrm{rf}^t$ and write the result to the ``state'' qubits,
$\ket{\psi}_\mathrm{st}^t$:
\begin{align}
\mathcal{W}_t\ket{11}_\mathrm{rf}^t\ket{000}_\mathrm{st}
&= \ket{11}_\mathrm{rf}^t\ket{001}_\mathrm{rf}^t\,, \\
\mathcal{W}_t\ket{00}_\mathrm{rf}^t\ket{000}_\mathrm{st}
&= \ket{00}_\mathrm{rf}^t\ket{010}_\mathrm{rf}^t\,, \\
\mathcal{W}_t\ket{01}_\mathrm{rf}^t\ket{000}_\mathrm{st}
&= \ket{01}_\mathrm{rf}^t\ket{100}_\mathrm{rf}^t\,,
\end{align}
see Fig.~\ref{fig:ir_D1} for its decomposition of $\mathcal{W}_t$ into $X$ and
\textsc{and} operators.
Before writing to the ``state'' qubits, the previous state needs to be erased;
therefore, the application of $\mathcal{W}_t$ is always preceded by the gate
$\mathcal{W}_{t - \delta t}^\dagger$ which resets the ``state'' qubits back to
$\ket{000}_\mathrm{st}$.
The inverses of the ``read'' and ``write'' gates, $\mathcal{R}_t^\dagger$ and
$\mathcal{W}_t^\dagger$, consist of the components of $\mathcal{R}_t$ and
$\mathcal{W}_t$ put in inverse order.
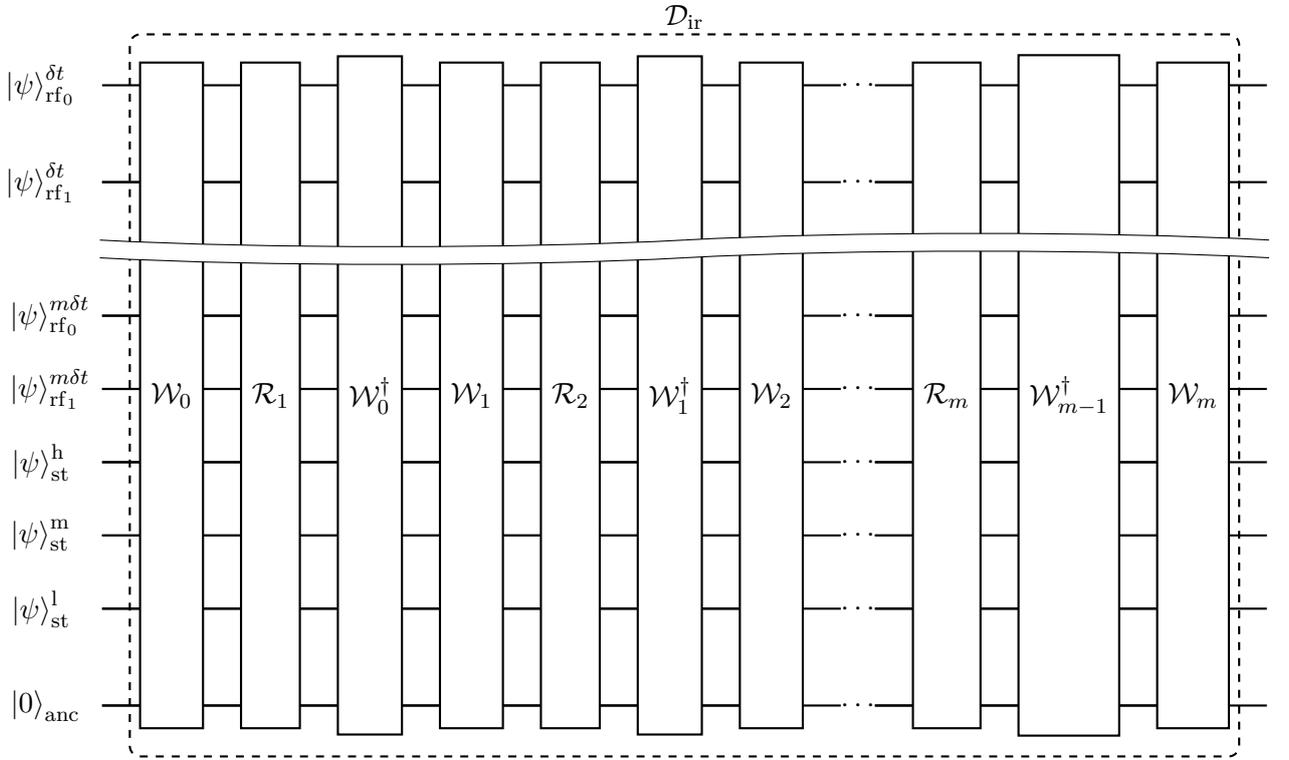
\begin{figure*}
\begin{center}
\begin{quantikz}[row sep=0.8cm]
  \lstick{$\ket{\psi}_{\mathrm{rf}_0}^{\delta t}\,\,\,$}
  & \gate[9]{\mathcal{W}_0}\gategroup[9, steps=11,
    style={dashed, rounded corners, inner xsep=0}]{$\mathcal{D}_\mathrm{ir}$}
  & \gate[9]{\mathcal{R}_1}
  & \gate[9]{\mathcal{W}_0^\dagger}
  & \gate[9]{\mathcal{W}_1}
  & \gate[9]{\mathcal{R}_2}
  & \gate[9]{\mathcal{W}_1^\dagger}
  & \gate[9]{\mathcal{W}_2}
  & \cdots\qw
  & \gate[9]{\mathcal{R}_m}
  & \gate[9]{\mathcal{W}_{m-1}^\dagger}
  & \gate[9]{\mathcal{W}_m}
  & \qw \\
\lstick{$\ket{\psi}_{\mathrm{rf}_1}^{\delta t}\,\,\,$}
  & \qw
  & \qw
  & \qw
  & \qw
  & \qw
  & \qw
  & \qw
  & \cdots\qw
  & \qw
  & \qw
  & \qw
  & \qw \\
\wave&&&&&&&&&&&&&&&& \\
\lstick{$\ket{\psi}_{\mathrm{rf}_0}^{m\delta t}$}
  & \qw
  & \qw
  & \qw
  & \qw
  & \qw
  & \qw
  & \qw
  & \cdots\qw
  & \qw
  & \qw
  & \qw
  & \qw \\
\lstick{$\ket{\psi}_{\mathrm{rf}_1}^{m\delta t}$}
  & \qw
  & \qw
  & \qw
  & \qw
  & \qw
  & \qw
  & \qw
  & \cdots\qw
  & \qw
  & \qw
  & \qw
  & \qw \\
\lstick{$\ket{\psi}_\mathrm{st}^\mathrm{h}\,\,\,\,$}
  & \qw
  & \qw
  & \qw
  & \qw
  & \qw
  & \qw
  & \qw
  & \cdots\qw
  & \qw
  & \qw
  & \qw
  & \qw \\
\lstick{$\ket{\psi}_\mathrm{st}^\mathrm{m}\,\,\,\,$}
  & \qw
  & \qw
  & \qw
  & \qw
  & \qw
  & \qw
  & \qw
  & \cdots\qw
  & \qw
  & \qw
  & \qw
  & \qw \\
\lstick{$\ket{\psi}_\mathrm{st}^\mathrm{l}\,\,\,\,$}
  & \qw
  & \qw
  & \qw
  & \qw
  & \qw
  & \qw
  & \qw
  & \cdots\qw
  & \qw
  & \qw
  & \qw
  & \qw \\
\lstick{$\ket{0}_\mathrm{anc}\,\,$}
  & \qw
  & \qw
  & \qw
  & \qw
  & \qw
  & \qw
  & \qw
  & \cdots\qw
  & \qw
  & \qw
  & \qw
  & \qw
\end{quantikz}
\end{center}
\caption{
  Gate $\mathcal{D}_\mathrm{ir}$ consists of a sequence of ``read'' and
  ``write'' gates. \label{fig:ir_D}}
\end{figure*}
Finally, the operator $\mathcal{D}_\mathrm{ir}$ can be assembled by a sequence
of ``read'' and ``write'' operators for each timestep as shown in
Fig.~\ref{fig:ir_D}.

\subsubsection{$\mathcal{M}_\mathrm{ir}$: risk measures $F(r_t)$}

For the implementation of the trinomial tree we consider as risk measure the
probability of the interest rate being equal to its long-term mean level after
$m$ timesteps.
In this case, the gate $\mathcal{M}_\mathrm{m}$ is simply:
\begin{center}
\begin{quantikz}
\lstick{$\ket{\psi}_\mathrm{st}^\mathrm{h}\,\,$}
  & \qw\gategroup[4, steps=1,
    style={dashed, rounded corners, inner xsep=0}]{$\mathcal{M}_\mathrm{m}$}
  & \qw\rstick{$\ket{\psi}_\mathrm{st}^\mathrm{h}$} \\
\lstick{$\ket{\psi}_\mathrm{st}^\mathrm{m}\,\,$}
  & \ctrl{2}
  & \qw\rstick{$\ket{\psi}_\mathrm{st}^\mathrm{m}$} \\
\lstick{$\ket{\psi}_\mathrm{st}^\mathrm{l}\,\,$}
  & \qw
  & \qw\rstick{$\ket{\psi}_\mathrm{st}^\mathrm{l}$}\\
\lstick{$\ket{0}_\mathrm{rm}\,$}
  & \targ{}
  & \qw\rstick{$\ket{\psi}_\mathrm{rm}$}
\end{quantikz}
\end{center}
with $\mathcal{M}_\mathrm{m}^\dagger = \mathcal{M}_\mathrm{m}$.

\subsubsection{Results}

\begin{table}
\begin{center}
\begin{tabular}{c|c}
Parameter  & Value           \\
\hline
$m$        & $3$             \\
$n$        & $1$-$9$         \\
$\delta r$ & $\sqrt{3V\!ar}$ \\
$\delta t$ & $3a/12$
\end{tabular}
\end{center}
\caption{
  List of the trinomial tree parameters for the interest rate risk factor
  evolution.
  \label{tab:ir_params}}
\end{table}
\begin{table}
\begin{center}
\begin{tabular}{c|ccc}
$q_{t\to t + \delta t}$ &                 & $r_{t+\delta t}$ &                \\
\hline
$r_t$                   &  high           & mid              & low            \\
\hline
high                    & $\frac{19}{24}$ & $\frac{4}{24}$   & $\frac{1}{24}$ \\
mid                     & $\frac{4}{24}$  & $\frac{16}{24}$  & $\frac{4}{24}$ \\
low                     & $\frac{1}{24}$  & $\frac{4}{24}$   & $\frac{19}{24}$\\
\end{tabular}
\end{center}
\caption{
  The transition probabilities of the trinomial tree.
  \label{tab:ir_p_params}}
\end{table}
Table~\ref{tab:ir_params} lists the choice of parameters for a quantum circuit
that implements trinomial trees for interest rate evolution (due to our
parametrization, we do not need to choose values for $a$ and $b$).
For each one of the $m = 3$ timesteps we need $2$ ``risk factor'' qubits, a
total of $6$: $\ket{\psi}_\mathrm{rf}$.
We need an additional $3$ qubits to store the interest rate state
$\ket{\psi}_\mathrm{st}$, $1$ qubit for the risk measure
$\ket{\psi}_\mathrm{rm}$, and $9$ ``ancilla'' qubits $\ket{\psi}_\mathrm{anc}$
for the $\mathcal{A}$ gates.
We can freely choose $\delta r$, which we set to a multiple of the standard
deviation.
Note that the definition $0 \leq q_{t \to t + \delta t} \leq 1$ constrains the
choice of $\delta t$ (see App.~\ref{app:ir_calcs}).
Having defined $\delta r$ and $\delta t$, we can calculate the transition
probabilities, see Table~\ref{tab:ir_params}.

When starting from $r_0 = b$, the probability of measuring the value
$r_{3\delta t} = b$ after $3$ timesteps can be calculated by adding up the
probabilities of all possible paths:
\begin{align}
P(r_{3\delta t} = b)
&= \sum_{s_1}\left[q_{\mathrm{m}s_1}\left(
  \sum_{s_2}q_{s_1s_2}\,q_{s_2\mathrm{m}}\right)\right]\,,
\end{align}
where $s_1,\,s_2\in\{\mathrm{h},\mathrm{m},\mathrm{l}\}$ are the nodes of the 
first and second timesteps, respectively.

\begin{figure}
  \centering
  \includegraphics{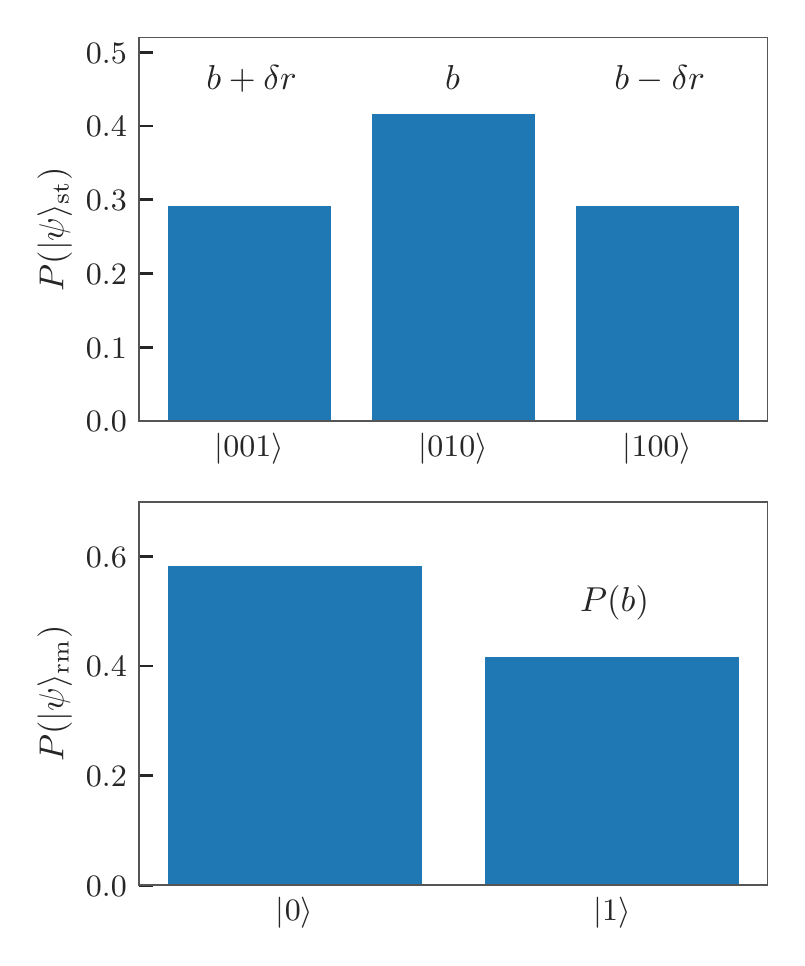}
  \caption{
    The distribution $P(r_{3\delta t})$ (top panel) and superposition of the
    ``risk factor'' qubit that encodes $P(r_{3\delta t} = b)$ (bottom panel).}
  \label{fig:ir}
\end{figure}
Figure~\ref{fig:ir} shows the probability distribution of the ``state'' qubits
(top panel) and the distribution of the ``risk measure'' qubit (bottom panel) at
$t = 3\delta t$ when measuring these qubits directly with $10,000$ shots.
\begin{figure}
  \centering
  \includegraphics{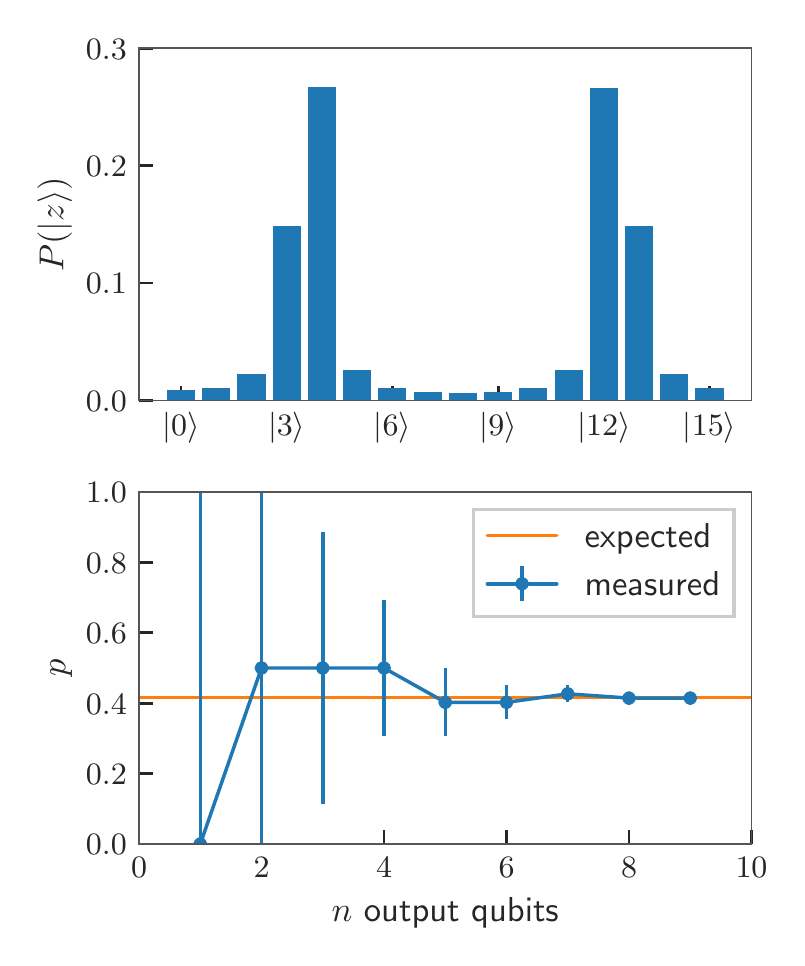}
  \caption{
    Top panel: the distribution $P(\ket{z})$ of the measured output state for
    $n = 4$ and $10,000$ shots.
    Bottom panel: the measured value of $p$ as a function of the number of
    output qubits for $1$ shot.}
  \label{fig:ir_results}
\end{figure}

Similar to the equity risk factor analysis, Fig.~\ref{fig:ir_results} shows the
distribution of the output qubits for $n = 4$ (top panel) and the convergence of
the measured value of $p = P(r_{3\delta t} = b)$ as a function of the number of
output qubits (bottom panel).

\subsection{Credit risk factors}
  \label{sec:credit}

This section focuses on simulating default probabilities of issuers.
Among the common approaches are structural and reduced-form credit risk
models, as well as credit rating migration.

\subsubsection{Structural credit risk models}

In structural credit risk models, e.g. the Merton model \cite{Merton1974},
default takes place when the assets of a company become less than its
liabilities.
As a simple example, we assume that the liabilities $D_t$ have maturity at time
$T$, and that the value of the assets $A_t$ follows the stochastic process:
\begin{align}
dA_t = \mu A_t dt + \sigma A_t dW_t\,,
\end{align}
where $\mu$ is the mean, $\sigma$ the volatility, and $W_t$ a Wiener process.
Default occurs if at time $t = T$ the value of the assets is less than, or equal
to, the liabilities, $A_T \leq D_T$; therefore, the probability of default is
$P(A_T \leq D_T)$.

\paragraph{$\mathcal{D}_\mathrm{def}$: the distribution $P(A_t)$}

Since this problem has a similar stochastic process to that of equity risk
factors, we can model the evolution with a binomial tree using the gate
$\mathcal{D}_\mathrm{def} = \mathcal{D}_\mathrm{eq}$ described in
Sect.~\ref{sec:equity}.
As an example, we consider the same binomial tree parameters as those listed in
Table~\ref{tab:eq_params} and set $D_T = A_0d^4$.
Based on the top panel of Fig.~\ref{fig:eq_rf}, default occurs if $j = 0$ or
$j = 1$ (a ``down'' price move at all timesteps or a single ``up'' move before
time $T$).

\paragraph{$\mathcal{M}_\mathrm{def}$: probability of default}

The risk measure here is the probability of default: $P(A_T \leq D_T)$.
While in the equity risk factor examples the gates $\mathcal{M}_\mathrm{max}$
and $\mathcal{M}_\mathrm{min}$ pick up a single value of
$\ket{\psi}_\mathrm{rf}$ ($\ket{11...1}_\mathrm{rf}$ and
$\ket{00...0}_\mathrm{rf}$, respectively), here we assemble the gate
$\mathcal{M}_\mathrm{def}$ such that it flips the ``risk factor'' qubit for any
value of $\ket{\psi}_\mathrm{rf}$ that satisfies $A_T \leq D_T$; namely, if/when
the ``risk factor'' qubits takes one of the following $7$ values:
$\ket{000000}$, $\ket{000001}$, $\ket{000010}$, ..., $\ket{100000}$.

For the implementation of the gate we also include ``count'' qubits,
$\ket{\psi}_\mathrm{c} = \ket{j}_\mathrm{c}$, in order to store the binary value
of $j$.
For $6$ ``risk factor'' qubits we have $j \in \{0,...,6\}$, and thus $3$
``count'' qubits are needed to encode these $7$ values.
We also include ``state'' qubits, $\ket{\psi}_\mathrm{st}$, to represent each of
the values of $j$ that correspond to default; since there are two such values
here, $j = 0$ and $j = 1$, we include two ``state'' qubits.

The gate $\mathcal{M}_\mathrm{def}$ can be decomposed into three parts.
The first part counts the number of up moves in $\ket{\psi}_\mathrm{rf}$, i.e.
it computes $j$, and writes it to $\ket{\psi}_\mathrm{c}$.
The second part reads the state $\ket{\psi}_\mathrm{c}$ and flips the
corresponding ``state'' qubit when that value is $j = 0$ or $j = 1$.
The third part consists of an \textsc{or} gate than takes as input the two
``state'' qubits and flips the ``risk measure'' qubit if any of them is in the
state $\ket{1}$.

The first part can be achieved with the gate
$\mathcal{C} = \mathcal{C}_m...\mathcal{C}_2\mathcal{C}_1$:
\begin{align}
\ket{j}_\mathrm{c} &= \mathcal{C}\ket{000}_\mathrm{c}\,,
\end{align}
where each operator $\mathcal{C}_l$ increments the count by one if the ``risk
factor'' qubit of timestep $l$ is an ``up'' move:
$\ket{\psi}_\mathrm{rf}^{l\delta t} = \ket{1}$.
\begin{center}
\begin{quantikz}[column sep=0.1cm, row sep=0.4cm]
\lstick{$\ket{\psi}_\mathrm{rf}^{l\delta t}\,\,$}
  & \ctrl{2}\gategroup[14, steps=14,
    style={dashed, rounded corners, inner xsep=0}]{$\mathcal{C}_l$}
  & \qw
  & \cdots\qw
  & \qw
  & \qw
  & \qw
  & \qw
  & \qw
  & \qw
  & \cdots\qw
  & \qw
  & \qw
  & \ctrl{2}
  & \ctrl{2}
  & \qw \\
\wave&&&&&&&&&&&&&&& \\
\lstick{$\ket{\psi}_\mathrm{c}^1\,\,\,\,\,$}
  & \ctrl{6}
  & \qw
  & \cdots\qw
  & \qw
  & \qw
  & \qw
  & \qw
  & \qw
  & \qw
  & \cdots\qw
  & \qw
  & \qw
  & \ctrl{6}
  & \targ{}
  & \qw \\
\lstick{$\ket{\psi}_\mathrm{c}^2\,\,\,\,\,$}
  & \qw
  & \ctrl{5}
  & \cdots\qw
  & \qw
  & \qw
  & \qw
  & \qw
  & \qw
  & \qw
  & \cdots\qw
  & \ctrl{5}
  & \targ{}
  & \qw
  & \qw
  & \qw \\
\wave&&&&&&&&&&&&&&& \\
\lstick{$\ket{\psi}_\mathrm{c}^{s-2}$}
  & \qw
  & \qw
  & \cdots\qw
  & \ctrl{6}
  & \qw
  & \qw
  & \qw
  & \qw
  & \ctrl{6}
  & \cdots\qw
  & \qw
  & \qw
  & \qw
  & \qw
  & \qw \\
\lstick{$\ket{\psi}_\mathrm{c}^{s-1}$}
  & \qw
  & \qw
  & \cdots\qw
  & \qw
  & \ctrl{6}
  & \qw
  & \ctrl{6}
  & \targ{}
  & \qw
  & \cdots\qw
  & \qw
  & \qw
  & \qw
  & \qw
  & \qw \\
\lstick{$\ket{\psi}_\mathrm{c}^s\,\,\,\,\,\,$}
  & \qw
  & \qw
  & \cdots\qw
  & \qw
  & \qw
  & \targ{}
  & \qw
  & \qw
  & \qw
  & \cdots\qw
  & \qw
  & \qw
  & \qw
  & \qw
  & \qw \\
\lstick{$\ket{0}_\mathrm{anc}^1\,\,$}
  & \targ{}
  & \ctrl{1}
  & \cdots\qw
  & \qw
  & \qw
  & \qw
  & \qw
  & \qw
  & \qw
  & \cdots\qw
  & \ctrl{1}
  & \ctrl{-5}
  & \targ{}
  & \qw
  & \qw \\
\lstick{$\ket{0}_\mathrm{anc}^2\,\,$}
  & \qw
  & \targ{}
  & \cdots\qw
  & \qw
  & \qw
  & \qw
  & \qw
  & \qw
  & \qw
  & \cdots\qw
  & \targ{}
  & \qw
  & \qw
  & \qw
  & \qw \\
\wave&&&&&&&&&&&&&&& \\
\lstick{$\ket{0}_\mathrm{anc}^{s-3}$}
  & \qw
  & \qw
  & \cdots\qw
  & \ctrl{1}
  & \qw
  & \qw
  & \qw
  & \qw
  & \ctrl{1}
  & \cdots\qw
  & \qw
  & \qw
  & \qw
  & \qw
  & \qw \\
\lstick{$\ket{0}_\mathrm{anc}^{s-2}$}
  & \qw
  & \qw
  & \cdots\qw
  & \targ{}
  & \ctrl{1}
  & \qw
  & \ctrl{1}
  & \ctrl{-6}
  & \targ{}
  & \cdots\qw
  & \qw
  & \qw
  & \qw
  & \qw
  & \qw \\
\lstick{$\ket{0}_\mathrm{anc}^{s-1}$}
  & \qw
  & \qw
  & \cdots\qw
  & \qw
  & \targ{}
  & \ctrl{-6}
  & \targ{}
  & \qw
  & \qw
  & \cdots\qw
  & \qw
  & \qw
  & \qw
  & \qw
  & \qw
\end{quantikz}
\end{center}

After the operator $\mathcal{C}$ counts the ``up'' moves ($j$ in total) and
writes them as a binary number to $\ket{\psi}_\mathrm{c}$, the operator
$\mathcal{J} = \mathcal{J}_m...\mathcal{J}_1\mathcal{J}_0$ reads $j$ and flips
the corresponding qubit of the ``state'' qubits $\ket{\psi}_\mathrm{st}$, see
Fig.~\ref{fig:jgate}.
\begin{figure*}
\begin{center}
\begin{quantikz}[column sep=0.2cm]
\lstick{$\ket{0}_\mathrm{c}^1\quad\,$}
  & \gate{X}\gategroup[12, steps=5,
    style={dashed, rounded corners, inner xsep=0}]{$\mathcal{J}_0$}
  & \gate[7]{\mathcal{\,A\,}}
  & \qw
  & \gate[7]{\mathcal{A}^\dagger}
  & \gate{X}
  & \qw\gategroup[12, steps=5,
    style={dashed, rounded corners, inner xsep=0}]{$\mathcal{J}_1$}
  & \gate[7]{\mathcal{\,A\,}}
  & \qw
  & \gate[7]{\mathcal{A}^\dagger}
  & \qw
  & \gate{X}\gategroup[12, steps=5,
    style={dashed, rounded corners, inner xsep=0}]{$\mathcal{J}_2$}
  & \gate[7]{\mathcal{\,A\,}}
  & \qw
  & \gate[7]{\mathcal{A}^\dagger}
  & \gate{X}
  & \cdots\qw
  & \gate[7]{\mathcal{\,A\,}}\gategroup[12, steps=3,
    style={dashed, rounded corners, inner xsep=0}]{$\mathcal{J}_m$}
  & \qw
  & \gate[7]{\mathcal{A}^\dagger}
  & \qw \\
\lstick{$\ket{0}_\mathrm{c}^2\quad\,$}
  & \gate{X}
  & \qw
  & \qw
  & \qw
  & \gate{X}
  & \gate{X}
  & \qw
  & \qw
  & \qw
  & \gate{X}
  & \qw
  & \qw
  & \qw
  & \qw
  & \qw
  & \cdots\qw
  & \qw
  & \qw
  & \qw
  & \qw \\
\wave&&&&&&&&&&&&&&&&&&&& \\
\lstick{$\ket{0}_\mathrm{c}^s\quad\,$}
  & \gate{X}
  & \qw
  & \qw
  & \qw
  & \gate{X}
  & \gate{X}
  & \qw
  & \qw
  & \qw
  & \gate{X}
  & \gate{X}
  & \qw
  & \qw
  & \qw
  & \gate{X}
  & \cdots\qw
  & \qw
  & \qw
  & \qw
  & \qw \\
\lstick{$\ket{0}_\mathrm{anc}^1\,\,$}
  & \qw
  & \qw
  & \qw
  & \qw
  & \qw
  & \qw
  & \qw
  & \qw
  & \qw
  & \qw
  & \qw
  & \qw
  & \qw
  & \qw
  & \qw
  & \cdots\qw
  & \qw
  & \qw
  & \qw
  & \qw \\
\wave&&&&&&&&&&&&&&&&&&&& \\
\lstick{$\ket{0}_\mathrm{anc}^{s-1}\,$}
  & \qw
  & \qw
  & \ctrl{1}
  & \qw
  & \qw
  & \qw
  & \qw
  & \ctrl{2}
  & \qw
  & \qw
  & \qw
  & \qw
  & \ctrl{3}
  & \qw
  & \qw
  & \cdots\qw
  & \qw
  & \ctrl{5}
  & \qw
  & \qw \\
\lstick{$\ket{0}_\mathrm{st}^{j=0}\,$}
  & \qw
  & \qw
  & \targ{}
  & \qw
  & \qw
  & \qw
  & \qw
  & \qw
  & \qw
  & \qw
  & \qw
  & \qw
  & \qw
  & \qw
  & \qw
  & \cdots\qw
  & \qw
  & \qw
  & \qw
  & \qw \\
\lstick{$\ket{0}_\mathrm{st}^{j=1}\,$}
  & \qw
  & \qw
  & \qw
  & \qw
  & \qw
  & \qw
  & \qw
  & \targ{}
  & \qw
  & \qw
  & \qw
  & \qw
  & \qw
  & \qw
  & \qw
  & \cdots\qw
  & \qw
  & \qw
  & \qw
  & \qw \\
\lstick{$\ket{0}_\mathrm{st}^{j=2}\,$}
  & \qw
  & \qw
  & \qw
  & \qw
  & \qw
  & \qw
  & \qw
  & \qw
  & \qw
  & \qw
  & \qw
  & \qw
  & \targ{}
  & \qw
  & \qw
  & \cdots\qw
  & \qw
  & \qw
  & \qw
  & \qw \\
\wave&&&&&&&&&&&&&&&&&&&& \\
\lstick{$\ket{0}_\mathrm{st}^{j=m}$}
  & \qw
  & \qw
  & \qw
  & \qw
  & \qw
  & \qw
  & \qw
  & \qw
  & \qw
  & \qw
  & \qw
  & \qw
  & \qw
  & \qw
  & \qw
  & \cdots\qw
  & \qw
  & \targ{}
  & \qw
  & \qw
\end{quantikz}
\end{center}
\caption{
  The gates $\mathcal{J}_l$ that flip the ``state'' qubit
  $\ket{0}_\mathrm{st}^{j=l}$ if $\ket{\psi}_\mathrm{c} = \ket{l}_\mathrm{c}$
  \label{fig:jgate}}
\end{figure*}
For our example, we are interested in the values $j = 0$ and $j = 1$ so we only
include gates $\mathcal{J}_0$ and $\mathcal{J}_1$.
Therefore, the components of gate $\mathcal{M}_\mathrm{def}$ are:
\begin{center}
\begin{quantikz}[column sep=0.2cm]
\lstick{$\ket{\psi}_\mathrm{rf}\,\,\,\,$}
  & \gate[3]{\mathcal{C}_1}\gategroup[6, steps=4,
    style={dashed, rounded corners, inner xsep=0}]{$\mathcal{C}$}
    \gategroup[6, steps=7,
      style={dashed, rounded corners, inner xsep=0, inner ysep=20pt}]
    {$\mathcal{M}_\mathrm{def}$}
    \qwbundle[alternate]{}
  & \gate[3]{\mathcal{C}_2}\qwbundle[alternate]{}
  & \cdots\qwbundle[alternate]{}
  & \gate[3]{\mathcal{C}_m}\qwbundle[alternate]{}
  & \qwbundle[alternate]{}\gategroup[6, steps=2,
    style={dashed, rounded corners, inner xsep=0}]{$\mathcal{J}$}
  & \qwbundle[alternate]{}
  & \qwbundle[alternate]{}
  & \qwbundle[alternate]{} \\
\lstick{$\ket{\psi}_\mathrm{c}\,\,\,\,\,$}
  & \qwbundle[alternate]{}
  & \qwbundle[alternate]{}
  & \cdots\qwbundle[alternate]{}
  & \qwbundle[alternate]{}
  & \gate[3]{\mathcal{J}_0}\qwbundle[alternate]{}
  & \gate[4]{\mathcal{J}_1}\qwbundle[alternate]{}
  & \qwbundle[alternate]{}
  & \qwbundle[alternate]{} \\
\lstick{$\ket{0}_\mathrm{anc}\,$}
  & \qwbundle[alternate]{}
  & \qwbundle[alternate]{}
  & \cdots\qwbundle[alternate]{}
  & \qwbundle[alternate]{}
  & \qwbundle[alternate]{}
  & \qwbundle[alternate]{}
  & \qwbundle[alternate]{}
  & \qwbundle[alternate]{} \\
\lstick{$\ket{0}_\mathrm{st}^{j=0}$}
  & \qw
  & \qw
  & \cdots\qw
  & \qw
  & \qw
  & \qw
  & \gate[3]{\mathrm{OR}}
  & \qw \\
\lstick{$\ket{0}_\mathrm{st}^{j=1}$}
  & \qw
  & \qw
  & \cdots\qw
  & \qw
  & \qw
  & \qw
  & \qw
  & \qw \\
\lstick{$\ket{0}_\mathrm{rm}\,\,$}
  & \qw
  & \qw
  & \cdots\qw
  & \qw
  & \qw
  & \qw
  & \qw
  & \qw
\end{quantikz}
\end{center}

\paragraph{Results}

\begin{figure}
  \centering
  \includegraphics{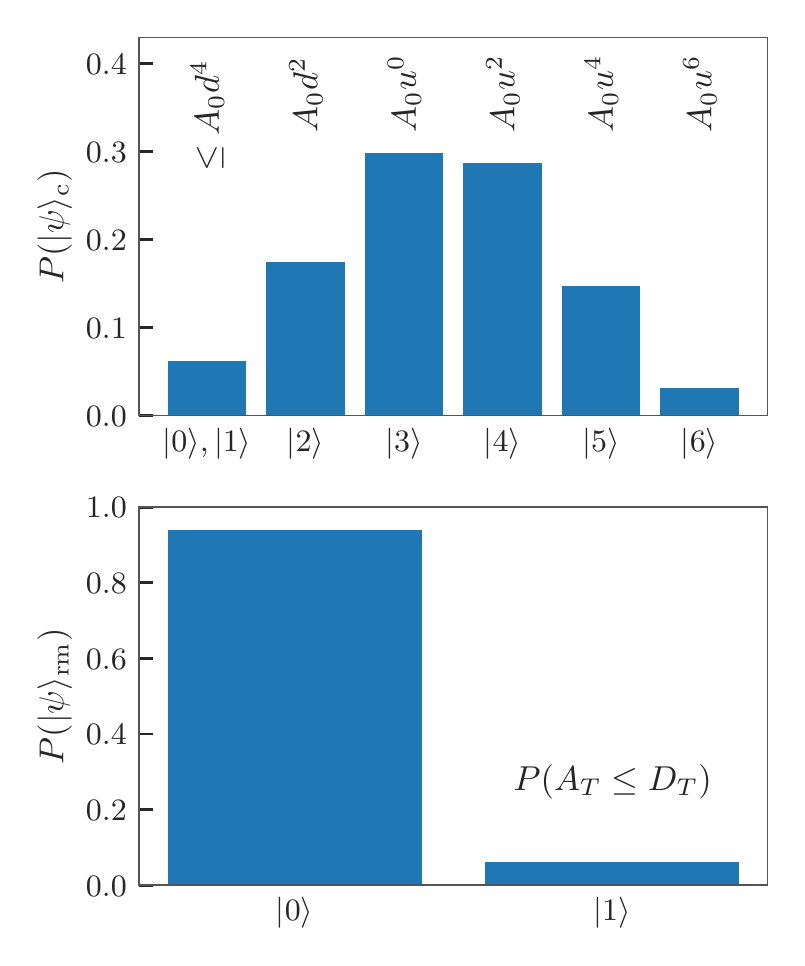}
  \caption{
    Top panel: the distribution of $A_T$ as generated from 10,000 shots; the
    first bar represents the  two ``default'' states.
    Bottom panel: the distribution of the ``risk measure'' qubit representing
    the probability of default.}
  \label{fig:pd}
\end{figure}
The top panel of Fig.~\ref{fig:pd} shows the probability distribution of the
``count'' qubits and the bottom panel that of the ``risk measure'' qubit.
Since the binomial tree parameters are the same as in the equity risk factor
example, the top panel is identical to Fig.~\ref{fig:eq_rf} apart from the
stacking of the ``default'' states $\ket{0}$ and $\ket{1}$ (the probability of
which is encoded in the ``risk measure'' qubit shown in the bottom panel).
\begin{figure}
  \centering
  \includegraphics{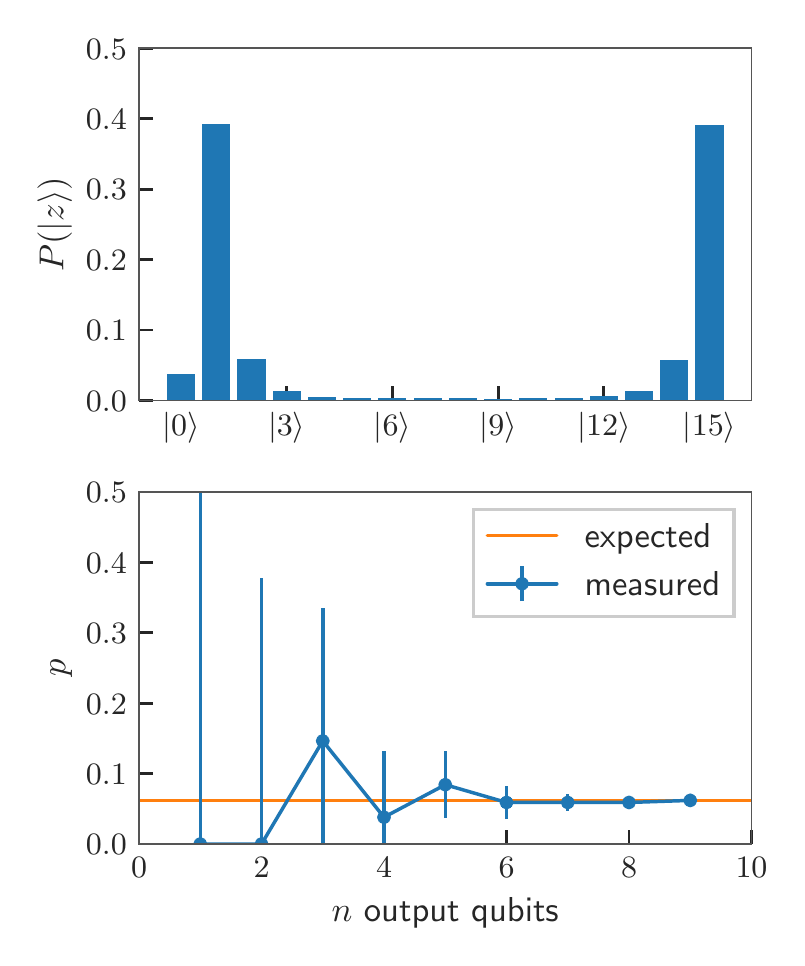}
  \caption{The measured probability distribution $P(\ket{z})$ (top panel) and
  the convergence of the estimated probability of default when using one shot
  (bottom panel).}
  \label{fig:pd_results}
\end{figure}
Figure~\ref{fig:pd_results} shows the measurement of the output qubits and the
convergence of the probability distribution when assembling all gates into a
quantum circuit.
The expected value of the probability is
\begin{align}
P(j = 0) &+ P(j = 1) = \nonumber\\
&= \sum_{j=0}^1 \frac{m!}{(m-j)!j!}q^j(1-q)^{m-j} \nonumber\\
&= (1-q)^m + mq(1-q)^{m-1}
\end{align}

\subsubsection{Reduced-form credit risk models}

Another approach to estimate probabilities of default is reduced-form credit
risk models, in which default is modelled as a statistical process.
The survival probability from time $t_0$ to time $t$ is given by:
\begin{align}
  P(t_0 + t) &= e^{-t/T_\mathrm{def}}\,,
\end{align}
where $T_\mathrm{def}$ is a characteristic timescale, the inverse of which is
called the hazard rate, $\lambda = 1/T_\mathrm{def}$.

To calculate the survival probability at time $T$ we can discretise the time
interval $t\in[0,T]$ with $m$ timesteps, $\delta t = T/m$, such that the
survival probability is:
\begin{align}
  P(T) = e^{-\sum\delta t/T} = \prod_{j = 1}^{m}P(\delta t)\,.
\end{align}
A visualisation of this process is:
\begin{center}
\begin{tikzpicture}[->, node distance = 2cm]
\node[state](1){$S_0$};
\node[state](2)[right of=1]{$S_{\delta t}$};
\node[state](3)[below of=2]{$D_{\delta t}$};
\node[state](4)[right of=2]{$S_{2\delta t}$};
\node[state](5)[below of=4]{$D_{2\delta t}$};
\node[state](6)[right of=4]{$S_{3\delta t}$};
\node[state](7)[below of=6]{$D_{3\delta t}$};
\path (1) edge [swap] node {}(2);
\path (1) edge [swap] node {}(3);
\path (2) edge [swap] node {}(4);
\path (2) edge [swap] node {}(5);
\path (3) edge [swap] node {}(5);
\path (4) edge [swap] node {}(6);
\path (4) edge [swap] node {}(7);
\path (5) edge [swap] node {}(7);
\end{tikzpicture}
\end{center}
where $S_t$ and $D_t$ denote the survival or default states at time $t$,
respectively.

\begin{table}
\begin{center}
\begin{tabular}{c|c}
Parameter                                  & Value        \\
\hline
$m$                                        & $6$          \\
$n$                                        & $1$-$9$      \\
$T$                                        & $1$          \\
$q_\mathrm{def}$                           & $2\%$        \\
$\theta_\mathrm{def}\frac{180^\circ}{\pi}$ & $\sim$$16.3^\circ$
\end{tabular}
\end{center}
\caption{
  List of parameters for the reduced-form credit risk model.
  \label{tab:surv_params}}
\end{table}

\paragraph{$\mathcal{D}_\mathrm{surv}$: the distribution $P(T)$}

At each of the $m$ timesteps we can represent survival with one ``risk factor''
qubit, such that $\ket{0}$ represents survival and $\ket{1}$ default.
Thus, if the company has survived until time $t = (l-1)\delta t$, then
$\ket{\psi}_\mathrm{rf}^{l\delta t} = \sqrt{1-q_\mathrm{def}}\ket{0}
+ \sqrt{q_\mathrm{def}}\ket{1}$, where $q_\mathrm{def} = 1 - P(\delta t)$.
If the company has defaulted at a previous timestep, then:
$\ket{\psi}_\mathrm{rf}^{l\delta t} = \ket{1}$.
For the example of $m = 3$:
\begin{center}
\begin{tikzpicture}[->, node distance = 2cm]
\node[state](1){$\ket{\emptyset\emptyset\emptyset}^0$};
\node[state](2)[right of=1]{$\ket{\emptyset\emptyset0}_\mathrm{rf}^{\delta t}$};
\node[state](3)[below of=2]{$\ket{\emptyset\emptyset1}_\mathrm{rf}^{\delta t}$};
\node[state](4)[right of=2]{$\ket{\emptyset00}_\mathrm{rf}^{2\delta t}$};
\node[state](5)[below of=4]{$\ket{\emptyset10}_\mathrm{rf}^{2\delta t}$};
\node[state](6)[right of=4]{$\ket{000}_\mathrm{rf}^{3\delta t}$};
\node[state](7)[below of=6]{$\ket{100}_\mathrm{rf}^{3\delta t}$};
\node[state](8)[below of=5]{$\ket{\emptyset11}_\mathrm{rf}^{2\delta t}$};
\node[state](9)[below of=7]{$\ket{110}_\mathrm{rf}^{3\delta t}$};
\node[state](10)[below of=9]{$\ket{111}_\mathrm{rf}^{3\delta t}$};
\path (1) edge [swap] node {}(2);
\path (1) edge [swap] node {}(3);
\path (2) edge [swap] node {}(4);
\path (2) edge [swap] node {}(5);
\path (4) edge [swap] node {}(6);
\path (4) edge [swap] node {}(7);
\path (3) edge [swap] node {}(8);
\path (5) edge [swap] node {}(9);
\path (8) edge [swap] node {}(10);
\end{tikzpicture}
\end{center}
Gate $\mathcal{D}_\mathrm{surv}$ can be assembled with rotation gates
$R_y(\theta_\mathrm{def})$ (shown below as $\theta_\mathrm{d}$ for brevity)
obtained from $q_\mathrm{def} = \sin(\theta_\mathrm{def}/2)$.
In addition, in order to ensure that in the event of default at time $t$ the
company will also be in default at $t + \delta t$, we include a controlled gate
$R_y(\theta_\mathrm{def}^\mathrm{c})$, where
$\theta_\mathrm{def}^\mathrm{c} + \theta_\mathrm{def} = \pi$: this breaks the
superposition of the qubit representing $t + \delta t$ and sets it to the state
$\ket{1}$.
Therefore, the gate $\mathcal{D}_\mathrm{surv}$ is:
\begin{center}
\begin{quantikz}[column sep=0.15cm]
\lstick{$\ket{0}_\mathrm{rf}^{\delta t}\,\,\,\,\,$}
  & \gate{R_y^{\theta_\mathrm{d}}}\gategroup[4, steps=7,
    style={dashed, rounded corners, inner xsep=0}]{$\mathcal{D}_\mathrm{surv}$}
  & \ctrl{1}
  & \qw
  & \qw
  & \cdots\qw
  & \qw
  & \qw
  & \qw \\
\lstick{$\ket{0}_\mathrm{rf}^{2\delta t}\,\,$}
  & \qw
  & \gate{R_y^{\theta_\mathrm{d}^\mathrm{c}}}
  & \gate{R_y^{\theta_\mathrm{d}}}
  & \ctrl{1}
  & \cdots\qw
  & \qw
  & \qw
  & \qw \\
\wave&&&&&&\ctrl{1}&& \\
\lstick{$\ket{0}_\mathrm{rf}^{m\delta t}$}
  & \qw
  & \qw
  & \qw
  & \qw
  & \cdots\qw
  & \gate{R_y^{\theta_\mathrm{d}^\mathrm{c}}}
  & \gate{R_y^{\theta_\mathrm{d}}}
  & \qw
\end{quantikz}
\end{center}
The inverse gate, $\mathcal{D}_\mathrm{surv}^\dagger$, has the rotation gates in
reverse order and the signs of the rotation angles flipped.

\paragraph{$\mathcal{M}_\mathrm{surv}$: survival probability}

The probability of survival at $t = T$ is represented by the state
$\ket{0...00}_\mathrm{rf}^T$.
Therefore, we can use the gate
$\mathcal{M}_\mathrm{surv} = \mathcal{M}_\mathrm{min}$ decribed in
Sect~\ref{sec:equity}, which flips the ``risk measure'' qubit if all ``risk
factor'' qubits are in the state $\ket{0}$.
The expected value of the probability of survival at time $T$ is
$(1 - q_\mathrm{def})^m$.

\paragraph{Results}

\begin{figure}
  \centering
  \includegraphics{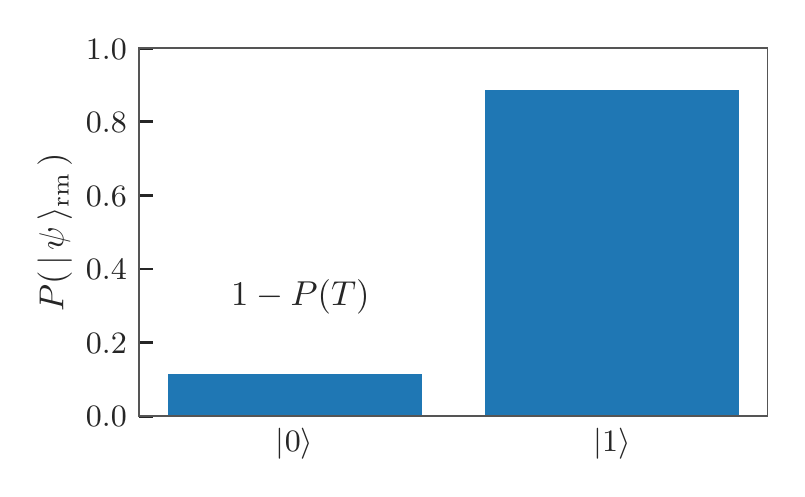}
  \caption{
    The probabilities of the ``risk measure'' qubit states: $\ket{1}$ denotes
    survival up to $t = T$ and $\ket{0}$ denotes default at any timestep before
    reaching $T$.}
  \label{fig:sp}
\end{figure}
Table~\ref{tab:surv_params} lists the parameters of the quantum circuit, and
Fig.~\ref{fig:sp} the survival ($\ket{1}_\mathrm{rm}$) and default
($\ket{0}_\mathrm{rm}$) probabilities encoded in the ``risk measure'' qubit.
\begin{figure}
  \centering
  \includegraphics{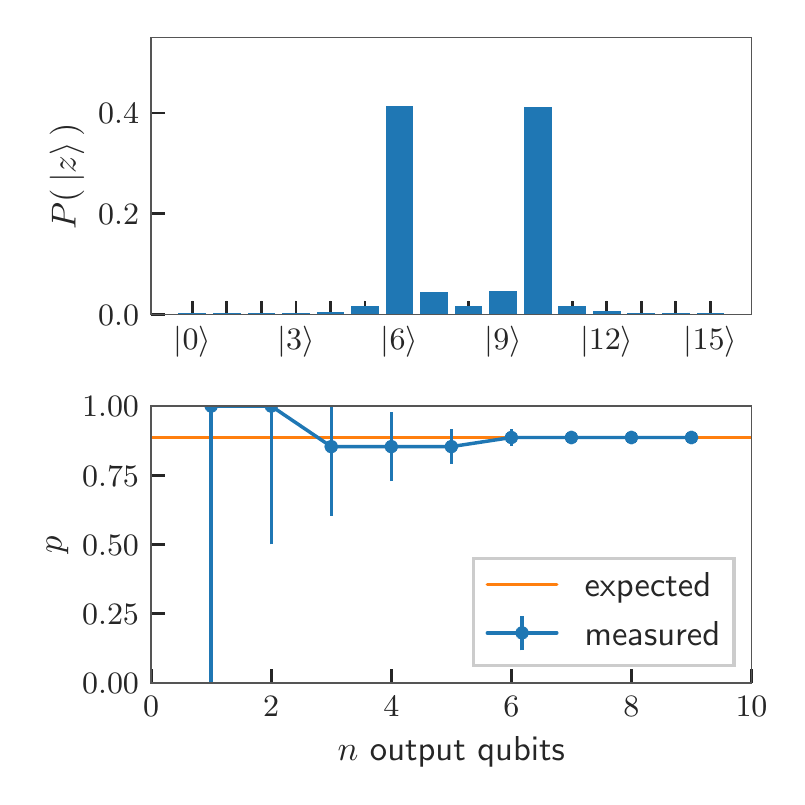}
  \caption{
    Top panel: the probability distribution of the output qubits for $m = 4$ and
    10,000 shots.
    Bottom panel: the estimated probability (one shot) and its convergence when
    increasing the number of output qubits.}
  \label{fig:sp_results}
\end{figure}
Figure~\ref{fig:sp_results} shows the measurement of the output state and its
convergence similarly to the previous use cases.

\subsubsection{Credit rating migration}

A more elaborate approach to simulate probabilities of default is with credit
rating migration matrices.
For example, consider the following three ratings: A (investment grade), B
(high yield), and D (defaulted), and assume that, at $t = 0$, an issuer has the
rating A$_0$.
The evolution of the credit rating can be described with multinomial trees, such
as this:
\begin{center}
\begin{tikzpicture}[->, node distance = 2cm]
\node[state](1){A$_0$};
\node[state](2)[right of=1]{A$_{\delta t}$};
\node[state](3)[below of=2]{B$_{\delta t}$};
\node[state](4)[below of=3]{D$_{\delta t}$};
\node[state](5)[right of=2]{A$_{2\delta t}$};
\node[state](6)[below of=5]{B$_{2\delta t}$};
\node[state](7)[below of=6]{D$_{2\delta t}$};
\node[state](8)[right of=5]{A$_{3\delta t}$};
\node[state](9)[below of=8]{B$_{3\delta t}$};
\node[state](10)[below of=9]{D$_{3\delta t}$};
\path (1) edge [swap] node {}(2);
\path (1) edge [swap] node {}(3);
\path (1) edge [swap] node {}(4);
\path (2) edge [swap] node {}(5);
\path (2) edge [swap] node {}(6);
\path (2) edge [swap] node {}(7);
\path (3) edge [swap] node {}(5);
\path (3) edge [swap] node {}(6);
\path (3) edge [swap] node {}(7);
\path (4) edge [swap] node {}(7);
\path (5) edge [swap] node {}(8);
\path (5) edge [swap] node {}(9);
\path (5) edge [swap] node {}(10);
\path (6) edge [swap] node {}(8);
\path (6) edge [swap] node {}(9);
\path (6) edge [swap] node {}(10);
\path (7) edge [swap] node {}(10);
\end{tikzpicture}
\end{center}
where each transition, e.g. $q_{\mathrm{A}\to\mathrm{B}}$ has a predefined
probability which is calibrated with historical data.

Since we have already described the implementation of such trees in
Sect.~\ref{sec:ir}, we do not provide a more detailed example here.
The credit rating evolution and probability of default at time $t$ can be
modelled with the gate $\mathcal{D}_\mathrm{migr} = \mathcal{D}_\mathrm{ir}$
(with a minor adjustment to set the initial rating to A, which would correspond
to the high interest rate level), and the default probability can be encoded
with $\mathcal{M}_\mathrm{def} = \mathcal{M}_\mathrm{l}$ (corresponding to
measuring the low interest rate level).

\section{Discussion}
  \label{sec:discussion}

We proceed to estimate the number of qubits required, and the resulting circuit
depth, for typical financial risk use cases.
We limit the analysis to scenario generation and ignore the additional qubits
and gates that would be needed when, for example, the pricing of a derivative is
also included in the circuit.

Since the number of ``risk factor'' qubits, $m_\mathrm{rf}$, determines the
number of time steps that discretises a time period, we can consider
$m_\mathrm{rf}$ to represent ``risk model accuracy''.
Similarly, because the number of output qubits, $n$, determines the error of the
estimate, we consider $n$ to represent ``measurement precision''.
The higher the number of input and output qubits is, the higher the accuracy and
precision are, respectively.

Financial risk models ignore intra-day changes, thus the smallest timestep is
one business day.
For scenario generation over long time horizons, such as those that span
decades, timesteps up to one month can be a sufficient as well as an efficient
choice.

Equity price scenarios are used in pricing exchange-traded equity derivatives,
most of which expire in less than $12$ months, with some extending up to $2$
years.
With $\sim$$260$ business days in a year, the $\mathcal{D}_\mathrm{eq}$ gate
would not require more than $\sim$$520$ ``risk factor'' qubits.
Equity scenarios are also used in equity indices derivatives, structural credit
risk models, convertible bonds, and over-the-counter equity derivatives, all of
which can have time horizons from few months to several years.
Even for periods as large as $50$ years, and by considering a timestep of one
month, the $\mathcal{D}_\mathrm{eq}$ gate has an upper limit of $600$ qubits.
Because the number of ``ancilla'' qubits, $m_\mathrm{anc}$, for the gates
$\mathcal{M}_\mathrm{min}$ and $\mathcal{M}_\mathrm{max}$ scales linearly with
$m_\mathrm{rf}$, these gates require roughly:
$m_\mathrm{anc} \simeq m_\mathrm{rf}$.
This is not the case for the gate $\mathcal{M}_\mathrm{def}$, which only needs
few additional qubits (``count'', ``state'', and ``ancilla'') as $m_\mathrm{rf}$
increases; e.g., only $8$ qubits are needed to count from $1$ to $256$.
Overall, the total number of input qubits (``risk factor'', ``risk measure'',
``ancilla'', ``count'', ``state'') for typical use cases of equity risk factor
scenarios has an upper limit of $\sim$$1,200$.

Reduced-form credit risk models can be used to simulate the survival probability
of an entity over several years, whether for credit default swaps or
counterparty credit risk use cases.
Similar to structural credit risk models, a timestep of one month can be
sufficient for long time horizons; this limits the number of ``risk factor''
qubits of the $\mathcal{D}_\mathrm{surv}$ gate to be a few hundred at most (e.g.
$600$ for $50$ years).
Because $\mathcal{M}_\mathrm{surv}$ needs a number of ``ancilla'' qubits similar
to $m_\mathrm{rf}$, the total number of input qubits for this type of circuits
would not exceed $\sim$$1,200$ either.

For interest rate scenarios, the number of ``risk factor'' qubits has a linear
dependence on the number of timesteps as in the case of equity scenarios.
Because interest rates do not fluctuate as much on a daily basis as equities, a
timestep of one month is often adequate to model their evolution.
On the other hand, by increasing the number of nodes per timestep, the number of
qubits needed to model the discrete interest rate values also increases (based
on the implementation of the $\mathcal{D}_\mathrm{ir}$ gate here).
The scaling is sublinear though, as each new additional qubit can double the
amount of interest rate values modelled.
Moreover, the $\mathcal{M}_\mathrm{m}$ gate of our example only needs one
additional qubit, the ``risk measure'' one.
While we refrain from a more detailed estimate, modelling a $50$-year period
would not require more than a thousand input qubits.
The same analysis also applies to multinomial trees for the modelling of credit
rating migration.

In terms of precision, we can require that
$\delta p \sim 1\,\mathrm{bp} = 0.01\% = 10^{-4}$,\footnote{
  ``bp'' stands for basis points.}
which can be achieved with $n \simeq 14$ output qubits (based on
Eq.~\ref{eq:dpqmc}).

Therefore, when accounting for both input and output qubits, scenario generation
for typical financial risk use cases would overall require around or a little
more than a thousand qubits for high-accuracy, high-precision results.

\begin{figure*}
\centering
\includegraphics{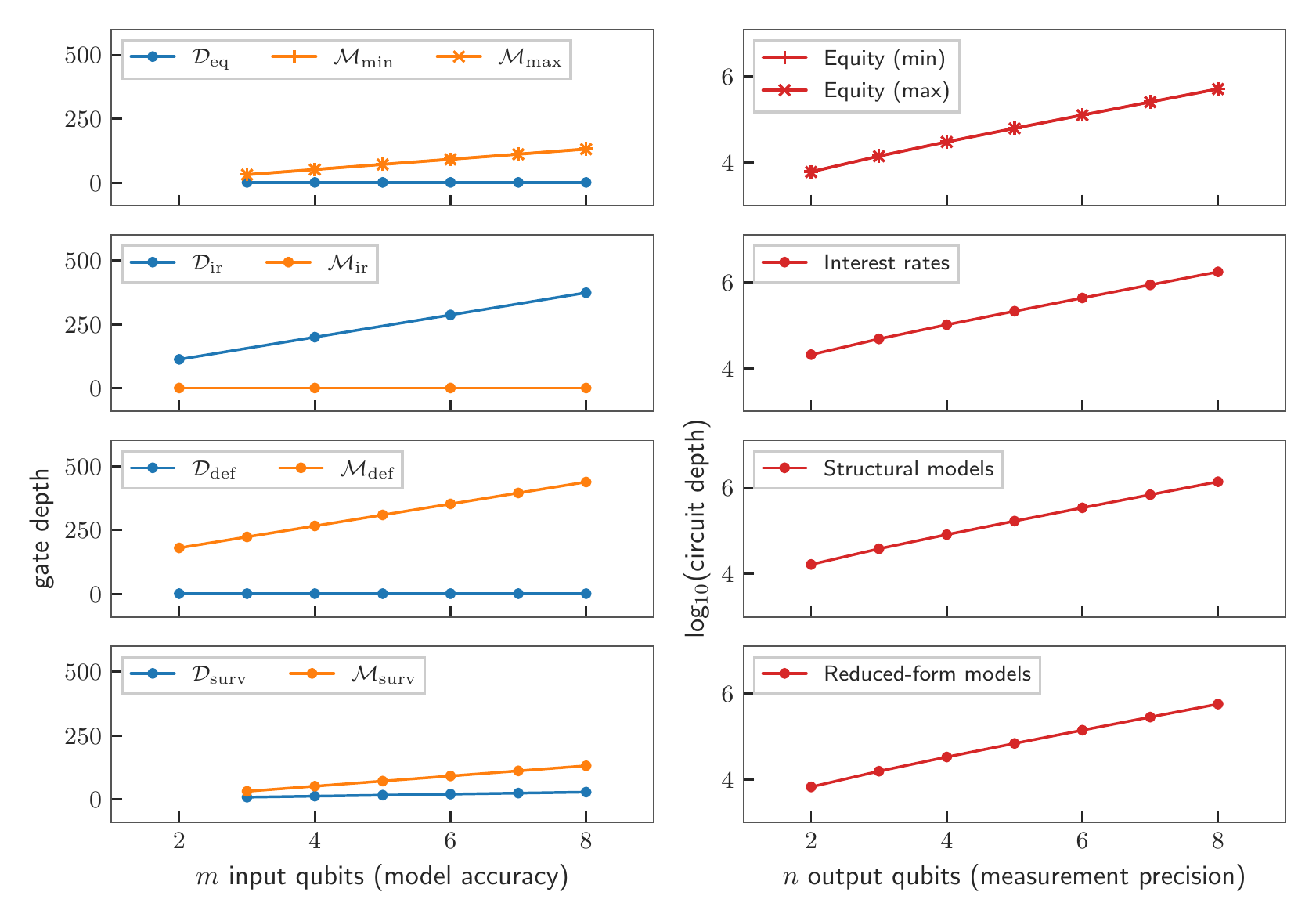}
\caption{
  Left panels: depth of the gates $\mathcal{D}$ and $\mathcal{M}$ as a function
  of the number of input qubits, $m$, which represents model accuracy.
  Right panels: depth (in log scale) of the entire quantum circuit as a
  function of the number of output qubits, $n$, which represents measurement
  precision.}
\label{fig:scaling}
\end{figure*}

Next, we proceed to estimate the circuit depth.
With current quantum computers, the higher the number of gates the more prone
the computations are to errors due to noise.
We calculate the depth numerically by decomposing the gates or circuits of
Sect.~\ref{sec:scenario_generation} into their constituent gates, a process that
we repeat until no further decomposition is possible.\footnote{
  For the numerical calculation of the gate or circuit depth we use the Qiskit
  functions \texttt{decompose()} and \texttt{depth()} \cite{Qiskit}.
  For example, the depth of a $2$-qubit circuit in which each qubit is going
  through an $H$ gate, $H\ket{0}\otimes H\ket{0} = \ket{++}$, has a gate count
  of $2$ and a circuit depth of $1$ because the two gates act in parallel.
  An example of decomposition is breaking down the Toffoli gate into the basic
  gates it consists of, which gives a gate depth of $11$.}
The left panels of Fig.~\ref{fig:scaling} show the depth of gates $\mathcal{D}$
and $\mathcal{M}$ as a function of the number of input qubits $m$ (model
accuracy); the depth is found to scale linearly with $m$.
Since for $m \simeq 10$ the depth is $\lesssim 500$, for $m \sim 1,000$ we
expect the gate depth to be on the order of $\lesssim 50,000$.

The right panels of Fig.~\ref{fig:scaling} show the total circuit depth (it
includes the gates $\mathcal{D}$, $\mathcal{M}$, $\prod\mathcal{Q}$,
$\mathrm{QFT}$, and $\mathrm{QFT}^\dagger$) in log scale as a function of the
number of output qubits $n$ (model precision).
The depth increases exponentially with respect to $n$ because every extra output
qubit doubles the number of times the $\mathcal{Q}$ gate is applied
(see Eq.~\ref{eq:piq}).
For $n = 14$, the total depth is projected to be on the order of
$\lesssim 10^8$.
Estimates from pricing exotic derivatives report qubit numbers and circuit
depths that are on the same order of magnitude, $\sim$$10^3$-$10^4$ and
$\sim$$10^8$, respectively \cite{Chakrabarti+2021, StamatopoulosZeng2023}.

While quantum computers are expected to reach $\sim$$1,000$ qubits in the next
few years, the estimated circuit depth prevents such use cases from being
feasible without significant progress on fault tolerance.
Moreover, note that all the above estimates apply when modelling a single risk
factor.
For a portfolio that depends on multiple risk factors, we would need that many
times of input qubits (parallel run), or to generate scenarios sequentially
(serial run) \cite{Dalzell+2023}.
Hardware limitation make the latter option more likely in the near term.

\section{Summary and conclusions}
  \label{sec:conclusions}

QAE algorithms have been shown to provide a quadratic speedup over their
classical counterparts, a result that has motivated several recent papers on
financial risk applications.
Most of these papers start from pre-computed risk factor distributions and focus
on the calculation of common risk measures, distribution loading, and QAE
optimisations.
In this paper, we extend these studies by consistently integrating scenario
generation --- which we call QMC --- into QAE quantum circuits.
Specifically, we assemble quantum gates that implement stochastic risk models
for the evolution of equity, interest rate, and credit risk factors.
For equities, we generate scenarios by discretising a geometric Brownian motion
with binomial trees, and for interest rates, by discretising mean-reversion
stochastic differential equations with bounded trinomial trees.
For credit risk factors, we calculate the default probability from structural
models based on binomial trees and the survival probability from reduced-form
models based on a Poisson process.
We also describe how multinomial tress can be used for the implementation of
credit rating migration matrices.
Moreover, we assemble quantum gates to encode generic risk measures, such as the
probabibilities of measuring the minimum, maximum, and the tail of a
distribution.
For each use case, we build end-to-end QMC/QAE circuits that incorporate: the
generation of risk factor scenarios (QMC), the encoding of the risk measure, and
the estimation of the risk measure value (QAE).
We then validate the quantum computation with simulated runs, and demonstrate
that the measured value converges to the expected value and that the error goes
to zero as the number of output qubits increases.

For the typical model accuracy and measurement precision of realistic financial
use cases, we estimate that risk factor evolution requires $\lesssim$$1,200$
qubits, which is within the capabilities of the quantum computers expected in
the next few years.
However, we estimate that the typical circuit depth is on the order of
$\lesssim$$10^8$, which poses the biggest challenge until fault-tolerant quantum
devices are available.

The probabilistic output of quantum gates combined with the quantum property of
superposition provide a natural framework for the implementation of stochastic
risk models.
On the one hand, quantum gates can model the output states of a random variable
by encoding the probabilities of the outcomes.
On the other hand, superposition enables the quantum circuit to simultaneously
model all possible paths of a time-dependent random variable, eliminating the
classical-computation constraint of iterating over paths.
We conclude that quantum financial risk applications can benefit from 
consistently incorporating scenario generation as part of QMC/QAE simulations,
further reducing their dependency on classical computers.

\acknowledgements{
  We are grateful to three anonymous referees for their detailed reports that
  helped improve the results and presentation of the paper.
  The quantum computations were performed on a classical computer using a
  zero-noise simulator from Qiskit (version \texttt{0.24.2}), an open-source
  framework for quantum computing \cite{Qiskit}.
}

\bibliographystyle{quantum}
\bibliography{paper}

\begin{thebibliography}{10}

\bibitem{Orus+2019}
Román Orús, Samuel Mugel, and Enrique Lizaso.
\newblock ``Quantum computing for finance: Overview and prospects''.
\newblock \href{https://dx.doi.org/10.1016/j.revip.2019.100028}{Reviews in
  Physics {\bf 4}, 100028}~(2019).

\bibitem{Egger+2020b}
Daniel~J. Egger, Claudio Gambella, Jakub Marecek, Scott McFaddin, Martin
  Mevissen, Rudy Raymond, Andrea Simonetto, Stefan Woerner, and Elena Yndurain.
\newblock ``Quantum computing for finance: State-of-the-art and future
  prospects''.
\newblock \href{https://dx.doi.org/10.1109/tqe.2020.3030314}{IEEE Transactions
  on Quantum Engineering {\bf 1}, 1--24}~(2020).

\bibitem{Gomez+2022}
Andrés Gómez, Alvaro Leitao~Rodriguez, Alberto Manzano, Maria Nogueiras,
  Gustavo Ordóñez, and Carlos Vázquez.
\newblock ``A survey on quantum computational finance for derivatives pricing
  and var''.
\newblock \href{https://dx.doi.org/10.1007/s11831-022-09732-9}{Archives of
  Computational Methods in Engineering {\bf 29}, 4137–4163}~(2022).

\bibitem{Herman+2022}
Dylan {Herman}, Cody {Googin}, Xiaoyuan {Liu}, Alexey {Galda}, Ilya {Safro},
  Yue {Sun}, Marco {Pistoia}, and Yuri {Alexeev}.
\newblock ``{A Survey of Quantum Computing for Finance}''~(2022).
\newblock  \href{http://arxiv.org/abs/2201.02773}{arXiv:2201.02773}.

\bibitem{WilkensMoorhouse2023}
Sascha Wilkens and Joe Moorhouse.
\newblock ``Quantum computing for financial risk measurement''.
\newblock \href{https://dx.doi.org/10.1007/s11128-022-03777-2}{Quantum
  Information Processing{\bf 22}}~(2023).

\bibitem{Intallura+2023}
Philip {Intallura}, Georgios {Korpas}, Sudeepto {Chakraborty}, Vyacheslav
  {Kungurtsev}, and Jakub {Marecek}.
\newblock ``{A Survey of Quantum Alternatives to Randomized Algorithms: Monte
  Carlo Integration and Beyond}''~(2023).
\newblock  \href{http://arxiv.org/abs/2303.04945}{arXiv:2303.04945}.

\bibitem{Dalzell+2023}
Alexander~M. {Dalzell}, Sam {McArdle}, Mario {Berta}, Przemyslaw {Bienias},
  Chi-Fang {Chen}, Andr{\'a}s {Gily{\'e}n}, Connor~T. {Hann}, Michael~J.
  {Kastoryano}, Emil~T. {Khabiboulline}, Aleksander {Kubica}, Grant {Salton},
  Samson {Wang}, and Fernando G.~S.~L. {Brand{\~a}o}.
\newblock ``{Quantum algorithms: A survey of applications and end-to-end
  complexities}''~(2023).
\newblock  \href{http://arxiv.org/abs/2310.03011}{arXiv:2310.03011}.

\bibitem{WoernerEgger2019}
Stefan Woerner and Daniel~J. Egger.
\newblock ``Quantum risk analysis''.
\newblock \href{https://dx.doi.org/10.1038/s41534-019-0130-6}{npj Quantum
  Information {\bf 5}, 15}~(2019).

\bibitem{Egger+2020a}
D.~J. Egger, R.~Garcia Gutierrez, J.~Cahue Mestre, and S.~Woerner.
\newblock ``Credit risk analysis using quantum computers''.
\newblock \href{https://dx.doi.org/10.1109/TC.2020.3038063}{IEEE Transactions
  on ComputersPages 1--1}~(5555).

\bibitem{Kaneko+2021}
Kazuya {Kaneko}, Koichi {Miyamoto}, Naoyuki {Takeda}, and Kazuyoshi {Yoshino}.
\newblock ``{Quantum speedup of Monte Carlo integration with respect to the
  number of dimensions and its application to finance}''.
\newblock \href{https://dx.doi.org/10.1007/s11128-021-03127-8}{Quantum
  Information Processing {\bf 20}, 185}~(2021).

\bibitem{Rebentrost+2018}
Patrick Rebentrost, Brajesh Gupt, and Thomas~R. Bromley.
\newblock ``Quantum computational finance: Monte carlo pricing of financial
  derivatives''.
\newblock \href{https://dx.doi.org/10.1103/PhysRevA.98.022321}{Phys. Rev. A
  {\bf 98}, 022321}~(2018).

\bibitem{Stamatopoulos+2020}
Nikitas Stamatopoulos, Daniel~J. Egger, Yue Sun, Christa Zoufal, Raban Iten,
  Ning Shen, and Stefan Woerner.
\newblock ``Option {P}ricing using {Q}uantum {C}omputers''.
\newblock \href{https://dx.doi.org/10.22331/q-2020-07-06-291}{{Quantum} {\bf
  4}, 291}~(2020).

\bibitem{CarreraVazquezWoerner2021}
Almudena Carrera~Vazquez and Stefan Woerner.
\newblock ``Efficient state preparation for quantum amplitude estimation''.
\newblock \href{https://dx.doi.org/10.1103/PhysRevApplied.15.034027}{Phys. Rev.
  Appl. {\bf 15}, 034027}~(2021).

\bibitem{Chakrabarti+2021}
Shouvanik Chakrabarti, Rajiv Krishnakumar, Guglielmo Mazzola, Nikitas
  Stamatopoulos, Stefan Woerner, and William~J. Zeng.
\newblock ``A {T}hreshold for {Q}uantum {A}dvantage in {D}erivative
  {P}ricing''.
\newblock \href{https://dx.doi.org/10.22331/q-2021-06-01-463}{{Quantum} {\bf
  5}, 463}~(2021).

\bibitem{Doriguello+2021}
Jo{\~a}o~F. {Doriguello}, Alessandro {Luongo}, Jinge {Bao}, Patrick
  {Rebentrost}, and Miklos {Santha}.
\newblock ``{Quantum algorithm for stochastic optimal stopping problems with
  applications in finance}''~(2021).
\newblock  \href{http://arxiv.org/abs/2111.15332}{arXiv:2111.15332}.

\bibitem{Tang+2020}
Hao {Tang}, Anurag {Pal}, Lu-Feng {Qiao}, Tian-Yu {Wang}, Jun {Gao}, and
  Xian-Min {Jin}.
\newblock ``{Quantum Computation for Pricing the Collateralized Debt
  Obligations}''~(2020).
\newblock  \href{http://arxiv.org/abs/2008.04110}{arXiv:2008.04110}.

\bibitem{Alcazar+2022}
Javier Alcazar, Andrea Cadarso, Amara Katabarwa, Marta Mauri, Borja Peropadre,
  Guoming Wang, and Yudong Cao.
\newblock ``Quantum algorithm for credit valuation adjustments''.
\newblock \href{https://dx.doi.org/10.1088/1367-2630/ac5003}{New Journal of
  Physics {\bf 24}, 023036}~(2022).

\bibitem{HanRebentrost2022}
Jeong~Yu {Han} and Patrick {Rebentrost}.
\newblock ``{Quantum advantage for multi-option portfolio pricing and valuation
  adjustments}''~(2022).
\newblock  \href{http://arxiv.org/abs/2203.04924}{arXiv:2203.04924}.

\bibitem{Stamatopoulos+2022}
Nikitas {Stamatopoulos}, Guglielmo {Mazzola}, Stefan {Woerner}, and William~J.
  {Zeng}.
\newblock ``{Towards Quantum Advantage in Financial Market Risk using Quantum
  Gradient Algorithms}''.
\newblock \href{https://dx.doi.org/10.22331/q-2022-07-20-770}{Quantum {\bf 6},
  770}~(2022).

\bibitem{Preskill2018}
John Preskill.
\newblock ``Quantum {C}omputing in the {NISQ} era and beyond''.
\newblock \href{https://dx.doi.org/10.22331/q-2018-08-06-79}{{Quantum} {\bf 2},
  79}~(2018).

\bibitem{Brassard+2002}
Gilles Brassard, Peter Høyer, Michele Mosca, and Alain Tapp.
\newblock ``Quantum amplitude amplification and estimation''.
\newblock \href{https://dx.doi.org/10.1090/conm/305/05215}{Quantum Computation
  and InformationPages 53--74}~(2002).

\bibitem{GroverRudolph2002}
Lov {Grover} and Terry {Rudolph}.
\newblock ``{Creating superpositions that correspond to efficiently integrable
  probability distributions}''~(2002).
\newblock
  \href{http://arxiv.org/abs/quant-ph/0208112}{arXiv:quant-ph/0208112}.

\bibitem{Herbert2021}
Steven Herbert.
\newblock ``No quantum speedup with grover-rudolph state preparation for
  quantum monte carlo integration''.
\newblock \href{https://dx.doi.org/10.1103/PhysRevE.103.063302}{Phys. Rev. E
  {\bf 103}, 063302}~(2021).

\bibitem{Zoufal+2019}
Christa Zoufal, Aur\'elien Lucchi, and Stefan Woerner.
\newblock ``Quantum generative adversarial networks for learning and loading
  random distributions''.
\newblock \href{https://dx.doi.org/10.1038/s41534-019-0223-2}{npj Quantum
  Information {\bf 1}, 103}~(2019).

\bibitem{LiKais2021}
Junxu {Li} and Sabre {Kais}.
\newblock ``{A universal quantum circuit design for periodical functions}''.
\newblock \href{https://dx.doi.org/10.1088/1367-2630/ac2cb4}{New Journal of
  Physics {\bf 23}, 103022}~(2021).

\bibitem{StamatopoulosZeng2023}
Nikitas {Stamatopoulos} and William~J. {Zeng}.
\newblock ``{Derivative Pricing using Quantum Signal Processing}''~(2023).
\newblock  \href{http://arxiv.org/abs/2307.14310}{arXiv:2307.14310}.

\bibitem{McArdle2022}
Sam {McArdle}, Andr{\'a}s {Gily{\'e}n}, and Mario {Berta}.
\newblock ``{Quantum state preparation without coherent arithmetic}''~(2022).
\newblock  \href{http://arxiv.org/abs/2210.14892}{arXiv:2210.14892}.

\bibitem{Montanaro2015}
Ashley Montanaro.
\newblock ``Quantum speedup of monte carlo methods''.
\newblock \href{https://dx.doi.org/10.1098/rspa.2015.0301}{Proceedings of the
  Royal Society A: Mathematical, Physical and Engineering Sciences {\bf 471},
  20150301}~(2015).

\bibitem{Giles2015}
Michael~B. Giles.
\newblock ``Multilevel monte carlo methods''.
\newblock \href{https://dx.doi.org/10.1017/S096249291500001X}{Acta Numerica
  {\bf 24}, 259–328}~(2015).

\bibitem{An+2021}
Dong An, Noah Linden, Jin-Peng Liu, Ashley Montanaro, Changpeng Shao, and Jiasu
  Wang.
\newblock ``Quantum-accelerated multilevel {M}onte {C}arlo methods for
  stochastic differential equations in mathematical finance''.
\newblock \href{https://dx.doi.org/10.22331/q-2021-06-24-481}{{Quantum} {\bf
  5}, 481}~(2021).

\bibitem{Hull2021}
{John C.} Hull.
\newblock ``Options, futures, and other derivatives''.
\newblock Pearson. ~(2021).
\newblock 11th ed., pearson global ed. edition.

\bibitem{Grover1996}
Lov~K. Grover.
\newblock ``A fast quantum mechanical algorithm for database search''.
\newblock In Gary~L. Miller, editor, Proceedings of the Twenty-Eighth Annual
  {ACM} Symposium on the Theory of Computing, Philadelphia, Pennsylvania, USA,
  May 22-24, 1996.
\newblock \href{https://dx.doi.org/10.1145/237814.237866}{Pages 212--219}.
\newblock {ACM}~(1996).

\bibitem{Suzuki+2020}
Yohichi Suzuki, Shumpei Uno, Rudy Raymond, Tomoki Tanaka, Tamiya Onodera, and
  Naoki Yamamoto.
\newblock ``Amplitude estimation without phase estimation''.
\newblock \href{https://dx.doi.org/10.1007/s11128-019-2565-2}{Quantum
  Information Processing{\bf 19}}~(2020).

\bibitem{Grinko+2021}
Dmitry Grinko, Julien Gacon, Christa Zoufal, and Stefan Woerner.
\newblock ``Iterative quantum amplitude estimation''.
\newblock \href{https://dx.doi.org/10.1038/s41534-021-00379-1}{npj Quantum
  Information{\bf 7}}~(2021).

\bibitem{Plekhanov+2022}
Kirill Plekhanov, Matthias Rosenkranz, Mattia Fiorentini, and Michael Lubasch.
\newblock ``Variational quantum amplitude estimation''.
\newblock \href{https://dx.doi.org/10.22331/q-2022-03-17-670}{{Quantum} {\bf
  6}, 670}~(2022).

\bibitem{Cox+1979}
John~C. Cox, Stephen~A. Ross, and Mark Rubinstein.
\newblock ``Option pricing: A simplified approach''.
\newblock \href{https://dx.doi.org/10.1016/0304-405X(79)90015-1}{Journal of
  Financial Economics {\bf 7}, 229--263}~(1979).

\bibitem{Vedral+1996}
Vlatko Vedral, Adriano Barenco, and Artur Ekert.
\newblock ``Quantum networks for elementary arithmetic operations''.
\newblock \href{https://dx.doi.org/10.1103/PhysRevA.54.147}{Phys. Rev. A {\bf
  54}, 147--153}~(1996).

\bibitem{OliveiraRamos2007}
David Oliveira and Rubens Ramos.
\newblock ``Quantum bit string comparator: Circuits and applications''.
\newblock Quantum Computers and Computing{\bf 7}~(2007).

\bibitem{QiskitBook2023}
Various authors.
\newblock ``Qiskit textbook''.
\newblock Github. ~(2023).
\newblock  url:~\url{github.com/Qiskit/textbook}.

\bibitem{Vasicek1977}
Oldrich Vasicek.
\newblock ``An equilibrium characterization of the term structure''.
\newblock \href{https://dx.doi.org/10.1016/0304-405X(77)90016-2}{Journal of
  Financial Economics {\bf 5}, 177--188}~(1977).

\bibitem{Merton1974}
Robert~C. Merton.
\newblock ``On the pricing of corporate debt: the risk structure of interest
  rates''.
\newblock \href{https://dx.doi.org/10.1111/j.1540-6261.1974.tb03058.x}{The
  Journal of Finance {\bf 29}, 449--470}~(1974).

\bibitem{Qiskit}
``Qiskit: An open-source framework for quantum computing''~(2021).

\bibitem{HullWhite1994}
John~C Hull and Alan~D White.
\newblock ``Numerical procedures for implementing term structure models i''.
\newblock \href{https://dx.doi.org/10.3905/jod.1994.407902}{The Journal of
  Derivatives {\bf 2}, 7--16}~(1994).

\end{thebibliography}

\appendix

\onecolumn

\section{Qubits, gates, and quantum circuits}
  \label{app:qc}

\begin{figure*}[t]
  \centering
  \includegraphics{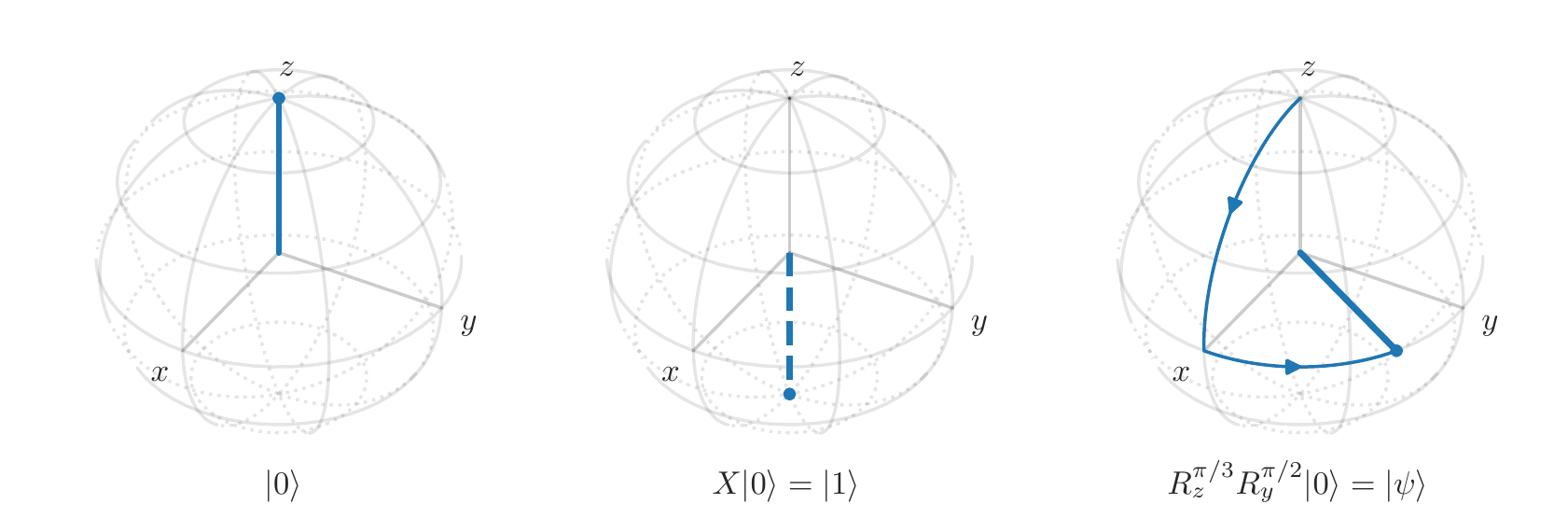}
  \caption{
    Left: the quantum state $\ket{0}$ on the Bloch sphere.
    Middle: the quantum state $\ket{1}$, which can be obtained by acting on
    $\ket{0}$ with the \textsc{not} operator, $X$.
    The dashed line indicates that the point is on the non-visible part of the
    sphere's surface.
    Right: the superposition
    $\ket{\psi} = \cos(\pi/4)\ket{0} + e^{i\pi/3}\sin(\pi/4)\ket{1}$ obtained by
    the operation of two rotation gates.}
  \label{fig:states}
\end{figure*}
Quantum computers consist of qubits, which are the equivalent of classical bits.
The quantum states $\ket{0}$ and $\ket{1}$, which represent the values of $0$
and $1$, are shown on the Bloch sphere in the left and middle panels of
Fig.~\ref{fig:states}, respectively.\footnote{
  $\ket{0}$ and $\ket{1}$ are orthonormal, i.e. $\langle 0|0\rangle = 1$,
  $\langle 1|1\rangle = 1$, and $\langle 0|1\rangle = \langle 1|0\rangle = 0$.}
The key property of qubits is that they can exist in any superposition of these
two states, i.e. $\ket{\psi} = a_0\ket{0} + a_1\ket{1}$, where $a_0$,
$a_1$ are complex numbers satisfying the condition $|a_0|^2 + |a_1|^2 = 1$.
When such a qubit is measured, it will give either the state $\ket{0}$ or
$\ket{1}$ with probabilities $a_0^2$ and $a_1^2$, respectively.\footnote{
  Measurement collapses the quantum state to a single state, destroying the
  superposition.
  If the qubit were to be measured again, it would be found in the same state.}
Due to the probabilistic nature of the measured outcome, repeated experiments
--- called shots --- are often needed to improve the precision of a result.
Quantum states of single qubits can be conveniently written using the
real-valued parameters $\theta\in[0,\pi]$ and $\phi\in[0,2\pi]$:
\begin{equation}
  \ket{\psi} = \cos(\theta/2)\ket{0} + e^{i\phi}\sin(\theta/2)\ket{1}\,,
\label{eq:one_qubit_state}
\end{equation}
where $\theta$ determines the probability of each state and $\phi$ is the
relative phase (we have ignored the global phase).

Quantum circuits are composed of quantum gates operating on qubits.
Because qubits can be put in a superposition of states, quantum gates process
multiple states simultaneously, when classical gates can only process one at a
time.\footnote{
  This is one of the key properties that can make quantum computers more
  efficient than classical ones.}
Single-qubit gates modify the state of a qubit --- for instance, see the middle
panel of Fig.~\ref{fig:states} that shows the $X$, or \textsc{not}, operator
flipping $\ket{0}$ to output $\ket{1}$.
Generally, single-qubit gates modify the parameters $\theta$ and $\phi$.
An example is shown in the right panel of Fig.~\ref{fig:states}, where the
operator $R_y^{\pi/2}$ rotates the qubit $\ket{0}$ by $\theta = \pi$ around the
$y$-axis, and then $R_z^{\pi/3}$ rotates it by $\phi = \pi/3$ around the
$z$-axis.
The quantum circuits for the three qubits of Fig.~\ref{fig:states} can be
written as
\begin{center}
\begin{quantikz}
\lstick{$\ket{0}$} & \qw & \qw & \meter{} \arrow[r] & \rstick{$\ket{0}$} \\
\lstick{$\ket{0}$} & \gate{X} & \qw & \meter{} \arrow[r]
  & \rstick{$\ket{1}$}\qw \\
\lstick{$\ket{0}$} & \gate{R_y^{\pi/2}} & \gate{R_z^{\pi/3}}
  & \meter{} \arrow[r] & \rstick{$\ket{0}$ or $\ket{1}$}\qw
\end{quantikz}
\end{center}
where the third qubit has a $50\%$ chance to be found in either the $\ket{0}$ or
$\ket{1}$ state.

\begin{figure*}[t]
  \centering
  \includegraphics{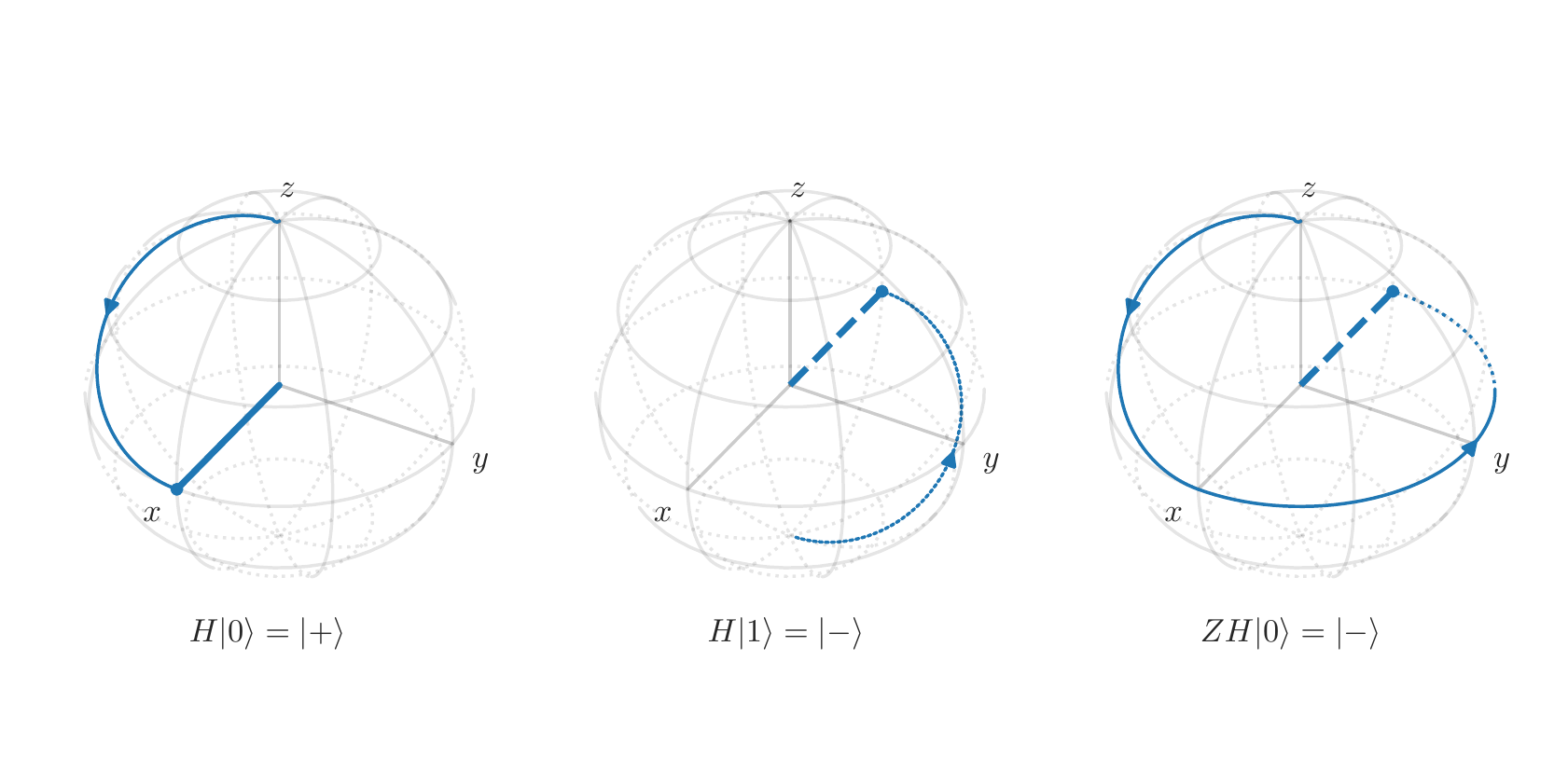}
  \caption{
    Left and middle: the quantum states $\ket{+} = H\ket{0}$ and
    $\ket{-} = H\ket{1}$, where $H$ is the Hadamard gate rotating the qubits by
    $\pi$ along the $x=z$ axis.
    Right: the quantum state $\ket{-} = ZH\ket{0}$, where $Z$ rotates the qubit 
    by $\pi$ along the $z$-axis, changing its state from $\ket{+}$ to
    $\ket{-}$.}
  \label{fig:fourier_basis}
\end{figure*}
Two-qubit gates leverage quantum entanglement, a property that binds the states
of two qubits together, and may alter the qubits depending on their initial
states.
This mechanism allows to implement logical operators similar to an ``if''
statement, e.g. the \textsc{cnot} (controlled \textsc{not}) gate below:
\begin{center}
\begin{quantikz}
\lstick{$\ket{0}$} & \ctrl{1} & \rstick{$\ket{0}$}\qw \\
\lstick{$\ket{0}$} & \targ{}  & \rstick{$\ket{0}$}\qw
\end{quantikz}
\begin{quantikz}
\lstick{$\ket{1}$} & \ctrl{1} & \rstick{$\ket{1}$}\qw \\
\lstick{$\ket{0}$} & \targ{}  & \rstick{$\ket{1}$}\qw
\end{quantikz}
\begin{quantikz}
\lstick{$\ket{0}$} & \ctrl{1} & \rstick{$\ket{0}$}\qw \\
\lstick{$\ket{1}$} & \targ{}  & \rstick{$\ket{1}$}\qw
\end{quantikz}
\begin{quantikz}
\lstick{$\ket{1}$} & \ctrl{1} & \rstick{$\ket{1}$}\qw \\
\lstick{$\ket{1}$} & \targ{}  & \rstick{$\ket{0}$}\qw
\end{quantikz}
\end{center}
When the ``control'' qubit (top) is in the state $\ket{1}$, the \textsc{cnot}
gate flips the ``target'' qubit (bottom).
If the ``control'' qubit is in the state $\ket{0}$, the ``target'' qubit remains
unchanged.
Another example with the \textsc{cnot} gate is the ``phase kickback'' mechanism,
where the phase of the target is ``kicked back'' to the controller, e.g.
\begin{center}
\begin{quantikz}
\lstick{$\ket{+}$} & \ctrl{1} & \rstick{$\ket{+}$}\qw \\
\lstick{$\ket{+}$} & \targ{}  & \rstick{$\ket{+}$}\qw
\end{quantikz}
\begin{quantikz}
\lstick{$\ket{+}$} & \ctrl{1} & \rstick{$\ket{-}$}\qw \\
\lstick{$\ket{-}$} & \targ{}  & \rstick{$\ket{-}$}\qw
\end{quantikz}
\begin{quantikz}
\lstick{$\ket{-}$} & \ctrl{1} & \rstick{$\ket{-}$}\qw \\
\lstick{$\ket{+}$} & \targ{}  & \rstick{$\ket{+}$}\qw
\end{quantikz}
\begin{quantikz}
\lstick{$\ket{-}$} & \ctrl{1} & \rstick{$\ket{+}$}\qw \\
\lstick{$\ket{-}$} & \targ{}  & \rstick{$\ket{-}$}\qw
\end{quantikz}
\end{center}
where $\ket{+}$ and $\ket{-}$ is another basis, with the qubits along the
$x$-axis, see Fig.~\ref{fig:fourier_basis}.\footnote{
  The relation of the two bases is:
  $\ket{+} = \frac{1}{\sqrt{2}}(\ket{0} + \ket{1})$ and
  $\ket{-} = \frac{1}{\sqrt{2}}(\ket{0} - \ket{1})$.}

The \textsc{cnot} gate is a building block of the three-qubit \textsc{and} gate
(also known as Toffoli or \textsc{ccnot} gate):
\begin{center}
\begin{quantikz}
\lstick{$\ket{0}$}
  & \ctrl{1}\gategroup[3,steps=1,
      style={dashed, rounded corners}]{AND}
  & \rstick{$\ket{0}$}\qw \\
\lstick{$\ket{0}$}
  & \ctrl{1}
  & \rstick{$\ket{0}$}\qw \\
\lstick{$\ket{0}$}
  & \targ{}
  & \rstick{$\ket{0}$}\qw
\end{quantikz}
\begin{quantikz}
\lstick{$\ket{1}$}
  & \gate[3]{\mathrm{AND}}
  & \rstick{$\ket{1}$}\qw \\
\lstick{$\ket{0}$}
  & \qw
  & \rstick{$\ket{0}$}\qw \\
\lstick{$\ket{0}$}
  & \qw
  & \rstick{$\ket{0}$}\qw
\end{quantikz}
\begin{quantikz}
\lstick{$\ket{0}$}
  & \gate[3]{\mathrm{AND}}
  & \rstick{$\ket{0}$}\qw \\
\lstick{$\ket{1}$}
  & \qw
  & \rstick{$\ket{1}$}\qw \\
\lstick{$\ket{0}$}
  & \targ{}
  & \rstick{$\ket{0}$}\qw
\end{quantikz}
\begin{quantikz}
\lstick{$\ket{1}$}
  & \gate[3]{\mathrm{AND}}
  & \rstick{$\ket{1}$}\qw \\
\lstick{$\ket{1}$}
  & \ctrl{1}
  & \rstick{$\ket{1}$}\qw \\
\lstick{$\ket{0}$}
  & \targ{}
  & \rstick{$\ket{1}$}\qw
\end{quantikz}
\end{center}
With the help of the \textsc{not} and \textsc{and} gates we can build the logic
for an \textsc{or} gate:
\begin{center}
\begin{quantikz}[column sep=0.3cm]
\lstick{$\ket{0}$}
    & \qw\gategroup[3, steps=6,
      style={dashed, rounded corners, inner xsep=0}]{OR}
    & \ctrl{1}
    & \gate{X}
    & \ctrl{1}
    & \gate{X}
    & \ctrl{1}
    & \rstick{$\ket{0}$}\qw \\
\lstick{$\ket{0}$}
    & \gate{X}
    & \ctrl{1}
    & \gate{X}
    & \ctrl{1}
    & \qw
    & \ctrl{1}
    & \rstick{$\ket{0}$}\qw \\
\lstick{$\ket{0}$}
    & \qw
    & \targ{}
    & \qw
    & \targ{}
    & \qw
    & \targ{}
    & \rstick{$\ket{0}$}\qw
\end{quantikz}
\begin{quantikz}
\lstick{$\ket{1}$}
    & \gate[3]{\mathrm{OR}}
    & \rstick{$\ket{1}$}\qw \\
\lstick{$\ket{0}$}
    & \qw
    & \rstick{$\ket{0}$}\qw \\
\lstick{$\ket{0}$}
    & \qw
    & \rstick{$\ket{1}$}\qw
\end{quantikz}
\begin{quantikz}
\lstick{$\ket{0}$}
    & \gate[3]{\mathrm{OR}}
    & \rstick{$\ket{0}$}\qw \\
\lstick{$\ket{1}$}
    & \qw
    & \rstick{$\ket{1}$}\qw \\
\lstick{$\ket{0}$}
    & \qw
    & \rstick{$\ket{1}$}\qw
\end{quantikz}
\begin{quantikz}
\lstick{$\ket{1}$}
    & \gate[3]{\mathrm{OR}}
    & \rstick{$\ket{1}$}\qw \\
\lstick{$\ket{1}$}
    & \qw
    & \rstick{$\ket{1}$}\qw \\
\lstick{$\ket{0}$}
    & \qw
    & \rstick{$\ket{1}$}\qw
\end{quantikz}
\end{center}
Other multi-qubit gates can be assembled as sub-circuits by combining basic
two-qubit gates, however, their depth is currently limited by hardware noise.

\section{Calculations}

In the below we drop the subscript ``in'' for brevity.

\subsection{Calculation of $\mathcal{Q}\ket{\psi}$}
  \label{app:Qpsi}

From Eq.~\eqref{eq:psi_in} notice that:
\begin{align}
\langle\psi_0|\psi_1\rangle &= 
  \langle\psi_0|\psi_1\rangle_\mathrm{rf}\langle0|1\rangle_\mathrm{rm} = 0\\
\langle\psi_1|\psi_0\rangle &= 
  \langle\psi_1|\psi_0\rangle_\mathrm{rf}\langle1|0\rangle_\mathrm{rm} = 0
\end{align}
and also
\begin{align}
\langle\psi|\psi\rangle &= 1 \nonumber\\
(1-p)\langle\psi_0|\psi_0\rangle
  + p\langle\psi_1|\psi_1\rangle &= 1 \nonumber\\
\langle\psi_0|\psi_0\rangle
  + p\left(\langle\psi_1|\psi_1\rangle
  - \langle\psi_0|\psi_0\rangle\right) &= 1
\end{align}
which, because it should hold for any $p$, implies:
\begin{align}
\langle\psi_0|\psi_0\rangle
  &= \langle\psi_1|\psi_1\rangle
  = 1
\end{align}
Applying $\mathcal{Q}$ (Eq.~\ref{eq:Q}) on $\ket{\psi}$ (Eq.~\ref{eq:psi_in})
gives:
\begin{align}
\mathcal{Q}\ket{\psi}
  &= Q_\psi
    \big(\openone - 2\ket{\psi_0}\bra{\psi_0}\big)
    \big(\sqrt{1-p}\ket{\psi_0} + \sqrt{p}\ket{\psi_1}\big)
  \nonumber\\
  &= \left[\openone - 2(1-p)\ket{\psi_0}\bra{\psi_0}
    - 2\sqrt{p(1-p)}\ket{\psi_0}\bra{\psi_1}
    - 2\sqrt{p(1-p)}\ket{\psi_1}\bra{\psi_0}
    - 2p\ket{\psi_1}\bra{\psi_1}\right] \nonumber\\
    &\qquad\times\left(-\sqrt{1-p}\ket{\psi_0} + \sqrt{p}\ket{\psi_1}\right)
  \nonumber\\
  &= -\sqrt{1-p}\ket{\psi_0} + \sqrt{p}\ket{\psi_1} 
    + 2(1-p)\sqrt{1-p}\ket{\psi_0} - 2p\sqrt{1-p}\ket{\psi_0}\nonumber\\
  &\qquad+ 2(1-p)\sqrt{p}\ket{\psi_1} - 2p\sqrt{p}\ket{\psi_1} \nonumber\\
  &=(1-4p)\sqrt{1-p}\ket{\psi_0} + (3-4p)\sqrt{p}\ket{\psi_1}\,.
\end{align}

\subsection{Calculation of $\mathcal{Q}\ket{\psi_\pm}$}
  \label{app:Qpsi_pm}

\begin{align}
\mathcal{Q}\ket{\psi_\pm}
  &= \frac{1}{\sqrt{2}}\mathcal{Q}_\psi
    \big(\openone - 2\ket{\psi_0}\bra{\psi_0}\big)
    \big(\ket{\psi_1} \pm i\ket{\psi_0}\big)
  \nonumber\\
  &= \frac{1}{\sqrt{2}}
    \left[\openone - 2(1-p)\ket{\psi_0}\bra{\psi_0}
      - 2\sqrt{p(1-p)}\ket{\psi_0}\bra{\psi_1}
    - 2\sqrt{p(1-p)}\ket{\psi_1}\bra{\psi_0} - 2p\ket{\psi_1}\bra{\psi_1}\right]
    \nonumber\\
  &\qquad\times\big(\ket{\psi_1} \mp i\ket{\psi_0}\big)
  \nonumber\\
  &= \frac{1}{\sqrt{2}}\left[\ket{\psi_1} \mp i\ket{\psi_0}
    \pm i2(1-p)\ket{\psi_0} - 2\sqrt{p(1-p)}\ket{\psi_0}
    \pm i2\sqrt{p(1-p)}\ket{\psi_1} - 2p\ket{\psi_1}\right] \nonumber\\
  &= \frac{1}{\sqrt{2}}
    \left[\pm i(1-2p)\ket{\psi_0} - 2\sqrt{p(1-p)}\ket{\psi_0}
    + (1 - 2p)\ket{\psi_1} \pm i2\sqrt{p(1-p)}\ket{\psi_1}\right]\,.
\end{align}
But
\begin{align}
1 - 2p &= 1 - 2\sin^2(\theta/2) = \cos\theta\,, \\
2\sqrt{p(1-p)} &= 2\sin(\theta/2)\cos(\theta/2) = \sin\theta\,.
\end{align}
Therefore,
\begin{align}
\mathcal{Q}\ket{\psi_\pm}
  &= \frac{1}{\sqrt{2}}\big(\pm i\cos\theta\ket{\psi_0} - \sin\theta\ket{\psi_0}
    + \cos\theta\ket{\psi_1} \pm i\sin\theta\ket{\psi_1}\big) \nonumber\\
  &= \frac{1}{\sqrt{2}}\cos\theta\big(\ket{\psi_1} \pm i\ket{\psi_0}\big)
    + \frac{1}{\sqrt{2}}\sin\theta\big(\pm i\ket{\psi_1} - \ket{\psi_0}\big) 
    \nonumber\\
  &= \cos\theta\ket{\psi_\pm} \pm i\sin\theta\ket{\psi_\pm} \nonumber\\
  &= e^{\pm i\theta}\ket{\psi_\pm}\,,
\end{align}
which implies that:
\begin{align}
\mathcal{Q}^k\ket{\psi_\pm}
  &= e^{\pm ik\theta}\ket{\psi_\pm}\,.
\end{align}

\subsection{Expressing $\ket{\psi}$ as a function of $\ket{\psi_\pm}$}
  \label{app:psi_psi_pm}

We can write $\ket{\psi_0}$ and $\ket{\psi_1}$ as functions of $\ket{\psi_\pm}$:
\begin{align}
\ket{\psi_0} &= -i\frac{1}{\sqrt{2}}\big(\ket{\psi_+} - \ket{\psi_-}\big)\,, \\
\ket{\psi_1} &= \frac{1}{\sqrt{2}}\big(\ket{\psi_+} + \ket{\psi_-}\big)\,.
\end{align}
Therefore,
\begin{align}
\ket{\psi}
  &= \sqrt{1-p}\ket{\psi_0} + \sqrt{p}\ket{\psi_1} \nonumber\\
  &= \frac{1}{\sqrt{2}}\sqrt{p}\big(\ket{\psi_+} + \ket{\psi_-}\big)
    -i\frac{1}{\sqrt{2}}\sqrt{1-p}\big(\ket{\psi_+} - \ket{\psi_-}\big)
    \nonumber\\
  &= -i\frac{1}{\sqrt{2}}
    \left[i\sin(\theta/2)\big(\ket{\psi_+} + \ket{\psi_-}\big)
    + \cos(\theta/2)\big(\ket{\psi_+} - \ket{\psi_-}\big)\right] \nonumber\\
  &= -i\frac{1}{\sqrt{2}}
    \left[\Big(\cos(\theta/2) + i\sin(\theta/2)\Big)\ket{\psi_+}
      - \Big(\cos(\theta/2) - i\sin(\theta/2)\Big)\ket{\psi_-}\right]\nonumber\\
  &= -i\frac{1}{\sqrt{2}}
    \big(e^{i\theta/2}\ket{\psi_+} - e^{-i\theta/2}\ket{\psi_-}\big)\,.
\end{align}

\subsection{The output qubits after the controlled gate $\mathcal{Q}$}
  \label{app:phase_kickback}

We set the output qubits --- which we label here with $l = 0, 1, ..., n-1$ ---
as controls to the gate $\mathcal{Q}$, and apply the gate $2^l$ times for each
qubit, respectively.
Their output state is:

\begin{align}
\prod_{l=0}^{n-1}\mathcal{Q}^{2^l}
  \ket{\psi}\ket{+}_\mathrm{out}^{\otimes n}
&=\ket{\psi}\bigotimes_{l=0}^{n-1}
  \left[\frac{1}{\sqrt{2}}
    \left(\ket{0} + e^{\pm i2^l\theta}\ket{1}\right)\right] \nonumber\\
&=\ket{\psi}\frac{1}{2^{n/2}}\bigotimes_{l=0}^{n-1}
  \left(e^{i2^l\theta\cdot0}\ket{0} + e^{\pm i2^l\theta\cdot1}\ket{1}\right)
  \nonumber\\
&=\ket{\psi}\frac{1}{2^{n/2}}\bigotimes_{l=0}^{n-1}
  \left(\sum_{b_l=0}^1e^{\pm i2^lb_l\theta}\ket{b_k}\right) \nonumber\\
&=\ket{\psi}\frac{1}{2^{n/2}}\sum_{b_0=0}^1\sum_{b_1=0}^1\cdots\sum_{b_{n-1}=0}^1
  \prod_{l=0}^{n-1}e^{\pm i2^lb_l\theta}\ket{b_0}\ket{b_1}...\ket{b_{n-1}}
  \nonumber\\
&=\ket{\psi}\frac{1}{2^{n/2}}\sum_{b_0=0}^1\cdots\sum_{b_{n-1}=0}^1
  e^{\pm\sum_{l=0}^{n-1}i2^lb_l\theta}\ket{b_0...b_{n-1}}
  \nonumber\\
&=\ket{\psi}\frac{1}{2^{n/2}}\sum_{x=0}^{2^n-1}e^{\pm ix\theta}\ket{x}\,.
\end{align}

\subsection{Inverse Quantum Fourier Transform}
  \label{app:QFT}

The inverse quantum Fourier transform of the state of the output qubits gives:
\begin{align}
\ket{\psi}_\mathrm{out}
&= \mathrm{QFT}^\dagger\left[
  \frac{1}{2^{n/2}}\sum_{x=0}^{2^n-1}e^{\pm ix\theta}\ket{x}\right]
  \nonumber\\
&=\frac{1}{2^{n/2}}\sum_{x=0}^{2^n-1}e^{\pm ix\theta}
  \mathrm{QFT}^\dagger\ket{x} \nonumber\\
&=\frac{1}{2^{n/2}}
  \sum_{x=0}^{2^n-1}e^{\pm ix\theta}\left(
    \frac{1}{2^{n/2}}\sum_{z=0}^{2^n-1}e^{-i2\pi xz/2^n}\ket{z}\right)
  \nonumber\\
&=\frac{1}{2^n}
  \sum_{z=0}^{2^n-1}\sum_{x=0}^{2^n-1}e^{ix(\pm\theta-2\pi z/2^n)}\ket{z}
  \nonumber\\
&=\sum_{z=0}^{2^n-1}a_z\ket{z}\,.
\end{align}

\subsection{The amplitudes $a_z$ when $2^n\theta/2\pi$ or
  $2^n(2\pi - \theta)/2\pi$ are integers}
  \label{app:az}

In the special case where either $2^n\theta/2\pi$ or $2^n(2\pi - \theta)/2\pi$
is an integer equal to $z_0$, the amplitude $a_{z_0}$ is:
\begin{align}
a_{z_0} &= \frac{1}{2^n}\sum_{x=0}^{2^n-1}1 = 1\,,
\end{align}
and for all other $a_z$:
\begin{align}
a_z
&= \frac{1}{2^n}\sum_{x=0}^{2^n-1}e^{ix(2\pi z_0/2^n-2\pi z/2^n)} \nonumber\\
&= \frac{1}{2^n}\sum_{x=0}^{2^n-1}\left(e^{i2\pi(z_0-z)/2^n}\right)^x\nonumber\\
&= \frac{1}{2^n}\left[\frac{1 - \left(e^{i2\pi(z_0-z)/2^n}\right)^{2^n}}
  {1 - e^{i2\pi(z_0-z)/2^n}}\right] \nonumber\\
&= \frac{1}{2^n}\left(\frac{1 - e^{i2\pi(z_0-z)}}
  {1 - e^{i2\pi(z_0-z)/2^n}}\right) \nonumber\\
&= 0
\end{align}
where $e^{i2\pi(z_0-z)} = 1$ because $z_0 - z$ is an integer.

\subsection{The general case for the amplitudes $|a_z|$}
  \label{app:az2}

If neither $2^n\theta/2\pi$ nor $2^n(2\pi - \theta)/2\pi$ is an integer, we can
find the closest integer such that $z_0 + \epsilon = 2^n\theta/2\pi$ or
$z_0 + \epsilon = 2^n(2\pi - \theta)/2\pi$, where $\epsilon\in(0,1/2]$.
Namely,
\begin{align}
a_z
&= \frac{1}{2^n}\sum_{x=0}^{2^n-1}
  e^{ix(2\pi z_0/2^n-2\pi z/2^n + 2\pi\epsilon/2^n)} \nonumber\\
&= \frac{1}{2^n}\sum_{x=0}^{2^n-1}
  \left(e^{i2\pi(z_0-z+\epsilon)/2^n}\right)^x\nonumber\\
&= \frac{1}{2^n}\left[
  \frac{1 - \left(e^{i2\pi(z_0-z+\epsilon)/2^n}\right)^{2^n}}
  {1 - e^{i2\pi(z_0-z+\epsilon)/2^n}}\right] \nonumber\\
&= \frac{1}{2^n}\left(\frac{1 - e^{i2\pi(z_0-z)}e^{i2\pi\epsilon}}
  {1 - e^{i2\pi(z_0-z+\epsilon)/2^n}}\right) \nonumber\\
&= \frac{1}{2^n}\left(\frac{1 - e^{i2\pi\epsilon}}
  {1 - e^{i2\pi(z_0-z+\epsilon)/2^n}}\right)
\end{align}
The probability $|a_z|^2$ is
\begin{align}
|a_z|^2
&= \frac{1}{2^{2n}}
  \left(\frac{1 - e^{i2\pi\epsilon}}{1 - e^{i2\pi(z_0-z+\epsilon)/2^n}}\right)
  \left(\frac{1 - e^{-i2\pi\epsilon}}{1 - e^{-i2\pi(z_0-z+\epsilon)/2^n}}\right)
  \nonumber\\
&= \frac{1}{2^{2n}}
  \left(\frac{2 - e^{i2\pi\epsilon} - e^{-i2\pi\epsilon}}
  {2 - e^{i2\pi(z_0-z+\epsilon)/2^n} - e^{-i2\pi(z_0-z+\epsilon)/2^n}}\right)
  \nonumber\\
&= \frac{1}{2^{2n}}
  \left(\frac{1 - \cos(2\pi\epsilon)}
  {1 - \cos[2\pi(z_0-z+\epsilon)/2^n]}\right)\,,
\end{align}
and, for large $n$, the probability of measuring the closest integer ($z = z_0$)
is:
\begin{align}
|a_{z_0}|^2
&= \frac{1}{2^{2n}}
  \left(\frac{1 - \cos(2\pi\epsilon)}
  {1 - \cos(2\pi\epsilon/2^n)}\right) \nonumber\\
&\simeq \frac{1}{2^{2n}}
  \left[1 - \left(1 - \frac{(2\pi\epsilon)^2}{2!}
    + \frac{(2\pi\epsilon)^4}{4!}\right)\right]
  \left[1 - \left(1 - \frac{(2\pi\epsilon)^2}{2^{2n}2!}\right)\right]^{-1}
  \nonumber\\
&= \frac{1}{2^{2n}}
  \left[\frac{(2\pi\epsilon)^2}{2!} - \frac{(2\pi\epsilon)^4}{4!}\right]
  \left[\frac{2^{2n}2!}{(2\pi\epsilon)^2}\right]
  \nonumber\\
&= 1 - \frac{\pi^2\epsilon^2}{3}
  \nonumber\\
&\geq 1 - \frac{\pi^2}{12}\,,
\end{align}
where in the last expression we took the maximum value of $\epsilon = 1/2$.

\subsection{Calculation of $\delta p$ for QAE}
  \label{app:dp}

The error in $p$ can be estimated as follows:
\begin{align}
\delta p
&= \delta\sin^2\left(\frac{\theta}{2}\right) \nonumber\\
&\simeq 2\sin\left(\frac{\theta}{2}\right)
  \cos\left(\frac{\theta}{2}\right)\frac{\delta\theta}{2} \nonumber\\
&= \sin\theta\frac{\pi}{2^n}\,,
\end{align}
because $\delta\theta = 2\pi/2^n$.

\section{Equity risk factor calculations}

\subsection{Binomial tree parameters}
  \label{app:binomial_tree}

By matching the mean of the continuous and discrete models we obtain the value
of $q$:
\begin{align}
E(S_{t+\delta t})
  &= qS_{t+\delta t}^\mathrm{u} + (1-q)S_{t+\delta t}^\mathrm{d} \nonumber\\
S_t e^{\mu \delta t}
  &= qS_tu + (1-q)S_td \nonumber\\
q &= \frac{e^{\mu \delta t} - d}{u - d} \nonumber\\
&= \frac{ue^{\mu \delta t} - 1}{u^2 - 1}
\end{align}
And from the variance, we obtain the value of $u$:
\begin{align}
V\!ar(S_{t+\delta t})
  &= E(S_{t+\delta t}^2) - E(S_{t+\delta t})^2 \nonumber\\
S_t^2e^{2\mu \delta t + \sigma^2\delta t} - S_t^2e^{2\mu \delta t}
  &= q(S_t^2u^2) + (1-q)(S_t^2d^2)
    - S_t^2e^{2\mu \delta t} \nonumber\\
e^{2\mu \delta t + \sigma^2\delta t}
&= qu^2 + (1-q)d^2 \nonumber\\
&= u^2\frac{ue^{\mu \delta t} - 1}{u^2 - 1}
    + \frac{1}{u^2}\frac{u^2 - ue^{\mu \delta t}}{u^2 - 1} \nonumber\\
&= \frac{1}{u(u^2 - 1)}\left(u^4e^{\mu \delta t} - u^3
    + u - e^{\mu \delta t} \right) \nonumber\\
&= \frac{1}{u(u^2 - 1)}\left[e^{\mu \delta t}(u^4 - 1) - u(u^2 - 1)\right]
  \nonumber\\
&= \frac{1}{u}\left[e^{\mu \delta t}(u^2 + 1) - u\right] \,.
\end{align}
We consider small timesteps, $\sigma\sqrt{\delta t} \ll 1$ and
$\mu\delta t \ll 1$, ignore terms higher than $\mathcal{O}(\delta t)$, and try
the solution $u = e^{b\sqrt{\delta t}}$.
With Taylor expansion, the terms $e^{\mu\delta t + \sigma^2\delta t}$, $u$,
$u^2$, and $u^{-1}$ are:
\begin{align}
e^{2\mu\delta t + \sigma^2\delta t}
&\simeq 1 + 2\mu\delta t + \sigma^2\delta t \\
u &\simeq 1 + b\sqrt{\delta t} + \frac{1}{2}b^2\delta t \\
u^2
&\simeq \left(1 + b\sqrt{\delta t} + \frac{1}{2}b^2\delta t\right)^2 \nonumber\\
&\simeq 1 + b\sqrt{\delta t} + \frac{1}{2}b^2\delta t
 + b\sqrt{\delta t} + b^2\delta t  + \frac{1}{2}b^2\delta t \nonumber\\
&= 1 + 2b\sqrt{\delta t} + 2b^2\delta t \\
u^{-1} &= e^{-b\sqrt{\delta t}} \nonumber\\
&\simeq 1 - b\sqrt{\delta t} + \frac{1}{2}b^2\delta t
\end{align}
Therefore,
\begin{align}
1 + 2\mu\delta t + \sigma^2\delta t
&\simeq
  \left(1 - b\sqrt{\delta t} + \frac{1}{2}b^2\delta t \right)
    \left[\left(1 + \mu \delta t\right)\left(
      1 + 2b\sqrt{\delta t} + 2b^2\delta t + 1\right)
  - \left(1 + b\sqrt{\delta t} + \frac{1}{2}b^2\delta t \right)
  \right] \nonumber\\
1 + 2\mu\delta t + \sigma^2\delta t
&\simeq
  \left(1 - b\sqrt{\delta t} + \frac{1}{2}b^2\delta t \right)
  \left(2 + 2b\sqrt{\delta t} + 2b^2\delta t + 2\mu\delta t
    - 1 - b\sqrt{\delta t} - \frac{1}{2}b^2\delta t \right) \nonumber\\
1 + 2\mu\delta t + \sigma^2\delta t
&\simeq
  \left(1 - b\sqrt{\delta t} + \frac{1}{2}b^2\delta t \right)
  \left(1 + b\sqrt{\delta t} + \frac{3}{2}b^2\delta t + 2\mu\delta t\right)
  \nonumber\\
1 + 2\mu\delta t + \sigma^2\delta t
&\simeq
  1 + b\sqrt{\delta t} + \frac{3}{2}b^2\delta t + 2\mu\delta t
  - b\sqrt{\delta t} - b^2\delta t  + \frac{1}{2}b^2\delta t \nonumber\\
b &\simeq \sigma
\end{align}

\subsection{Decomposition of the $\mathcal{Q}_\psi$ operator}
  \label{app:q_decomp}

The operator that applies a reflection to the state $\ket{\psi}_\mathrm{in}$ can
be written as:
\begin{align}
\mathcal{Q}_\psi
&= \openone - 2\ket{\psi}_\mathrm{in}\bra{\psi}_\mathrm{in}
  \nonumber\\
&= \openone - 2\mathcal{M}\mathcal{D}\ket{0}_\mathrm{in}
    \bra{0}_\mathrm{in}\mathcal{D}^\dagger\mathcal{M}^\dagger
  \nonumber\\
&= \mathcal{M}\mathcal{D}\big(\openone - 2\ket{0}_\mathrm{in}
    \bra{0}_\mathrm{in}\big)\mathcal{D}^\dagger\mathcal{M}^\dagger
  \nonumber\\
&= \mathcal{M}\mathcal{D}Q_{00}\mathcal{D}^\dagger\mathcal{M}^\dagger\,.
\end{align}

\subsection{Example circuit with $m = 2$ and $n = 3$}
  \label{app:ex_equity}

Below is an example of the gates $\mathcal{D}_\mathrm{eq}$,
$\mathcal{M}_\mathrm{max}$, $\mathcal{Q}$, QFT, and QFT$^\dagger$ for a quantum
circuit that estimates the probability of measuring the maximum value,
$S_\mathrm{max}$.
The parameters adopted for the binomial tree are $q = 0.3827$, $\mu = 0$,
$\sigma = 0$, $T = 1$, and for the quantum circuit are $m = 2$ and $n = 3$.
\begin{center}
\begin{quantikz}[column sep=0.4cm]
\lstick{$\ket{0}_\mathrm{rf}^0\,\,\,$}
  & \gate{R_y^{\theta_\mathrm{u}}}
    \gategroup[5,steps=1,style={dotted, rounded corners,
      inner xsep=0,inner ysep=0.2pt}]{$\mathcal{D}$}
  & \ctrl{1}
    \gategroup[5,steps=1,style={dotted, rounded corners,
      inner xsep=0,inner ysep=0.2pt}]{$\mathcal{M}$}
  & \qw
    \gategroup[5,steps=3,style={dotted, rounded corners,
      inner xsep=0,inner ysep=0.2pt}]{$\mathcal{Q}_{\psi0}$}
    \gategroup[5,steps=14,style={dotted, rounded corners,
      inner xsep=2,inner ysep=16pt}]{$\mathcal{Q}$}
  & \qw
  & \qw
  & \ctrl{1}
    \gategroup[5,steps=1,style={dotted, rounded corners,
      inner xsep=0,inner ysep=0.2pt}]{$\mathcal{M}^\dagger$}
  & \gate{R_y^{-\theta_\mathrm{u}}}
    \gategroup[5,steps=1,style={dotted, rounded corners,
      inner xsep=0,inner ysep=0.2pt}]{$\mathcal{D}^\dagger$}
  & \gate{X}
    \gategroup[5,steps=7,style={dotted, rounded corners,
      inner xsep=0,inner ysep=0.2pt}]{$\mathcal{Q}_{00}$}
  & \ctrl{1}
  & \qw
  & \qw
  & \qw
  & \ctrl{1}
  & \gate{X}
  & \gate{R_y^{\theta_\mathrm{u}}}
    \gategroup[5,steps=1,style={dotted, rounded corners,
      inner xsep=0,inner ysep=0.2pt}]{$\mathcal{D}$}
  & \ctrl{1}
    \gategroup[5,steps=1,style={dotted, rounded corners,
      inner xsep=0,inner ysep=0.2pt}]{$\mathcal{M}$}
  & \qw \\
\lstick{$\ket{0}_\mathrm{rf}^1\,\,\,$}
  & \gate{R_y^{\theta_\mathrm{u}}}
  & \ctrl{1}
  & \qw
  & \qw
  & \qw
  & \ctrl{1}
  & \gate{R_y^{-\theta_\mathrm{u}}}
  & \gate{X}
  & \ctrl{2}
  & \qw
  & \qw
  & \qw
  & \ctrl{2}
  & \gate{X}
  & \gate{R_y^{\theta_\mathrm{u}}}
  & \ctrl{1}
  & \qw \\
\lstick{$\ket{0}_\mathrm{rm}\,$}
  & \qw
  & \targ{}
  & \gate{X}
  & \gate{Z}
  & \gate{X}
  & \targ{}
  & \qw
  & \gate{X}
  & \qw
  & \ctrl{1}
  & \gate{Z}
  & \ctrl{1}
  & \qw
  & \gate{X}
  & \qw
  & \targ{}
  & \qw \\
\lstick{$\ket{0}_\mathrm{anc}^1$}
  & \qw
  & \qw
  & \qw
  & \qw
  & \qw
  & \qw
  & \qw
  & \qw
  & \targ{}
  & \ctrl{1}
  & \qw
  & \ctrl{1}
  & \targ{}
  & \qw
  & \qw
  & \qw
  & \qw \\
\lstick{$\ket{0}_\mathrm{anc}^2$}
  & \qw
  & \qw
  & \qw
  & \qw
  & \qw
  & \qw
  & \qw
  & \qw
  & \qw
  & \targ{}
  & \ctrl{-2}
  & \targ{}
  & \qw
  & \qw
  & \qw
  & \qw
  & \qw
\end{quantikz}
\end{center}

\begin{center}
\begin{quantikz}[column sep=0.2cm]
\lstick{$\ket{0}_\mathrm{in}^{\otimes 5}$}
  & \gate{\mathcal{D}}\qwbundle[alternate]{}
  & \gate{\mathcal{M}}\qwbundle[alternate]{}
  & \gate{\mathcal{Q}}\qwbundle[alternate]{}
  & \gate{\mathcal{Q}}\qwbundle[alternate]{}
  & \gate{\mathcal{Q}}\qwbundle[alternate]{}
  & \gate{\mathcal{Q}}\qwbundle[alternate]{}
  & \gate{\mathcal{Q}}\qwbundle[alternate]{}
  & \gate{\mathcal{Q}}\qwbundle[alternate]{}
  & \gate{\mathcal{Q}}\qwbundle[alternate]{}
  & \qwbundle[alternate]{}
  & \qwbundle[alternate]{}
  & \qwbundle[alternate]{}
  & \qwbundle[alternate]{}
  & \qwbundle[alternate]{}
  & \qwbundle[alternate]{}
  & \qwbundle[alternate]{}
  & \qwbundle[alternate]{} \\
\lstick{$\ket{0}_\mathrm{out}^0$}
  & \qw
  & \gate{H}\gategroup[3,steps=1,
      style={dotted, rounded corners, inner xsep=0,inner ysep=0.2pt},
      label style={label position=below, anchor=north, yshift=-0.2cm}]{QFT}
  & \ctrl{-1}
  & \qw
  & \qw
  & \qw
  & \qw
  & \qw
  & \qw
  & \swap{2}\gategroup[3,steps=7,
      style={dotted, rounded corners, inner xsep=0,inner ysep=0.2pt},
      label style={label position=below, anchor=north, yshift=-0.2cm}]
      {QFT$^\dagger$}
  & \gate{H}
  & \ctrl{1}
  & \qw
  & \ctrl{2}
  & \qw
  & \qw
  & \meter[]{}\qw  \\
\lstick{$\ket{0}_\mathrm{out}^1$}
  & \qw
  & \gate{H}
  & \qw
  & \ctrl{-2}
  & \ctrl{-2}
  & \qw
  & \qw
  & \qw
  & \qw
  & \qw
  & \qw
  & \gate{R_z^{-\frac{\pi}{2}}}
  & \gate{H}
  & \qw
  & \ctrl{1}
  & \qw
  & \meter[]{}\qw \\
\lstick{$\ket{0}_\mathrm{out}^2$}
  & \qw
  & \gate{H}
  & \qw
  & \qw
  & \qw
  & \ctrl{-3}
  & \ctrl{-3}
  & \ctrl{-3}
  & \ctrl{-3}
  & \swap{}
  & \qw
  & \qw
  & \qw
  & \gate{R_z^{-\frac{\pi}{4}}}
  & \gate{R_z^{-\frac{\pi}{2}}}
  & \gate{H}
  & \meter[]{}\qw
\end{quantikz}
\end{center}

\begin{figure}[t]
  \centering
  \includegraphics{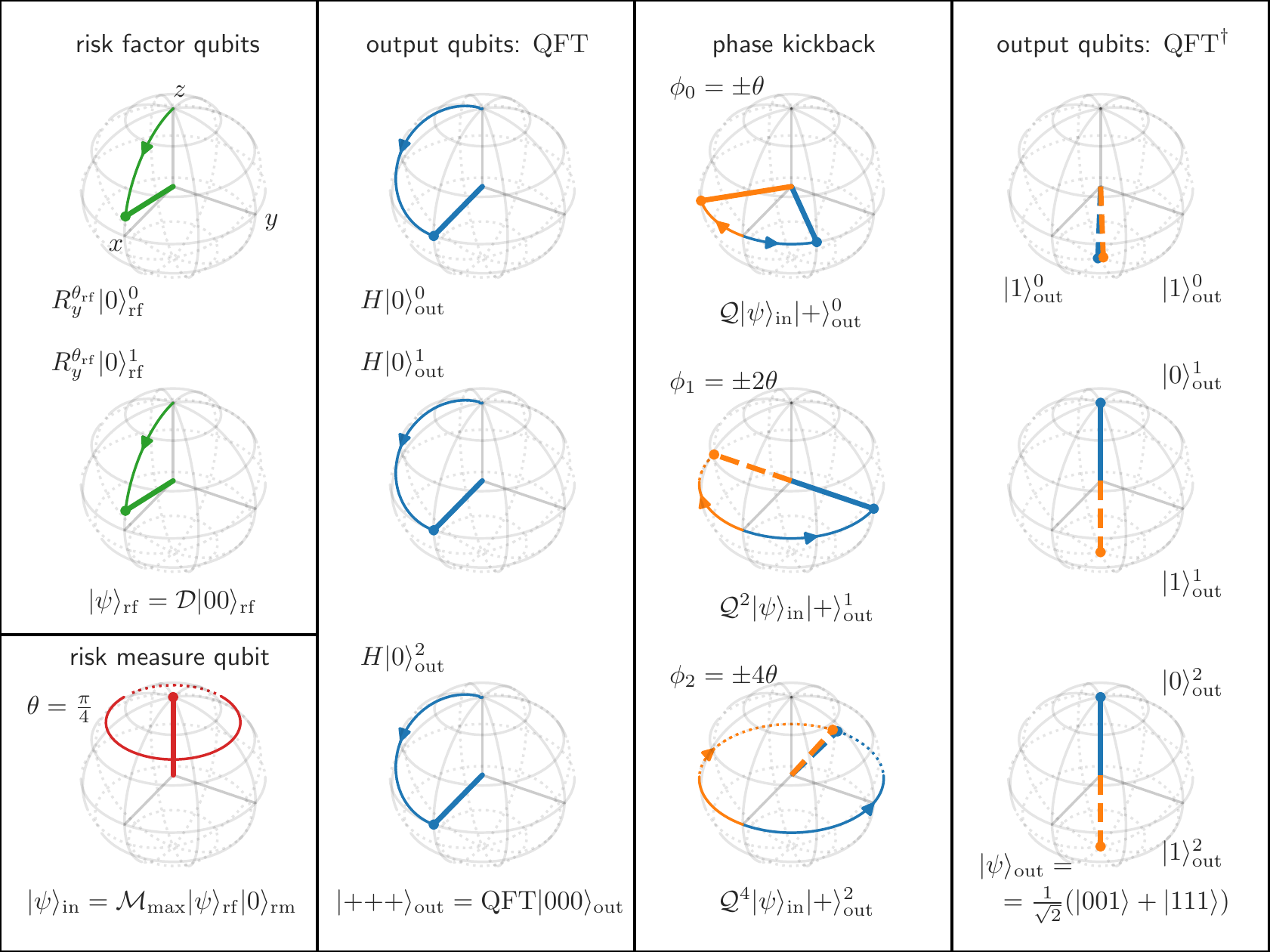}
  \caption{
    The qubit states before and after the operation of the quantum gates.
    Top left: gate $\mathcal{D}$ rotates the ``risk factor'' qubits (green)
    around the $y$-axis.
    Bottom left: the controlled gate $\mathcal{M}$ reads the ``risk factor''
    qubits and sets the ``risk measure'' qubit in superposition (red), encoding
    the probability that both ``risk factor'' qubits are in the state $\ket{1}$
   (since no rotation is involved, the phase of the ``risk measure'' qubit does
    not have a specific value and we draw the state as a circle).
    Left vertical panel: the output qubits (blue) before and after the
    application of the QFT gate, which consists of a Hadamard gate, $H$, applied
    to each qubit.
    Middle vertical panel: the controlled gate $\mathcal{Q}$ imprints the angle
    $\theta$ of the ``risk measure'' qubit as a positive ($+\phi$, blue) or
    negative ($-\phi$, orange) phase onto the first output qubit (top); the
    controlled gate $\mathcal{Q}^2$ imprints the angle $2\theta$ onto the phase
    of the second qubit (middle); the controlled gate $\mathcal{Q}^4$ imprints
    the angle $4\theta$ onto the phase of the third qubit (bottom).
    Right vertical panel: the inverse quantum Fourier transform gate,
    QFT$^\dagger$, leverages quantum interference to transform the phases of the
    output qubits to a binary number expressed with $\ket{0}$ and $\ket{1}$
    states.}
  \label{fig:eq_example_qubits}
\end{figure}
The transformation of the qubit states by the quantum gates is shown in
Fig.~\ref{fig:eq_example_qubits}.
Specifically, the ``risk factor'' qubits ($\ket{0}_\mathrm{rf}^0$ and
$\ket{0}_\mathrm{rf}^1$) before and after the operation of the gate
$\mathcal{D}$ are shown in green in the top left panel.
The application of gate $\mathcal{M}$ on the ``risk measure'' qubit
($\ket{0}_\mathrm{rm}$) is shown in red in the bottom left panel.
From left to right, the three vertical panels show the transformation of the
output qubits (in blue and orange; $\ket{0}_\mathrm{out}^0$,
$\ket{0}_\mathrm{out}^1$, and $\ket{0}_\mathrm{out}^2$) by the QFT,
$\prod\mathcal{Q}$, and QFT$^\dagger$ gates, respectively.

\begin{figure}[t]
  \centering
  \includegraphics[width=0.85\textwidth]{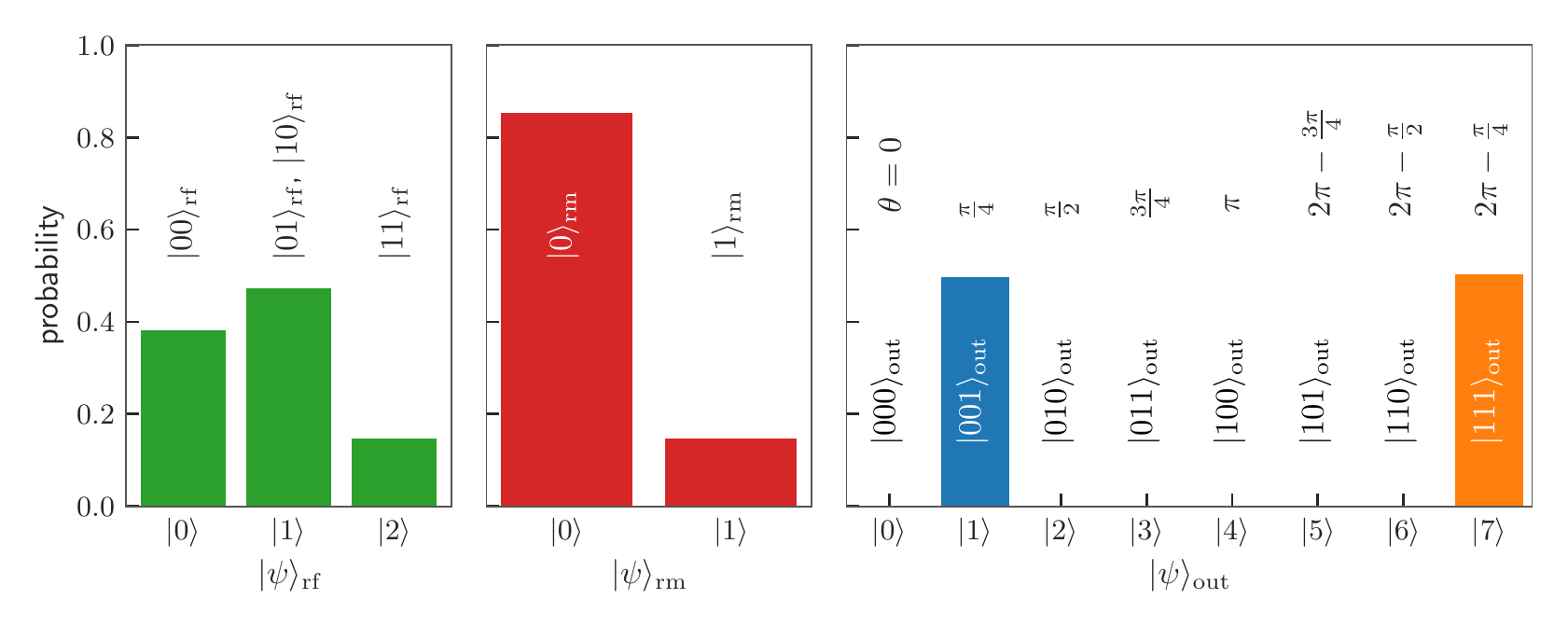}
  \caption{
    The probability distributions of the risk factor (left), risk measure
    (middle), and output qubits (right).}
  \label{fig:eq_example_results}
\end{figure}
Figure~\ref{fig:eq_example_results} shows the probability distribution of the
``risk factor'' (left, in green), ``risk measure'' (middle, in red), and output
(right, in blue/orange) qubits.
The distributions of the input qubits are only shown for illustrative purposes,
they are not measured in the circuit because this would collapse their states.
The measurement of the output qubit gives either the state $\ket{1}$ or
$\ket{7}$; from Eqs.~(\ref{eq:z0_1}) and (\ref{eq:z0_2}), these imply
\begin{align}
1 &= 2^3\frac{\theta}{2\pi} \\
7 &= 2^3\frac{2\pi - \theta}{2\pi}
\end{align}
both of which give $\theta = \pi/4$.
Therefore, from Eq.~(\ref{eq:psin2}) we have $p = \sin(\pi/8)^2 = 0.1464$, and
since $p = q^2$ (because we require two up moves), we validate the computation
by recovering the input value $q = 0.3827$.

\section{Interest rate risk factor calculations}
  \label{app:ir_calcs}

The sum of the probabilities satisfy: $q_{wu} + q_{wm} + q_{wd} = 1$, so
\begin{equation}
q_{wm} = 1 - q_{wu} - q_{wd}\,.
\end{equation}
The expected value of the Vasicek model should match that of the trinomial tree.
We can parametrise the three possible values of $r_t$ ($b - \delta r$, $b$, and
$b + \delta r$), as $r_t = b + c_t\delta r$, where $c_t = \{-1, 0, 1\}$.
Therefore,
\begin{align}
E(r_{t + \delta t})
  &= q_{wu}\left(b + \delta r\right)
  + q_{wm}b + q_{wd}\left(b - \delta r\right) \nonumber\\
(b + c_t\delta r)e^{-a\delta t} + b(1 - e^{-a\delta t})
  &= q_{wu}b + q_{wu}\delta r
  + q_{wm}b + q_{wd}b - q_{wd}\delta r \nonumber\\
b + c_te^{-a\delta t}\delta r
  &= b + (q_{wu} - q_{wd})\delta r \nonumber\\
q_{wd}
  &= q_{wu} - c_te^{-a\delta t}
\end{align}
Similarly, the variance of the continuous and discrete models should match:
\begin{align}
V\!ar(r_{t + \delta t})
  &= E(r_{t + \delta t}^2) - E(r_{t + \delta t})^2 \nonumber\\
\frac{\sigma^2}{2a}(1 - e^{-2a\delta t})
  &= q_{wu}\left(b + \delta r\right)^2
  + q_{wm}b^2 + q_{wd}\left(b - \delta r\right)^2
  - (b + c_te^{-a\delta t}\delta r)^2 \nonumber\\
\frac{\sigma^2}{2a}(1 - e^{-2a\delta t})
&= q_{wu}b^2 + 2q_{wu}b\delta r + q_{wu}\delta r^2
  + q_{wm}b^2 + q_{wd}b^2 - 2q_{wd}b\delta r + q_{wd}\delta r^2 \nonumber\\
&\quad - b^2 - 2bc_te^{-a\delta t}\delta r - c_t^2e^{-2a\delta t}\delta r^2
  \nonumber\\
\frac{\sigma^2}{2a}(1 - e^{-2a\delta t})
&= 2(q_{wu} - q_{wd})b\delta r + (q_{wu} + q_{wd})\delta r^2
  - 2bc_te^{-a\delta t}\delta r - c_t^2e^{-2a\delta t}\delta r^2
  \nonumber\\
\frac{\sigma^2}{2a}(1 - e^{-2a\delta t})
&= (q_{wu} + q_{wd} - c_t^2e^{-2a\delta t})\delta r^2
\end{align}
By setting $\delta r^2 = 3V\!ar(r_{t + \delta t})$ \cite[e.g.][]{HullWhite1994}
we obtain
\begin{align}
\frac{1}{3}
&= q_{wu} + q_{wd} - c_t^2e^{-2a\delta t} \nonumber\\
\frac{1}{3}
&= 2q_{wu} - c_te^{-a\delta t} - c_t^2e^{-2a\delta t} \nonumber\\
q_{wu}
&= \frac{1}{6} + \frac{1}{2}c_te^{-a\delta t}
  + \frac{1}{2}c_t^2e^{-2a\delta t} \nonumber\\
&\simeq \frac{1}{6}
  + \frac{1}{2}\left[c_t(1 - a\delta t) + c_t^2(1 -2a\delta t)\right] \,.
\end{align}
For $c_t = 0$,
\begin{align}
q_{mu} &= \frac{1}{6} \\
q_{mm} &= \frac{2}{3} \\
q_{md} &= q_{wu}
\end{align}
For $c_t = 1$,
\begin{align}
q_{uu}
&\simeq \frac{1}{6} + 1 - \frac{3}{2}a\delta t \nonumber\\
&\simeq \frac{7}{6} - \frac{3}{2}a\delta t \\
q_{ud}
&\simeq q_{wu} - 1 + a\delta t \nonumber\\
&\simeq \frac{1}{6} - \frac{1}{2}a\delta t \\
q_{um} &\simeq 1 - \frac{8}{6} + 2a\delta t \nonumber\\
&\simeq - \frac{1}{3} + 2a\delta t
\end{align}
And for $c_t = -1$,
\begin{align}
q_{du}
&\simeq \frac{1}{6} - \frac{1}{2}a\delta t \\
q_{dd}
&\simeq q_{wu} + 1 - a\delta t \nonumber\\
&\simeq \frac{7}{6} - \frac{3}{2}a\delta t \\
q_{dm} &\simeq 1 - \frac{8}{6} + 2a\delta t \nonumber\\
&\simeq - \frac{1}{3} + 2a\delta t
\end{align}

When $c_t = \pm 1$ we need to ensure positive probabilities, therefore:
\begin{align}
0 &< \frac{1}{6} - \frac{1}{2}a\delta t < 1 \\
0 &< \frac{7}{6} - \frac{3}{2}a\delta t < 1 \\
0 &< - \frac{1}{3} + 2a\delta t < 1
\end{align}
or
\begin{align}
-\frac{5}{6} &< a\delta t < \frac{1}{3} \\
\frac{1}{9} &< a\delta t < \frac{7}{9} \\
\frac{1}{6} &< a\delta t < \frac{2}{3}
\end{align}
which implies $\frac{1}{6} < a\delta t < \frac{2}{6}$.
We choose $\delta t = \frac{3}{12}a$.
For $c_t = 1$, this gives
\begin{align}
q_{uu} &\simeq \frac{19}{24} \\
q_{ud} &\simeq \frac{1}{24} \\
q_{um} &\simeq \frac{4}{24}
\end{align}
and for $c_t = -1$,
\begin{align}
q_{du} &\simeq \frac{1}{24} \\
q_{dd} &\simeq \frac{19}{24} \\
q_{dm} &\simeq \frac{4}{24}
\end{align}

\end{document}